\documentclass[11pt,letter,reqno]{amsart}
\usepackage[utf8]{inputenc}

\usepackage{mathtools}
\usepackage{amsthm,amssymb, amsmath, amsfonts, amssymb}
\usepackage{dsfont}
\usepackage{bm}
\usepackage{soul}

\usepackage[shortlabels]{enumitem}
\usepackage{hyperref}
\hypersetup{colorlinks = true, linkcolor = blue, citecolor = blue, urlcolor = blue}
\usepackage{cite}
\allowdisplaybreaks
\usepackage[usenames,dvipsnames]{xcolor}

\usepackage[margin=1in]{geometry}
\usepackage[textsize=small]{todonotes}

\usepackage{bookmark}

\usepackage{graphicx}
\usepackage{tikz}
\usepackage{tikz-cd}
\usetikzlibrary{positioning}
\usepackage{pgfplots}
\usepgfplotslibrary{external,fillbetween}

\usepackage[all]{xy}

\usepackage{cleveref}

\theoremstyle{plain}
\newtheorem{theorem}{Theorem}[section]
\newtheorem*{theorem*}{Theorem}
\newtheorem{conjecture}[theorem]{Conjecture}
\newtheorem{corollary}[theorem]{Corollary}
\newtheorem{lemma}[theorem]{Lemma}
\newtheorem{observation}[theorem]{Observation}
\newtheorem{proposition}[theorem]{Proposition}
\newtheorem{definition}[theorem]{Definition}
\newtheorem{claim}[theorem]{Claim}
\newtheorem*{claim*}{Claim}

\newtheorem{question}[theorem]{Question}
\newtheorem{condition}[theorem]{Condition}
\newtheorem{remark}[theorem]{Remark}

\newenvironment{poc}{\begin{proof}[Proof of Claim]}{\end{proof}}

\usepackage{thmtools}
\newtheorem{innercustomthm}{Theorem}
\newenvironment{customthm}[1]
  {\renewcommand\theinnercustomthm{#1}\innercustomthm}
  {\endinnercustomthm}
  
\newtheorem{innercustomlem}{Lemma}
\newenvironment{customlem}[1]{\renewcommand\theinnercustomlem{#1}\innercustomlem}
  {\endinnercustomlem}
  
\newtheorem{innercustomcor}{Corollary}
\newenvironment{customcor}[1]
{\renewcommand\theinnercustomcor{#1}\innercustomcor}
  {\endinnercustomcor}
  
\newtheorem{innercustomprop}{Proposition}
\newenvironment{customprop}[1]
{\renewcommand\theinnercustomprop{#1}\innercustomprop}
  {\endinnercustomprop}
  
\newtheorem{innercustomcond}{Condition}
\newenvironment{customcond}[1]
{\renewcommand\theinnercustomcond{#1}\innercustomcond}
  {\endinnercustomcond}

  \newtheorem{innercustomclaim}{Claim}
\newenvironment{customclaim}[1]
{\renewcommand\theinnercustomclaim{#1}\innercustomclaim}
  {\endinnercustomclaim}

\newcommand{\kawasaki}{\mathcal{K}}

\newcommand{\plusDownUp}{\mathcal{P}}
\newcommand{\artanh}{\mathrm{artanh}}

\newcommand{\E}{\mathbb{E}}
\newcommand{\Var}{\mathrm{Var}}
\newcommand{\R}{\mathbb{R}}
\newcommand{\N}{\mathbb{N}}
\newcommand{\Z}{\mathbb{Z}}
\newcommand{\C}{\mathbb{C}}
\newcommand{\ind}[1]{\mathds{1}_{#1}}
\newcommand{\Prob}{\mathbb{P}}

\def\gam{\gamma}
\def\lam{\lambda}
\def\sig{\sigma}
\def\al{\alpha}
\def\eps{\epsilon}

\newcommand{\tv}[1]{\|#1\|_{\mathrm{TV}}}
\newcommand{\size}[1]{\left\lvert #1 \right\rvert}
\newcommand{\absolute}[1]{\left\lvert #1 \right\rvert}
\newcommand{\gap}{\mathfrak{gap}}
\newcommand{\complex}{\mathcal{X}}

\def\tmix{\tau_{\mathrm{mix}}}

\def\G{\mathcal G}
\def\cG{{\bf G}}
\def\T{\mathbb T}
\newcommand{\bsig}{\boldsymbol{\sigma}}

\def\upeta{\eta_+}
\def\downeta{\eta_-}
\def\poly{\mathrm{poly}}

\title{Fast and Slow Mixing of the Kawasaki Dynamics on Bounded-Degree Graphs}

\author{Aiya Kuchukova}
\address{School of Mathematics \\ Georgia Institute of Technology}
\email{aiya.kuchukova@math.gatech.edu}

\author{Marcus Pappik}
\address{Hasso Plattner Institute \\ University of Potsdam}
\email{marcus.pappik@hpi.de}

\author{Will Perkins}
\address{School of Computer Science \\ Georgia Institute of Technology}
\email{math@willperkins.org}

\author{Corrine Yap}
\address{School of Mathematics \\ Georgia Institute of Technology}
\email{math@corrineyap.com}

\subjclass[2020]{Primary: 60J10, Secondary: 82B20, 68W20}

\begin{document}

\maketitle 

\begin{abstract}
    We study the worst-case mixing time of the global Kawasaki dynamics for the fixed-magnetization Ising model on the class of graphs of maximum degree $\Delta$.  Proving a conjecture of Carlson, Davies, Kolla, and Perkins, we show that below the tree uniqueness threshold, the Kawasaki dynamics mix rapidly for all magnetizations. Disproving a conjecture of Carlson, Davies, Kolla, and Perkins, we show that the regime of fast mixing does not extend throughout the regime of tractability for this model: there is a range of parameters for which there exist efficient sampling algorithms for the fixed-magnetization Ising model on max-degree $\Delta$ graphs, but the  Kawasaki dynamics can take exponential time to mix. Our techniques involve showing spectral independence in the fixed-magnetization Ising model and proving a sharp threshold for the existence of multiple metastable states in the Ising model with external field on random regular graphs.
\end{abstract}

\section{Introduction}
\label{sec:Intro}

The Ising model on a finite graph $G = (V, E)$ is  the following probability distribution on $\Omega = \{+1, -1\}^V$:
\begin{equation}   
\label{eqIsingDef}
\mu_{G, \beta, \lam}(\sig)= \frac{\lam^{|\sig|^+}e^{\beta m_G(\sig)}}{Z_G(\beta, \lam)}
\end{equation}
 where $|\sig|^+ = |\{\sig^{-1}(+1)\}|$ is the number of vertices assigned a $+1$ spin under $\sig$ which we call the {\em size} of $\sig$, and $m_G(\sig)$ is the number of monochromatic edges in $G$ under the $2$-coloring given by $\sig \in \Omega$. 
 The measure $\mu_{G, \beta, \lam}$ is called the {\em Gibbs measure} on $G$ with {\em inverse temperature}  $\beta \geq 0$ and {\em external field} $\lam \geq 0$. 
The normalizing constant $Z_G(\beta, \lam) = \sum_{\sig \in \Omega}\lam^{|\sig|^+} e^{\beta m_G(\sig)}$ is the {\em partition function} of the Ising model. Throughout this paper, we focus on the {\em ferromagnetic} case, $\beta \ge 0$, in which  agreeing spins on edges are preferred.

Spin models on graphs are the source of many interesting computational problems. Questions about the tractability of approximate counting (estimating the partition function) and approximate sampling (from the Gibbs distribution) are  studied extensively. 

In the case of the ferromagnetic Ising model, Jerrum and Sinclair~\cite{JS93} showed that there is a polynomial-time approximation algorithm on all graphs at all temperatures, and Randall and Wilson~\cite{randall1999sampling} gave an efficient sampling algorithm.  

In other cases, such as the anti-ferromagnetic Ising model ($\beta <0$) and the hard-core model of weighted independent sets, approximate counting and sampling can be computationally hard (e.g., no polynomial-time algorithm exists unless NP$=$RP).
For the class $\G_{\Delta}$ of graphs of maximum degree $\Delta$, these two models exhibit {\em computational thresholds}: as the activity or external field parameter $\lam$ varies, there is a sharp threshold between tractability (efficient approximate counting and sampling) and intractability (NP-hardness)~\cite{weitz2006counting,sly2010computational,sly2014counting,galanis2016inapproximability}. 
Moreover, the critical value $\lam_c = \lam_c(\Delta,\beta)$ is the phase transition point of the corresponding model on the infinite $\Delta$-regular tree $\mathbb T_\Delta$ (more precisely, it is the threshold for the {\em uniqueness of Gibbs measure} on $\T_\Delta$, a notion which we discuss shortly). 
Thus there is a remarkable connection between computational thresholds and statistical physics phase transitions.  Even further, the threshold $\lam_c$ has also been recently been shown to be a {\em dynamical threshold}: it is the threshold for rapid mixing of the Glauber dynamics,  a natural Markov chain for sampling from spin models like the Ising or hard-core models, on graphs in $\G_\Delta$~\cite{mossel2009hardness,weitz2006counting,anari2021spectral,chen2021optimal}.  So in these cases, three different thresholds (computational, dynamical, uniqueness on the tree) coincide. 

A very similar picture has emerged for the model of a uniformly random independent set of a given size. For the class of graphs $\G_\Delta$, there is a critical density $\alpha_c(\Delta)$ so that if $\alpha < \alpha_c$,  there are efficient algorithms to approximately count and sample independent sets of density $\alpha$, while if $\alpha>\alpha_c$ no such algorithms exist unless NP$=$RP~\cite{davies2023approximately}.  Jain, Michelen, Pham, and Vuong~\cite{JMPV23} recently proved that this computational threshold $\alpha_c$ also marks the dynamical threshold---for $\alpha< \alpha_c$, the natural ``down-up'' random walk on independent sets of a given size mixes rapidly. The threshold $\alpha_c(\Delta)$ is closely connected to a uniqueness threshold on the tree: it is the smallest expected density of an independent set in the hard-core model on $G \in \G_\Delta$ at activity $\lam_c(\Delta)$.   

The main focus of this paper is on dynamical thresholds of the {\em fixed-magnetization} Ising model with inverse temperature parameter $\beta$ and magnetization $\eta$. The magnetization (per vertex) of an Ising configuration $\sigma$ is  $\eta(\sig):= \frac{ \sum_{v \in V(G)} \sigma(v)}{|V(G)|}$.  A configuration $\sigma$ of magnetization $\eta$ has size (number of $+1$ spins) exactly $k=\lfloor n \frac{\eta+1}{2}\rfloor$.   Denote by $\Omega_k$ the set of all configurations of size $k$. 

The fixed-magnetization Ising  model with inverse temperature $\beta \ge 0$ and magnetization $\eta \in [-1,1]$ is then a probability distribution defined similarly to~\eqref{eqIsingDef} but on $\Omega_k$, where $k=\lfloor n\frac{\eta + 1}{2}\rfloor $, as
$$
    \hat{\mu}_{G, \beta, \eta}(\sigma) = \frac{e^{\beta m_{G}(\sigma)}}{\hat Z_{G, \eta}(\beta)} , 
$$
where 
$$
    \hat Z_{G, \eta}(\beta) = \sum_{\sigma \in \Omega_{k}} e^{\beta m_{G}(\sigma)} 
$$
is the fixed-magnetization partition function.  Here we use floors to avoid restricting to values of $\eta$ where $n\frac{\eta + 1}{2}$ is an integer.
  The distribution $\hat{\mu}_{G, \beta, \eta}$ is exactly that of $\mu_{G,\beta,\lam}$ conditioned on the event $\{\sigma \in \Omega_k\}$. Note that the external field plays no role in the fixed-magnetization model since $\lam^{|\sig|^+}$ is constant on $\Omega_k$.  

In statistical physics, the fixed-magnetization Ising model is the {\em canonical ensemble} while the Ising model is the {\em grand canonical ensemble}. In order to define the critical parameters in the fixed-magnetization setting, we must first discuss those of the ferromagnetic Ising model.
While the Ising model has no computational threshold (there are efficient algorithms for all parameters) one can still ask about the relationship between uniqueness and  dynamical thresholds.  The natural dynamics in this setting are the {\em Glauber dynamics}, a Markov chain on the state space $\Omega$ with stationary distribution $\mu_{G,\beta,\lam}$  which at each step chooses a uniformly random vertex and updates its spin according to the conditional distribution given  the spins of its neighbors. For the case $\lam =1$ (``no external field'') the dynamical threshold has been identified, and it coincides with the uniqueness threshold. For $\Delta \geq 3$, let the {\em critical inverse temperature} of the Ising model on $\T_\Delta$ be denoted by
$$
    \beta_u(\Delta) := \ln\left(\frac{\Delta}{\Delta - 2}\right) \,.
$$
The value $\beta_u(\Delta)$ is the Gibbs uniqueness threshold for the Ising model (with $\lam=1$) on $\T_\Delta$ (see e.g.~\cite{baxter} and below in Section~\ref{secIsingTree} for a precise definition).
Mossel and Sly~\cite{mossel2013exact} proved that for $ 0 \le \beta < \beta_u$ and any $\lam$, the Glauber dynamics are rapidly mixing for any $G \in \G_\Delta$. This threshold in $\beta$ is sharp due to the analysis of the random $\Delta$-regular graph in~\cite{gerschenfeld2007reconstruction,dembo2010ising}: for $\beta > \beta_u$ and $\lam=1$, the Glauber dynamics for the Ising model take exponential time to mix.

For general $\lam \geq 0$, in the regime $\beta > \beta_u$, the threshold landscape is not as well understood. Note that the model is symmetric around $\lam=1$ by swapping the role of $+$ and $-$ spins and so for each threshold, its inverse is also a threshold; for clarity we will define thresholds for the case $\lam \ge 1$. Let $\lam_u (\Delta,\beta)$ be the  Gibbs uniqueness threshold of the ferromagnetic Ising model on $\T_\Delta$; that is, $\lam_u$ is the smallest $\lam_0 \ge 1$ so that there is a unique Gibbs measure for the Ising model on $\T_\Delta$ with inverse temperature $\beta$ and external field $\lam$, for all $\lam > \lam_0$ (again see~\cite{baxter} and Section~\ref{secIsingTree} for details).  The value of $\lam_u$ can be given implicitly as the solution to an equation involving $\Delta, \beta$, and $\lam$. Unlike in the above mentioned examples, while $\lam_u$ marks a phase transition on the tree, it does not mark a computational transition (since sampling from the ferromagnetic Ising model is tractable on all graphs and all parameters) and it has not been  established as a dynamical threshold (though this also has not been ruled out).   Below in Theorem~\ref{thm:GlauberRG} we show that the worst-case mixing time of Glauber dynamics over $\G_\Delta$ is exponential when $|\log \lam| < \log \lam_u$.

The complementary result (fast mixing of the Glauber dynamics for $ G \in \G_\Delta$ when $|\log \lam| > \log \lam_u$) is  not known to hold. Instead, sufficient conditions for fast mixing have been given that require $\lam$ to be somewhat larger than $\lam_u$. An interesting insight is that upper bounds on the dynamical threshold are often connected to zero-freeness of the map $\lambda \mapsto Z_G(\beta, \lambda)$ considered as a complex polynomial. 
Throughout this paper, we particularly focus on the \emph{analytic threshold} $\lambda_a(\Delta, \beta)$, defined by the following requirement: for all $G \in \mathcal{G}_{\Delta}$, every compact $D \subset (\lambda_a(\Delta, \beta), \infty)$ and every partial spin assignment $\tau_U: U \to \{-1, +1\}$, $U \subset V$ it holds that $Z^{\tau_U}_G(\beta, \lambda)$ (the partition function restricted to configurations that are consistent with $\tau_U$) is non-zero for all $\lambda$ in some uniform complex neighborhood of $D$. 
A formal definition of $\lambda_a$ is given in \Cref{sec:thresholds}.
In contrast to the uniqueness threshold, $\lam_a(\Delta,\beta)$ has not been determined.  It was previously known that $\lam_a (\Delta, \beta) \ge \lam_u(\Delta,\beta)$ and the best known upper bound  is
\begin{equation} \label{eq:lambda_a}
     \lambda_a (\Delta, \beta) \leq \min\left\{\frac{(\Delta - 2) e^{2\beta} - \Delta}{e^{\beta(2-\Delta)}}, e^{\beta \Delta}\right\} =: \bar{\lam_a} \,.
\end{equation}
The first expression in the minimum of \eqref{eq:lambda_a} was proven by Shao and Sun~\cite{shao2021contraction}, and the second bound of $e^{\beta \Delta}$ (which is smaller than the first expression for $\Delta \ge 4$ and $\beta$ large enough) was proven by Shao and Ye~\cite{shao2024zero}. In \Cref{sec:thresholds}, we show that $\lambda_a \neq \lambda_u$ in general.

It turns out that this analytic threshold $\lambda_a$ is closely related to the dynamical threshold.
More precisely, Chen, Liu, and Vigoda~\cite{chen2023rapid} proved that the first bound in \eqref{eq:lambda_a} can be used to define a regime in which the ferromagnetic Ising model satisfies $\ell_{\infty}$-independence (see \Cref{sec:prelim-mixing}), a stronger version of spectral independence that implies rapid mixing of Glauber dynamics.
Their derivation of the threshold used techniques similar to those of Shao and Sun~\cite{shao2021contraction} which resulted in coinciding bounds, but 
a more systematic connection was provided by Chen, Liu and Vigoda in \cite{chen2022spectral}.
They showed that for a broad class of spin systems, sufficiently strong zero-freeness assumptions imply $\ell_{\infty}$-independence.
With small adjustments, we use their technique to argue that the ferromagnetic Ising model satisfies $\ell_{\infty}$-independence for all $|\log \lambda| > \log \lambda_a(\Delta, \beta)$ (see \Cref{thm:zero_free_SI}).

Returning to the fixed-magnetization model, on lattices this model is studied in, e.g.,~\cite{dobrushin1992wulff,cerf2000wulff}, where interesting geometric behavior is described; the behavior of the Kawasaki dynamics (the natural analogue of Glauber dynamics) on $\Z^d$ has been studied extensively in, e.g.,~\cite{lu1993spectral,cancrini1999spectral,cancrini2000spectral,cancrini2002logarithmic}.  Here we focus on dynamical behavior over the class of all graphs of maximum degree $\Delta$.  

To understand algorithmic and dynamical thresholds in the fixed-magnetization Ising model, we need to define some further parameters using those described above for the Ising model.
The mean magnetization of the $+$ measure on $\T_\Delta$ (explained in detail in Section~\ref{secIsingTree} but informally is the measure induced by all $+1$ boundary conditions) is 
$$
    \eta^{+}_{\Delta, \beta, \lambda} \coloneqq \tanh\left(L^{*} + \artanh(\tanh(L^{*})\tanh(\beta/2))\right)
$$
where $L^{*}$ is the largest solution to 
\[
    L = \log(\lambda) + (\Delta - 1) \artanh(\tanh(L) \tanh(\beta/2))\, .
\]
We are specifically interested in the following three quantities:
 \begin{align*}
 \eta_c(\Delta, \beta) &=  \eta^{+}_{\Delta, \beta, 1} \\ 
     \eta_u (\Delta,\beta) &= \eta^{+}_{\Delta, \beta, \lambda_u} \\
     \eta_a(\Delta, \beta) &=  \eta^{+}_{\Delta, \beta, \lambda_a}  \,.
 \end{align*}
 For $\beta> \beta_u$, we have $0 < \eta_c < \eta_u < \eta_a$. 

Carlson, Davies, Kolla, and Perkins~\cite{CDKP21} showed recently that the fixed-magnetization Ising model exhibits quite different algorithmic behavior than the Ising model: it exhibits a computational threshold. In particular, for $\beta< \beta_u$ and any $\eta$, as well as for $\beta > \beta_u$ and $|\eta|>\eta_c$, there are efficient approximate counting and sampling algorithms for the Ising model at fixed mean magnetization $\eta$ on $\G_\Delta$, while for $\beta>\beta_u$ and $|\eta| < \eta_c$, there are no such algorithms unless NP$=$RP. Thus $\beta_u$ and $\eta_c$ mark the computational threshold in the fixed-magnetization Ising model.

We consider dynamical thresholds for the fixed-magnetization Ising model on $\G_\Delta$. 
Given a distribution, one candidate for an efficient approximate sampling algorithm is a Markov chain whose stationary distribution is our target distribution, but the efficiency of this algorithm depends on the mixing time. Recall that the mixing time of a Markov chain is the number of steps, in the worst-case over initial distribution, required for a Markov chain to reach $1/4$ total variation distance of its stationary distribution (see Section~\ref{sec:prelim-mixing} for a formal definition). As mentioned above, the natural dynamics associated to the fixed-magnetization Ising model are the {\em Kawasaki dynamics}, which is a reversible Markov chain on $\Omega_k$. At each step of the chain, a $+1$ vertex and a $-1$ vertex are chosen uniformly at random and have their spins swapped with a probability depending on the ratio of the Ising probabilities  of the two configurations. This is sometimes referred to as the {\em global} Kawasaki dynamics, whereas the {\em local} Kawasaki dynamics restrict to swapping spins of neighboring vertices. 

Our main contributions concern the {\em mixing time} of the Kawasaki dynamics. 
Taking $\tv{\mu-\nu} := \sup_{A \in \mathcal A}|\mu(A) - \nu(A)|$ to be the  total variation distance between probability distributions $\mu$ and $\nu$ on a probability space $(\Omega, \mathcal A)$, the mixing time of a Markov chain on $\Omega$ that has transition matrix $P$ and stationary distribution $\pi$ is
$$\tau_{mix} := \inf \left \{t: \max_{x \in \Omega} \tv{P^{t}(x, \cdot) - \pi} \leq \frac14  \right \} \,.$$

Resolving one conjecture of Carlson, Davies, Kolla, and Perkins and disproving another (part (i) and (ii) respectively of~\cite[Conjecture 1]{CDKP21}), we establish thresholds in the mean magnetization for fast and slow mixing  of the Kawasaki dynamics on $\G_\Delta$.

\begin{theorem}\label{thm:main-mixing}
Fix $\Delta \geq 3, \beta \geq 0$, and $\eta \in [-1, 1]$. For the Kawasaki dynamics, the following two statements hold:
\begin{enumerate}
    \item If $0 \le \beta < \beta_u$ or if $\beta \geq \beta_u$ and $|\eta| > \eta_{a}$,  
    then the Kawasaki dynamics for $\hat{\mu}_{G, \beta, \eta}$ have mixing time $O(|V(G)|^2)$ for all $G \in \G_{\Delta}$. \label{thm:main_mixing:rapid}
    \item If $\beta > \beta_u$ and $|\eta| < \eta_u$, then there exists a sequence of graphs $G_n \in \G_{\Delta}$ with $|V(G_n)|\to \infty$ such that the Kawasaki dynamics for $\hat{\mu}_{G_n, \beta, \eta}$ have mixing time $\exp \left( \Omega(|V(G_n)|) \right)$ on $G_n$.\label{thm:main_mixing:slow}
\end{enumerate}
\end{theorem}

Fast mixing of the dynamics for all $\eta$ when $\beta<\beta_u$ was conjectured in~\cite{CDKP21}.  The slow mixing for some $\eta > \eta_c$ disproves the conjecture from~\cite{CDKP21} asserting the coincidence of the algorithmic and dynamical thresholds. Indeed, this tells us there is an interval of magnetizations for which there exists an efficient sampling algorithm, but for which the Kawasaki dynamics can take exponential time to mix. The algorithm provided in \cite{CDKP21} is that of rejection sampling from the Ising model at a suitably chosen value of $\lambda$: the algorithm outputs samples from $\mu_{G, \beta, \lam}$, rejecting them until one with the desired magnetization is output. The sampling algorithm for $\mu_{G,\lam}$ is not the Glauber dynamics (which can be slow mixing in general), but rather one based on the even-subgraph representation of the Ising model partition function~\cite{JS93,randall1999sampling}.  Further, in using an Ising model sampler, one can change magnetizations and thus  avoid bottlenecks for the Kawasaki dynamics of the kind we create in construction our slow-mixing example.  

If it were established that $\lam_a(\Delta,\beta)= \lam_u(\Delta, \beta)$ for some choice of $\Delta$ and $\beta$, then Theorem~\ref{thm:main-mixing} would give the sharp dynamical threshold for the fixed-magnetization model with these parameters.  It is an interesting question to understand the dynamical threshold in both the Ising model and fixed-magnetization Ising model if instead it holds that $\lam_u(\Delta, \beta) < \lam_a(\Delta, \beta)$; indeed, we show in \Cref{lemma:zeros} that we cannot have $\lam_a = \lam_u$ in general.

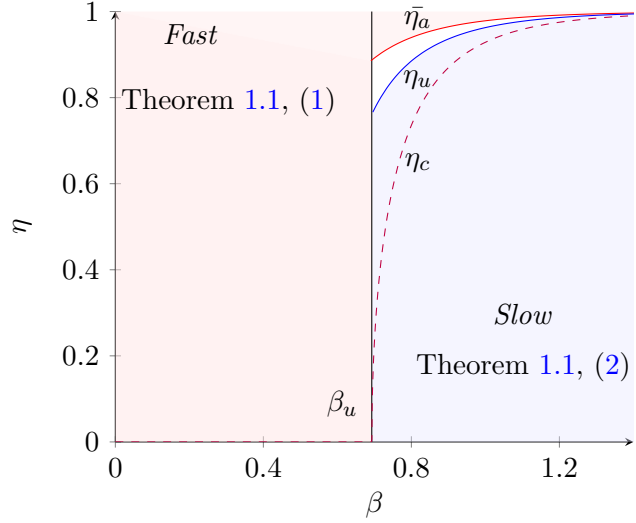
\begin{figure}[h]
\centering
\begin{tikzpicture}
    \begin{axis}[axis lines = left, xlabel = \(\beta\), ylabel={\(\eta\)}, xmin=0, xmax=1.4, ymin=0, ymax=1, xtick distance = 0.4]
    \draw[color=black] ({axis cs:0.693,0}|-{rel axis cs:0,0}) -- ({axis cs:0.693,0}|-{rel axis cs:0,1});
    \addplot[draw=none, domain=0:1.6, name path=zero]
    {0};
    \addplot[draw=none, name path=max]
    {1};
    \addplot [restrict x to domain=0.92:1.64, color=red, shift=({-23,0}), name path=etas] table {plot-data.tex};
    \addplot[red!10, opacity=0.4] fill between[of=etas and max, soft clip={domain=0:1.5}];
    \addplot[draw=none, name path = diagonal]{x*(1-0.885)/(-0.693)+1};
    \addplot[red!10, opacity=0.5] fill between[of=diagonal and zero, soft clip={domain=0:0.693}];
    \addplot [color=purple,shift=({0,0}), dashed, name path=etac] table {plot-data.tex};
    \addplot [restrict x to domain=0.815:2, color=blue,shift=({-12,0}), name path=etau] table {plot-data.tex};
    \addplot[blue!10, opacity=0.4] fill between[of=etau and zero, soft clip={domain=0.693:1.5}];
    \end{axis}
    \node at (4,5.6) {$\bar{\eta_a}$};
    \node at (4,4.8) {$\eta_u$};
    \node at (4,3.7) {$\eta_c$};
    \node at (3,.5) {$\beta_u$};
    \node at (1,5.4) {\emph{Fast}};
    \node at (1.5,4.5) {Theorem~\ref{thm:main-mixing},~(\ref{thm:main_mixing:rapid})};
    \node at (5.4,1.7) {\emph{Slow}};
    \node at (5.4,1) {Theorem~\ref{thm:main-mixing},~(\ref{thm:main_mixing:slow})};
\end{tikzpicture}
\label{fig:fixed-diagram}
\caption{Sketch of the phase space for the fixed-magnetization model on $\mathcal G_\Delta$ when $\Delta=4$, where $\bar{\eta_a} = \eta_{\Delta, \beta, \bar{\lam}_a}$}
\end{figure}

A diagram of the computational and dynamical thresholds for the fixed-magnetization Ising model is given in Figure~\ref{fig:fixed-diagram}.  
Interestingly, while the graph on which we show slow mixing is the union of random regular graphs, the behavior of the Kawasaki dynamics on a single copy of the random regular graph can be very different. Recently, Bauerschmidt, Bodineau, and Dagallier \cite{bauerschmidt2023kawasaki} (see also~\cite{bauerschmidt2023stochastic}) showed that  the local Kawasaki dynamics for the fixed-magnetization Ising model mixes in time $O(n \log^6 n)$ on random $\Delta$-regular graphs at all magnetizations when $\beta < 1/(8\sqrt{\Delta-1})$. In particular, when $\Delta $ is sufficiently large this regime of fast mixing includes parameters outside the tree uniqueness phase, i.e. inside the range of parameters for which we prove exponentially-slow mixing in the worst case over graphs in $\G_\Delta$.

Towards the proof of Theorem~\ref{thm:main-mixing},(\ref{thm:main_mixing:slow}), we establish that the Glauber dynamics for the Ising model on the random $\Delta$-regular graph takes exponential time to mix when $\beta>\beta_u$ and $\lam$ is in the non-uniqueness regime for $\T_\Delta$.  
\begin{theorem}
\label{thm:GlauberRG}
Fix $\Delta \ge 3$, $\beta > \beta_u(\Delta)$, and $ | \log \lam| <  \log \lam_u(\Delta,\beta)$.  Let $G$ be a uniformly random $\Delta$-regular graph on $n$ vertices.  Then with high probability as $n \to \infty$, the mixing time of the Glauber dynamics for the Ising model on $G$ is $e^{\Theta(n)}$.
\end{theorem}

This theorem complements the result of Can, van der Hofstad, and Kumagai~\cite{CvdHK21} showing that when $| \log \lam| > \log \lam_u$, with high probability over the random regular graph the mixing time of the Glauber dynamics is $O(n \log n)$; they conjectured that the mixing time is exponential when $| \log \lam| <  \log \lam_u$, which Theorem~\ref{thm:GlauberRG} confirms.

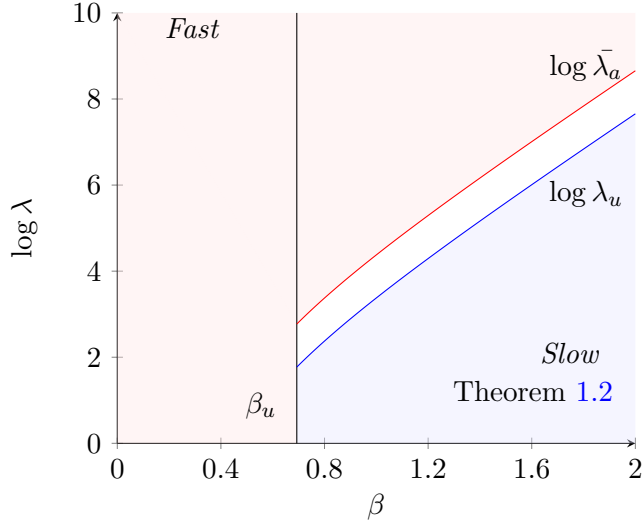
\begin{figure}[h]
\centering
\begin{tikzpicture}
    \begin{axis}[axis lines = left, xlabel = \(\beta\), ylabel={\(\log \lambda\)}, xmin=0, xmax=2, ymin=0, ymax=10, xtick distance = 0.4]
    \draw[color=black] ({axis cs:0.693,0}|-{rel axis cs:0,0}) -- ({axis cs:0.693,0}|-{rel axis cs:0,1});
    \addplot[draw=none, name path=zero]
    {0};
    \addplot[draw=none, name path=max]
    {10};
    \addplot[domain=0.693:2, samples=100,color=red,name path=lambdas]{ln((2*exp(2*x)-4)/(exp(-2*x)))};
    \addplot[red!10, opacity=0.4] fill between[of=lambdas and max, soft clip={domain=0:2}];
    \addplot[draw=none, name path = diagonal]{(x*(10-2.773)/(-0.693)+10)};
    \addplot[red!10, opacity=0.4] fill between[of=diagonal and zero, soft clip={domain=0:0.693}];
    \addplot[domain=0.693:2, samples=100,color=blue,name path=lambdau]{ln((2*exp(2*(x))-4)/(exp(-2*(x))))-1};
    \addplot[blue!10, opacity=0.4] fill between[of=lambdau and zero, soft clip={domain=0.693:2}];
    \end{axis}
    \node at (6.2,5) {$\log \bar{\lambda_a}$};
    \node at (6.2,3.3) {$\log \lambda_u$};
    \node at (1.9,.5) {$\beta_u$};
    \node at (1,5.5) {\emph{Fast}};
    \node at (6,1.2) {\emph{Slow}};
    \node at (5.5,0.7) {Theorem~\ref{thm:GlauberRG}};
\end{tikzpicture}
\caption{Sketch of the phase space for the Ising model Glauber dynamics on $\mathcal G_{\Delta}$ when $\Delta=4$.}
\label{fig:Ising-diagram}
\end{figure}

Theorem~\ref{thm:GlauberRG} also fills in more of the picture for dynamical thresholds in the Ising model on graphs in $\G_{\Delta}$; see Figure~\ref{fig:Ising-diagram}. 

Before we give an overview of our proof techniques, we state some open questions.
Our first question is concerned with the gap between the analytic threshold and the uniqueness threshold for the Ising model.

\begin{question} \label{q:analytic_uniqueness}
    What is the mixing time for $\beta > \beta_u$ and $\lam_u \leq \lam \leq \lam_a$, at the values where $\lam_u < \lam_a$? 
\end{question}

Next we conjecture the following improvement of part (\ref{thm:main_mixing:rapid}) of \Cref{thm:main-mixing}.
\begin{conjecture}
    If $0 \le \beta < \beta_u$ or if $\beta > \beta_u$ and $|\eta| > \eta_a$, then the Kawasaki dynamics for $\hat{\mu}_{G, \beta, \eta}$ are optimally mixing: the mixing time is in $O(|V(G)| \cdot \log(|V(G)|))$ for all $G \in \G_{\Delta}$.
\end{conjecture}
The analogous statement for independent sets is proved in~\cite{JMPV23} by proving a log-Sobolev inequality for the down-up walk with constant $\Omega(1/n)$.

\subsection{Overview of Techniques}
\label{secTechOverview}

The proofs of Theorems~\ref{thm:main-mixing} and~\ref{thm:GlauberRG} involve several different ingredients, including local central limit theorems,  spectral independence, and first- and second-moment methods for spin models on random graph.  We give an overview of the techniques here.

\subsubsection{Fast Mixing}
At a high level, the proof of Theorem~\ref{thm:main-mixing}, (\ref{thm:main_mixing:rapid}) follows the strategy used by Jain, Michelen, Pham, and Vuong~\cite{JMPV23} to show fast mixing for the down-up walk on independent sets of density less than $\alpha_c(\Delta)$. 

In order to derive an upper bound on the mixing time of the Kawasaki dynamics for the fixed-magnetization Ising model, we prove that the spectral gap of the associated transition matrix is bounded below by $\Omega(1/n)$.
To achieve this, we study a related down-up Ising  walk on $\Omega_k$ while arguing that the respective spectral gaps of the Kawasaki dynamics and the down-up walk are within a constant factor of each other.
This allows us to make use of recent literature that relates the spectral gap of  a down-up walk to spectral independence \cite{alev2020improved,anari2021spectral,chen2022localization}. 

Informally speaking, spectral independence captures the idea that for most pairs of vertices $v, w \in V$, the spins assigned to $v$ and $w$ by a random configuration from $\hat{\mu}_{G, \beta, \eta}$  are almost independent.
While spectral independence for the Ising model has been studied before by Chen, Liu, and Vigoda \cite{chen2023rapid}, no comparable result exists for the fixed-magnetization model. 
To derive the required spectral independence property, we follow an approach introduced in~\cite{JMPV23} to analyze the down-up walk for fixed-size independent sets. The idea is to choose $\lambda$ such that a random configuration from $\mu_{G, \beta, \lambda}$ has expected magnetization per vertex close to $\eta$. We then view $\hat{\mu}_{G, \beta, \eta}$ as $\mu_{G, \beta, \lambda}$ conditioned on the desired magnetization.

Using this perspective, we show that $\hat{\mu}_{G, \beta, \eta}$ satisfies $\ell_{\infty}$-independence with the following two steps.
\begin{enumerate}[(1)]
\item An extremal combinatorics result on the  magnetization of the Ising model  from \cite{CDKP21} shows that for any $G \in \mathcal{G}_{\Delta}$, the value of $\lambda$ that achieves expected magnetization $\eta$ satisfies $|\log\lam| > \log\lam_a$ if $|\eta|>\eta_a$. This allows us to use an approach by Chen, Liu, and Vigoda \cite{chen2022spectral} to derive $O(1)$-$\ell_{\infty}$-independence for the  Ising model for all such $\lambda$ based on our zero-freeness assumption.

\item We next show that the probability under $\mu_{G, \beta, \lambda}$ of drawing a configuration with exactly the correct magnetization   is sufficiently large, and that this probability does not change significantly after conditioning on the spin of a vertex.
For the former, a lower bound of $\Theta(1/\sqrt{n})$ can be derived from existing local central limit theorems for the expected number of $+1$ spins \cite{CDKP21}. 
For the latter, we perform a similar analysis to \cite{JMPV23} and use an Edgeworth expansion to prove that conditioning on the spin of a vertex changes this probability by at most $O(n^{-3/2})$.
For both results it is crucial that the Ising model satisfies sufficiently strong zero-freeness assumptions for all considered $\lambda$. 
\end{enumerate}

The above discussion indicates  how we obtain spectral independence for $\hat{\mu}_{G, \beta, \eta}$. The bulk of our work comes from leveraging this to derive a lower bound on the spectral gap of the down-up walk. This requires us to prove that spectral independence also holds when an arbitrary vertex set $U \subset V$ with $|U| < k$ is fixed (or \emph{pinned}) to have spin $+1$.
Such pinnings interfere with the proof strategy above for several reasons.
First of all, pinning vertices to $+1$ decreases the $\lambda$ that we need to choose to obtain the desired magnetization $\eta$. 
In particular, if we aim for $\eta > \eta_a$, this might cause the required value of $\lambda$ to leave the regime in which zero-freeness (and $\ell_{\infty}$-independence) for the Ising model is guaranteed.
We circumvent this by observing that the Kawasaki dynamics is symmetric under swapping $+1$ and $-1$ spins.
Hence, it suffices to consider $\eta < - \eta_a$, and an application of the FKG inequality ensures that we only need to consider $\lambda < 1/\lambda_a(\Delta, \beta)$ for all relevant pinnings.

The second difficulty is that once the number of free vertices $k - \size{U}$ becomes sub-linear in $n$, both the local central limit theorem and the Edgeworth expansion can fail.
Similar to \cite{JMPV23}, we solve this issue by using the localization framework by Chen and Eldan \cite{chen2022localization}, which allows us to factorize the spectral gap of the down-up walk into the spectral gaps of two Markov chains that are easier to analyze.
The first chain is a generalization of the down-up walk that updates $\Theta(n)$ vertices in each step, and we can analyze its spectral gap based on the spectral independence result described above using the local-to-global framework for local spectral expanders \cite{alev2020improved,anari2021spectral,chen2021optimal,chen2023rapid}.
The second walk is a simple down-up walk but with a set of vertices $U \subset V$ pinned to $+1$.
In particular, we need to show that there is some $\alpha > 0$ (depending on $\beta$ and $\Delta$) such that for $k - \size{U} \le \alpha n$, the spectral gap of such a pinned down-up walk is bounded below by $\Omega(1/n)$.

For bounding the spectral gap of the pinned walk, we use a coupling argument. 
Specifically, we construct a suitable metric on the state space such that the distance between two coupled copies of the Markov chain contracts in expectation in each step.
For the independent set model studied in \cite{JMPV23}, such a contracting coupling is well known, appearing in the original ``path coupling'' paper of Bubley and Dyer~\cite{bubley1997path}.
In contrast, for the fixed-magnetization Ising model, no such result exists, and the default choice of coupling (sometimes called the identity coupling) and metric (the number of vertices on which both configurations differ) does not exhibit the desired contraction. 
Roughly speaking, this is because the ferromagnetism can cause certain types of disagreements to increase the probability that new disagreements are created. 
We overcome this problem by studying a refined metric, which assigns different weights to ``good'' and ``bad'' disagreements
in a way that guarantees that distances under this new metric decrease in expectation under the coupling, thus establishing the desired bound on the spectral gap.

\subsubsection{Slow Mixing}

For the slow mixing results, we leverage the connection between the Ising model on the infinite tree  $\mathbb T_\Delta$ and the behavior of the model on a uniformly random  $\Delta$-regular graph.  In the relevant range of parameters ($\beta>\beta_u$, $1< \lam < \lam_u$) there are two distinct Ising Gibbs measures on $\mathbb T_\Delta$, the ``plus measure'' and the ``minus measure.''  On the random graph these two Gibbs measures manifest themselves as a dominant and subdominant metastable state: sets of configurations for which the Glauber dynamics take exponential time to escape from.   The existence of multiple metastable states immediately shows slow mixing of the Glauber dynamics (Theorem~\ref{thm:GlauberRG}), and we then use this to construct a graph on which the Kawasaki dynamics is slow mixing, proving  Theorem~\ref{thm:main-mixing},(\ref{thm:main_mixing:slow}).

To do this,  we exhibit the existence of a bottleneck in the state space of the model on a $\Delta$-regular graph $H$ constructed as the disjoint union of several copies of a random $\Delta$-regular graph. 
We define two different subsets of configurations of the fixed-magnetization Ising model on $H$: in the set of configurations $S_1$, each copy of the random graph comprising $H$ has magnetization $\eta$; in the set $S_2$, some copies have magnetization approximately $\eta_+ > \eta$ and some copies have magnetization approximately $\eta_- < \eta$ (chosen in such a way that their average is $\eta$). We then show that a third set $S_3$ separates $S_1$ and $S_2$ (under single-step updates of the Kawasaki dynamics) and carries exponentially less probability mass in the fixed-magnetization Ising model than either $S_1$ or $S_2$. Via a standard conductance argument this proves exponentially slow mixing of the Kawasaki dynamics. 

 Bounds on the weights of the sets $S_1, S_2$, and $S_3$  will follow from the existence of the metastable states on the random  graph.   One metastable state consists of configurations with magnetization close to $\eta_{\Delta, \beta,\lam}^+$ and the other consists of configurations with magnetization close to $\eta_{\Delta, \beta,\lam}^-$.  That is, the two metastable states are in correspondence with the two distinct extremal Gibbs measures on $\T_{\Delta}$ (which is why $\lam < \lam_u$ is crucial).

Identifying the metastable states follows from determining which states (organized according to their magnetizations) contribute significantly to the partition function $Z_G(\beta, \lam)$ of the Ising model on the random $\Delta$-regular graph.  A first guess about how much each state contributes to $Z_G(\beta, \lam)$ would be to take the expected contribution.  The exponential order of this expectation is captured by a function $f_{\Delta, \beta, \lam}(\eta)$.  From~\cite{GSVY16}, we know that the critical points of this function correspond to fixed points of a  recursion on $\mathbb T_{\Delta}$, and that the second-moment method can be used to lower bound the contribution of the state with magnetization $\eta$, where $\eta$ is the maximum of $f_{\Delta, \beta, \lam}(\eta)$.  This suffices to determine the dominant state of the Ising model on the random graph (as was done in much greater generality by Dembo and Montanari in~\cite{dembo2010ising}).  

To identify subdominant metastable states, however, we need to analyze the contribution of states with magnetization $\eta$ when $\eta$ is a local maximum of $f_{\Delta, \beta, \lam}(\eta)$.
For this we follow the approach of~\cite{Coja-Oghlan22} utilizing non-reconstruction in planted models. While their setting is the $q$-state Potts model for $q \geq 3$, many of their results can be translated to our context of the external-field Ising model. We discuss their techniques in greater detail in Section~\ref{sec:metastable-proof}.

When we construct the graph $H$ as the union of random graphs, we also must understand how the behavior of the fixed-magnetization Ising model relates to that of the Ising model. To do this, we give a new and simple argument in Section~\ref{secKawasakislow} to bound the probability of hitting a given magnetization in the Ising model.

\subsection{Outline}
In Section~\ref{secPrelim}, we collect preliminary results that will be used in our proofs. In Section~\ref{sec:fast}, we prove our fast-mixing result, Theorem~\ref{thm:main-mixing},(\ref{thm:main_mixing:rapid}). In Section~\ref{sec:slow} we prove our slow-mixing results, Theorem~\ref{thm:main-mixing},(\ref{thm:main_mixing:slow}) and Theorem~\ref{thm:GlauberRG}. In the appendices, we provide additional proofs that are not included in the main body; Appendices~\ref{appendix:zero_freeness_SI} and \ref{appendix:fast} are relevant to fast mixing and Appendix~\ref{appendix:slow} concerns slow mixing.

\section{Preliminaries}
\label{secPrelim}

Throughout the paper and unless otherwise stated, we will make the following assumptions: $\Delta \geq 3$ is fixed, $\beta \geq 0$, $G = (V, E) \in \mathcal G_{\Delta}$, and $n = |V|$.

We will often switch between notation of $\eta$ for the magnetization per vertex and $k =\lfloor \frac{\eta + 1}{2} n \rfloor $ for the number of $+1$ spins in such a configuration. We will thus abuse notation and write $\hat \mu_{G,\beta,k}$ for $\hat \mu_{G,\beta,\eta}$ and $\hat Z_{G,k}(\beta)$ for $\hat Z_{G,\eta}(\beta)$ when it makes things more clear. We will also on occasion drop $G$ and $\beta$ from the subscripts of our Gibbs measure notation as well as the subscripts and argument of our partition function notation when $G$ and $\beta$ do not play a role in the proofs.

\subsection{Ising Model on the Infinite Tree}
\label{secIsingTree}

Let $\T_{\Delta}$ denote the infinite $\Delta$-regular tree.  Since it  has infinitely many vertices, one cannot define the Ising model on $\T_{\Delta}$ via~\eqref{eqIsingDef}.  Instead, the Dobrushin-Lanford-Ruelle equations can be used to define ``infinite-volume Gibbs measures'' for the Ising model and other spin models on infinite graphs.  This approach says that a probability measure $\mu$ on $\{\pm 1\}^{V(\T_{\Delta})}$ is a Gibbs measure for the Ising model at inverse temperature $\beta$ and external field $\lam$ if 
the conditional measure on any finite set of vertices given a  configuration on the complement is the Ising model defined by~\eqref{eqIsingDef} with the appropriate boundary conditions.  See~\cite{georgii2011gibbs} for more details. 

A main question about Gibbs measures on infinite graphs is whether for a given specification of parameters (i.e. $\beta$ and $\lam$ in the Ising case) and a given infinite graph $G$ there is a unique Gibbs measure or multiple distinct Gibbs measures. The transition between uniqueness and non-uniqueness as a parameter varies marks a phase transition.

Understanding uniqueness and non-uniqueness of the Ising model on $\T_\Delta$ is relatively simple because of monotonicity and the FKG inequality.  There are two extreme infinite-volume Gibbs measures in the sense of maximizing or minimizing the probability that a fixed vertex of $\T_\Delta$ gets a $+1$ spin: the ``$+$ measure'' on $\T_\Delta$ is the Gibbs measure realized by taking a weak limit of finite-volume Gibbs measures on depth $N$ truncations of $\T_\Delta$ with boundary vertices assigned $+1$ spins; the ``$-$ measure'' is the weak limit of finite-volume measures with boundary vertices receiving $-1$ spins.

The quantities $\eta^+_{\Delta,\beta,\lam}$ and $\eta^-_{\Delta,\beta,\lam}$ are the respective expectations of $\sigma(v)$ (for any fixed $v$ in $\T_{\Delta}$) under these two Gibbs measures.  The quantities can be calculated as solutions to fixed point equations (see e.g.~\cite{baxter}), giving
$$\eta^{+}_{\Delta, \beta, \lambda} = \tanh\left(L^{*} + \artanh(\tanh(L^{*})\tanh(\beta/2))\right)$$
where $L^{*}$ is the largest solution to 
\[
    L = \log(\lambda) + (\Delta - 1) \artanh(\tanh(L) \tanh(\beta/2))\, .
\]

The following proposition summarizes information about $\eta^+_{\Delta,\beta,\lam}$, $\eta^-_{\Delta,\beta,\lam}$ and Gibbs uniqueness that we will use (all follow from the results in~\cite{baxter}).  

\newpage

\begin{proposition}
    \label{propIsingTree}
    Fix $\Delta \ge 3$.
    \begin{itemize}
         \item There is uniqueness of Gibbs measure for the Ising model with parameters $\beta,\lam$ on $\T_\Delta$ if and only if $\eta^+_{\Delta,\beta,\lam}=\eta^-_{\Delta,\beta,\lam}$.
         \item For $\beta\le \beta_u(\Delta)=\ln\left(\frac{\Delta}{\Delta -2}\right) $, there is uniqueness for all $\lam$.
         \item For $\beta>\beta_u(\Delta)$ there is $\lam_u >1$ so that there is uniqueness if and only if $|\log \lam| > \log \lam_u$. 
         \item $\eta^+_{\Delta,\beta,\lam}$ is continuous and strictly increasing in $\lam$ on the interval $[1, \infty)$. 
 In particular, recall that $\eta_c (\Delta,\beta) = \eta^+_{\Delta,\beta,1}$ and $\eta_u(\Delta, \beta) = \eta^+_{\Delta,\beta,\lam_u}$; then for every $\eta \in [\eta_c, \eta_u]$ there is $\lam \in [1,\lam_u]$ so that  $\eta^+_{\Delta,\beta,\lam} = \eta$. 
    \end{itemize}

\end{proposition}

Finally, it will be important to bound the expected magnetization in the Ising model for given $\beta,\lam$ and any $G \in \mathcal G_\Delta$. The bound is an extremal result proved in~\cite{CDKP21}.
\begin{theorem}[\hspace{1sp}{\cite[Theorem 3]{CDKP21}}]\label{thm:extremalmag}
    For $G \in \G_{\Delta}$, $\lam \geq 1$, and $\beta \geq 0$, 
    $$\E_{\sig\sim\mu_{G,\beta,\lam}}[\eta(\sig)] \leq \eta^+_{\Delta, \beta, \lam} \,.$$
\end{theorem}

\subsection{Pinned Models}

For the fast-mixing argument, we will frequently consider pinned versions of our models, meaning conditioned on some subset of vertices having been assigned a particular spin.
For $U \subset V$, we call a function $\tau_U: U \to \{+1, -1\}$ a {\em pinning} on $U$.
We write $\Omega^{\tau_U} = \{\sigma \in \Omega \mid \forall u \in U: \sigma(u) = \tau_U(u)\}$ for the set of Ising configurations on $G$ that agree with $\tau_U$ on $U$.
The {\em Ising  partition function with pinning $\tau_U$} is defined as 
\[
    Z_{G}^{\tau_U} (\beta, \lambda) = \sum_{\sigma \in \Omega^{\tau_U}} \lambda^{\size{\sigma}^{+}} e^{\beta m_{G}(\sigma)},
\]
and the {\em Ising model under pinning $\tau_U$} is defined by Gibbs measure
\[
    \mu_{G, \beta, \lambda}^{\tau_U}(\sigma) = \frac{\ind{\sigma \in \Omega^{\tau_U}} \lambda^{|\sig|^+} e^{\beta m_{G}(\sigma)}}{Z_{G}^{\tau_U} (\beta, \lambda)} .
\] 
Note that for $\lambda > 0$, it holds that $\mu_{\beta, \lambda}^{\tau_U}$ is a well-defined probability distribution with support $\Omega^{\tau_U}$.
We allow for the case $U = \emptyset$, which is equivalent to the unpinned Ising model.
Often, $\tau_U$ will be the constant $+1$ function on $U$, in which case we write $\Omega^{U}$, $Z_{G}^{U}$ and $\mu_{\beta, \lambda}^{U}$.

Analogously to the  Ising model, we will also impose pinnings on the fixed-magnetization model.
To this end, set $\Omega_{k}^{\tau_U} = \{\sigma \in \Omega_{k}\,:\, \forall u \in U: \sigma(u) = \tau_U(u)\}$
and define the {\em fixed-magnetization partition function with pinning $\tau_U$} as
\[
    \hat{Z}_{G, k}^{\tau_U} (\beta) = \sum_{\sigma \in \Omega_{k}^{\tau_U}} e^{\beta m_{G}(\sigma)}.
\]
The {\em fixed-magnetization Ising model under pinning $\tau_U$} is a probability measure with support $\Omega_{k}^{\tau_U}$ defined by
\[
    \hat{\mu}_{G, \beta, k}^{\tau_U}(\sigma) = \frac{\ind{\sigma \in \Omega_{k}^{\tau_U}} e^{\beta m_{G}(\sigma)}}{\hat{Z}_{G, k}^{\tau_U}(\beta)} .
\]
Throughout the paper, we assume $\size{\tau_U}^{+} \le k$ so that the expression above is well-defined. 
As with the Ising model, we write $\Omega_k^U$, $\hat{Z}_{G, k}^{U}$ and $\hat{\mu}_{G, \beta, k}^{U}$ when $\tau_U$ is the constant $+1$ function.

\subsection{Kawasaki Dynamics, Down-up Walk, and Glauber Dynamics}
Here we formally define the three Markov chains that we will analyze. Our main object of study is the Kawasaki dynamics for the fixed-magnetization Ising model. For this, we fix a size $k$ where $1 \le k \le \size{V}-1$. Recall that configurations are functions from $V(G) \to \{+1, -1\}$, so $X_t^{-1}(+1)$ refers to the set of vertices assigned to $+1$.
\begin{definition}[Kawasaki dynamics] \label{def:kawasaki}
    The Kawasaki dynamics on $\Omega_k$ is a Markov chain $\kawasaki_{\beta, k} = (X_t)_{t \geq 0}$ given by the following update rule: 
    \begin{enumerate}
        \item Pick $u \in X_{t}^{-1} (+1)$ and $w \in X_{t}^{-1} (-1)$ uniformly at random, and set $X \in \Omega_{k}$ such that $X(v) = X_t(w), X(w) = X_t(v)$, and $X(u) = X_t(u)$ for $u \neq v, w$.
        \item Set $X_{t+1} = X$ with probability $\min\left\{1, \frac{\hat{\mu}_{G, \beta, k}(X)}{\hat{\mu}_{G, \beta, k}(X_t)}\right\}$, and set $X_{t+1} = X_t$ otherwise.
    \end{enumerate}
\end{definition}

In other words, the Kawasaki dynamics chooses two vertices with opposite spins and swaps their spins with probability proportional to the change in monochromatic edges.

For proving fast mixing of the Kawasaki dynamics, we use the down-up walk on the $+1$ spins as a proxy for our analysis. 
Here we will also need to consider the Markov chain under plus pinnings.
\begin{definition}[Down-up walk with plus pinnings] \label{def:plusDownUp}
    For $U \subset V$ and with $\size{U} < k$ we define the {\em $+1$-down-up walk} on $\Omega_k^{U}$ as a Markov chain $\plusDownUp_{\beta, k}^{U} = (Y_t)_{t \geq 0}$, given by the following update rule:
    \begin{enumerate}
        \item Pick $v \in Y_{t}^{-1} (+1) \setminus U$ uniformly at random and set $W = Y_{t}^{-1} (+1) \setminus \{v\}$.
        \item Draw $Y_{t+1}$ from $\hat{\mu}_{G, \beta, k}^{W}$.
    \end{enumerate}
    We write $\plusDownUp_{\beta, k}$ if $U = \emptyset$. 
\end{definition}

The following observation is easy to check.
\begin{observation} \label{obs:ergodicity}
    $\kawasaki_{\beta, k}$ and $\plusDownUp_{\beta, k}$ are ergodic and reversible with respect to $\hat{\mu}_{\beta, k}$.
    Moreover, there is a constant $C \ge 1$ that only depends on $\Delta$ and $\beta$ such that for all $\sigma_1 \neq \sigma_2$
    \[
        \frac{1}{C} \cdot \plusDownUp_{\beta, k}(\sigma_1, \sigma_2)
        \le \kawasaki_{\beta, k} (\sigma_1, \sigma_2)
        \le C \cdot \plusDownUp_{\beta, k}(\sigma_1, \sigma_2).
\]
\end{observation}

Lastly, we also consider the Glauber dynamics for the Ising model.
\begin{definition}[Glauber dynamics] The Glauber dynamics on $\Omega$ is a Markov chain $(X_t)_{t \geq 0}$, given by the following update rule:
\begin{enumerate}
    \item Pick $v \in V(G)$ uniformly at random.
    \item For $u \neq v$, set $X_{t+1}(u) = X_t(u)$, and sample $X_{t+1}(v)$ from the marginal distribution at $v$ conditioned on $X_{t+1}(N(v))$. 
\end{enumerate}
    
\end{definition}

\subsection{Mixing Times}\label{sec:prelim-mixing}
Our goal in analyzing the Kawasaki dynamics is to understand the {\em mixing time} of this Markov chain, as defined in the previous section.
We will use several different techniques 
which we now describe.

\subsubsection{Upper Bounds on Mixing Time}

A common way to upper-bound the mixing time of a reversible Markov chain $P$ is by lower-bounding its spectral gap, which can be defined via the following variational characterization.
\begin{definition} \label{def:poincare}
    Let $P$ be a transition matrix that is reversible with respect to $\pi$. 
    We denote by $\gap(P)$ the {\em spectral gap} (or Poincar\'e constant) of $P$, which is defined as the largest constant $\gamma$ such that 
    \[
        \gamma \Var_{\pi}(f) \leq \mathcal{E}_P(f,f)
    \]
    for any function $f: \Omega \rightarrow \mathbb{R}$, where $\Var_{\pi}(f)$ is the variance of $f$ with respect to $\pi$ and $\mathcal{E}_{P}$ is the Dirichlet form of $P$, given by
    \[
        \mathcal{E}_P (f, g) = \frac{1}{2} \sum_{x, y \in \Omega} (f(x) - f(y))(g(x) - g(y)) P(x, y) \pi(x) \hspace{2em} f, g: \Omega \to \R. 
    \]
\end{definition}
Using this characterization of the spectral gap, we have the following observation.
\begin{observation} \label{obs:poincare_comparison}
    Suppose $P_1$ and $P_2$ are transition matrices that are both reversible with respect to $\pi$.
    If there are constants $\alpha_1, \alpha_2 > 0$ such that $\alpha_1 \cdot P_1(x, y) \le P_2(x, y) \le \alpha_2 \cdot P_1(x, y)$ for all $x \neq y$, then $\alpha_1 \cdot \gap(P_1) \le \gap(P_2) \le \alpha_2 \cdot \gap(P_1)$.
\end{observation}
On account of \Cref{obs:ergodicity}, this allows us to study the spectral gap of the down-up walk $\plusDownUp_{\beta, k}$ instead of the Kawasaki dynamics $\kawasaki_{\beta, k}$.

An upper bound on the mixing time of an ergodic, reversible Markov chain with transition matrix $P$ can be obtained from its spectral gap via the following standard relationship (see \cite[Theorem $12.4$]{levin2017markov}):
\begin{align*}
    \tmix \le \gap(P)^{-1} \cdot \log\left(\frac{4}{\min_{x \in \Omega} \pi(x)}\right) .
\end{align*} 

There are various ways to obtain bounds on the spectral gap of a Markov chain, one of which is to construct a contracting coupling.
For a transition matrix $P$, we say that a Markov chain $(X_t, Y_t)_{t \ge 0}$ on $\Omega \times \Omega$ is a {\em coupling} of $P$ with itself if each of the marginal processes $(X_t)_{t \ge 0}$ and $(Y_t)_{t\geq0}$ is a Markov chain with transition matrix $P$. 
We use this notion to bound the spectral gap.
\begin{theorem}[\hspace{1sp}{\cite[Theorem $13.1$]{levin2017markov}}] \label{thm:gap_from_contraction}
    Suppose $\Omega$ is finite and let $(X_t, Y_t)_{t \ge 0}$ be a coupling of $P$ with itself.
    If there is a constant $c > 0$ and a function $\rho: \Omega \times \Omega \to \R_{\ge 0}$ such that $\rho(x, y) = 0$ if and only if $x = y$, and for all $t \in \Z_{\geq 0}$ it holds that
    \[
        \E[\rho(X_{t+1}, Y_{t+1}) \mid X_t, Y_t] \le (1 - c) \rho(X_t, Y_t) , 
    \]
    then the spectral gap of $P$ is at least $c$.
\end{theorem} 
We will use \Cref{thm:gap_from_contraction} to show that the down-up walk $\plusDownUp^{U}_{\beta, k}$ has a spectral gap of $\Omega(1/k)$ whenever $k - \size{U} \le \alpha n$ for some $\alpha$ depending on $\Delta$ and $\beta$. 
In particular, by the symmetry of the Kawasaki dynamics under swapping all spins, this proves a spectral gap of $\Omega(1/k)$ for $\kawasaki_{\beta, k}$ if $k \le \alpha n$ or $k \ge (1-\alpha) n$, but it does not cover the full regime of \Cref{thm:main-mixing},(\ref{thm:main_mixing:rapid}).

To prove the full result of \Cref{thm:main-mixing},(\ref{thm:main_mixing:rapid}), we prove that $\hat{\mu}^{U}_{\beta, k}$ satisfies spectral independence for suitable $k \in \N$ and sets $U \subset V$.
Spectral independence is a property of the stationary distribution $\pi$ of a Markov chain, and it was recently used to bound the spectral gap and prove rapid mixing of various chains \cite{alev2020improved,anari2021spectral,chen2021optimal,chen2023rapid,chen2022localization,JMPV23}. 
For the following discussion of spectral independence, we restrict ourselves to distributions on $\Omega = 2^V$ where $V$ is some finite set (e.g., the vertices of a graph).
Note that this encompasses both the fixed-magnetization Ising model as well as the Ising model, by associating $S \in \Omega$ with the Ising configuration that maps all vertices in $S$ to $+1$.
In this setting, we adopt the following notation: for a distribution $\pi$ on $\Omega$, a subset $S$ drawn from $\pi$, and $v \in V$, we write $\pi(v)$ for the probability that $v \in S$
and $\pi(\overline{v})$ for the probability that $v \notin S$.
We extend this to conditional probabilities, writing for example $\pi(v \mid \overline{u})$ for the probability that $v \in S$ given $u \notin S$.

\begin{definition} \label{def:influence}
The {\em influence matrix} of a distribution $\pi$ on $2^V$ is the matrix $M_{\pi} \in \R^{n \times n}$ with entries 
$$M_{\pi}[u, v]= \begin{cases}
    0 & \text{ if } \pi(u) = 0\\
    \pi(v \mid u) - \pi(v) & \text{ otherwise}
\end{cases} $$
\end{definition}

Using this definition of $M_{\pi}$, the definition of spectral independence of $\pi$ is as follows.

\begin{definition}
A probability distribution $\pi$ on $2^V$ is called \emph{$C$-spectrally independent} (for $C \ge 0$) if the largest eigenvalue of $M_{\pi}$ is at most $C$.  
\end{definition}

Since directly bounding the largest eigenvalue of $M_{\pi}$ is usually challenging, a common approach is to bound the $\ell_{\infty}$-norm of $M_{\pi}$ instead.
This leads to the stronger notion of $\ell_{\infty}$-independence.

\begin{definition}
A probability distribution $\pi$ on $2^V$ is \emph{$C$-$\ell_{\infty}$-independent} (for $C \geq 0$) if $$\lVert M_{\pi} \rVert_{\infty} \coloneqq \max_{u \in V} \sum_{v \in V} |M_{\pi}[u, v]| \leq C\ .$$
\end{definition}

\begin{remark}
    There are various definitions of the pairwise influence matrix in the literature \cite{anari2021spectral,chen2023rapid,chen2021optimal}.
    For spin systems with two possible states for each vertex (such as the Ising model), pairwise influence is commonly defined as $M_{\pi}[u, v] = \pi(v \mid u) - \pi(v \mid \overline{u})$.
    However, note that switching between the two definitions only changes the spectral radius by some constant factor, provided that $\pi(v)$ is uniformly bounded away from $0$ and $1$.
    Since this is the case for the Ising model, given that $\lambda > 0$, existing spectral independence results such as \cite{chen2023rapid} carry over to our definition.
    Moreover, we will see later that \Cref{def:influence} is more natural for canonical ensembles, such as the fixed-magnetization Ising model, as it relates more directly to local spectral expansion of the associated simplicial complex (see \Cref{lemma:SE_from_SI} for more details).
\end{remark}

There are different ways to derive bounds on the spectral gap of a Markov chain from spectral independence.
The most popular approach is the use of \emph{local-to-global theorems}, which are applicable whenever the Markov chain in question can be represented as a down-up walk on a suitable weighted simplicial complex \cite{alev2020improved,anari2021spectral,chen2023rapid,chen2021optimal}.
Local-to-global theorems allow us to express the spectral gap of the down-up walk in terms of spectral gaps of local walks on the complex, which can then be related to the spectral radius of the pairwise influence matrix. 

A more recent framework was introduced by Chen and Eldan \cite{chen2022localization} and uses \textit{localization schemes}.
A localization scheme maps a probability distribution $\pi$ on $\Omega$ to a localization process---a random sequence of probability measures that interpolates between $\pi$ and a random Dirac measure.
Via the localization process, a localization scheme gives rise to a Markov chain with stationary distribution $\pi$.
Provided that the localization process exhibits a property called ``approximate conservation of variance,'' this can be used to bound the spectral gap of the associated Markov chain.
For a broad class of localization schemes, approximate conservation of variance follows if all measures along the localization process exhibit sufficiently strong spectral independence. 
Since we are studying the fixed-magnetization Ising model, we are particularly concerned with distributions $\pi$ on $\Omega_k$.
In this setting, the canonical choice for a localization scheme is the subset simplicial-complex localization (see \cite[Example 5]{chen2022localization}), and the natural associated Markov chain is the down-up walk $\plusDownUp_{\beta, k}$.

The main difficulty in applying the above frameworks in our setting is that they usually assume $O(1)$-spectral independence of the pinned distributions $\hat{\mu}^{U}_{\beta, k}$ for all $U \subset V$ with $0 \le \size{U} \le k - 1$. 
Unfortunately, we will not be able to derive  spectral independence for all such $U$.
Moreover, for the localization framework, it is not clear if the subset simplicial-complex localization allows us to derive approximate conservation of variance from spectral independence.
To overcome these difficulties, we use an argument similar to that of Jain, Michelen, Pham and Vuong \cite{JMPV23}.
We combine the techniques above as follows: 
first, we use a localization scheme to show that for any $\ell \le k-1$, the spectral gap of $\plusDownUp_{\beta, k}$ is bounded below by the product of the spectral gap of the pinned down-up walk $\plusDownUp^{U}_{\beta, k}$ for any $U \subset V$ with $\size{U} = \ell$ and the spectral gap of the $(k, \ell)$-down-up walk, a modified version of $\plusDownUp_{\beta, k}$ that resamples $k - \ell$ plus spins in each step.
Choosing $\ell$ such that $k - \ell \le \alpha  n$ for some suitable $\alpha > 0$, we can use a coupling argument as discussed before to show that $\gap(\plusDownUp^{U}_{\beta, k}) \in \Omega(1/k)$ for every $U \subset V$ with $\size{U} = \ell$.
To lower-bound the spectral gap of the $(k, \ell)$-down-up walk, we use a local-to-global theorem by Chen, Liu and Vigoda \cite{chen2021optimal}.
This only requires us to show that $\hat{\mu}^{W}_{\beta, k}$ satisfies $O(1)$-spectral independence for all $W \subset V$ with $k - \size{W} \ge \alpha' n$ for some $0 < \alpha' < \alpha$.
The range of $k$ for which we can show this $O(1)$-spectral independence leads to the magnetization range given in \Cref{thm:main-mixing},(\ref{thm:main_mixing:rapid}).

\subsubsection{Lower Bounds on Mixing Time}
To prove slow mixing, we exhibit the existence of a bottleneck in the state space, a set of configurations which separates two parts of the state space and carries an exponentially smaller probability in the stationary distribution than either of the two parts. The following lemma captures a simple form of this argument, often phrased in terms of {\em conductance}, for proving lower bounds on the mixing times of Markov chains.
\begin{lemma}
\label{lem:bottleneck}
    Let $(X_t)_{t \geq 0}$ be a Markov chain on the state space $\Omega$ with stationary distribution $\pi$. Suppose there exists disjoint sets $S_1, S_2, S_3 \subset \Omega$ so that the following hold:
    \begin{itemize}
        \item For the chain to pass from $S_2$ to $S_1$ it must pass through $S_3$;
        \item $\pi(S_1) \ge \pi(S_2)$
        \item $\pi(S_3) \le e^{-\Omega(n)} \pi(S_2)$.
    \end{itemize}
    Then the mixing time of the chain $(X_t)$ is $\exp (\Omega(n))$.
\end{lemma} 
The statement is an immediate corollary of, e.g.,~\cite[Claim 2.3] {dyer2002counting}.

To prove Theorem~\ref{thm:GlauberRG}, we define $S_1, S_2, S_3$ to be sets of configurations with certain magnetizations.  $S_1$ will be those configurations whose magnetization per vertex is close to that of the plus measure on $\mathbb T_{\Delta}$ (when $\lam >1$); $S_2$ will be those  whose magnetization per vertex is close to that of the minus measure; and $S_3$ will be configurations whose magnetization is just larger than that of $S_2$.

To prove Theorem~\ref{thm:main-mixing},(\ref{thm:main_mixing:slow}), we consider a graph $H$ made up of disjoint copies of a random regular graph.  We define $S_1$ to be the set of configurations with magnetization $\eta$ on each copy; $S_2$ will be a set of configurations with magnetization $\eta_+$ on some copies and $\eta_-$ on other copies, with $\eta_-< \eta < \eta_+$, in such a way that the overall magnetization is $\eta$.  Again $S_3$ will be a neighborhood of $S_2$.  In both cases, the main work will be in verifying the conditions of Lemma~\ref{lem:bottleneck}.

\subsection{Thresholds for Zero-Freeness and Spectral Independence}\label{sec:thresholds} 

The definition of $\lambda_a(\Delta, \beta)$ is based on viewing the Ising partition function as a polynomial in the (complex) variable $\lambda$.
For $z \in \C$ and $\delta > 0$, we write $\mathcal{N}(z, \delta)$ for the open ball of radius $\delta$ around $z$ in $\C$. 
\begin{definition}[Absolute zero-freeness]
\label{def:absolute_zerofree}
    Given $\beta \ge 0$, $\Delta \in \N$, $\lambda > 0$ and $\delta > 0$, we say that the Ising model is \emph{absolutely $\delta$-zero-free at activity $\lambda$} if for all graphs $G \in \mathcal{G}_{\Delta}$, all pinnings $\tau_U$ with $U \subseteq V$ and all $\lambda' \in \mathcal{N}(\lambda, \delta)$ it holds that $Z^{\tau_U}_{G}(\beta, \lambda') \neq 0$.
\end{definition}

We now define $\lambda_a(\Delta, \beta)$ as follows.  
\begin{definition}
    For $\Delta \in \N$ and $\beta \ge \beta_u(\Delta)$ we set $\lambda_a(\Delta, \beta)$ to be the smallest $\lambda_a \ge 1$ such that for every compact set $D \subset (\lambda_a, \infty)$ there is some $\delta > 0$ such that for all $\lambda \in D$ the Ising model is absolutely $\delta$-zero-free at $\lambda$.
\end{definition}

An important implication of absolute zero-freeness is given in the following theorem.
Its proof follows a similar argument to those in \cite{chen2022spectral} while using the ferromagnetism of the model and Montel's theorem (see \Cref{sec:stability}) to avoid the requirement of multivariate zero-freeness.
A full proof can be found in \Cref{appendix:zero_freeness_SI}.
\begin{theorem} \label{thm:zero_free_SI}
    Fix $\beta \ge 0$ and $\Delta \in \N$. 
    Let $D \subset \R_{>0}$ be compact and assume there is some $\delta > 0$ such that the ferromagnetic Ising model is absolutely $\delta$-zero-free at every $\lambda \in D$. 
    Then, there is some constant $C > 0$, only depending on $D$, $\lambda$, $\beta$ and $\Delta$, such that for all $\lambda \in D$, $G \in \mathcal{G}_{\Delta}$ and all pinnings $\tau_U$ it holds that $\hat{\mu}^{\tau_U}_{G, \beta, \lambda}$ is $C$-$\ell_{\infty}$-independent.
\end{theorem}

Lastly, we note that the two thresholds $\lam_a$ and $\lam_u$ cannot be equal to one another in general:

\begin{lemma} \label{lemma:zeros}
    There are $\Delta \ge 3$ and $\beta > \beta_u(\Delta)$ such that $\lambda_u(\Delta, \beta) < \lambda_a(\Delta, \beta)$.
\end{lemma}

\begin{proof}
    Suppose the opposite was true.
    Then, by definition of $\lambda_a$ and symmetry of the fixed-magnetization model, it holds that for all $\Delta \ge 3$, $\beta > \beta_u(\Delta)$ and compact $D \subset (0, 1/\lambda_u(\Delta, \beta))$ there is some $\delta = \delta(\Delta, \beta, D)$ such that for all $\lambda \in \mathcal{N}(D, \delta)$, all $G \in \mathcal{G}_{\Delta}$ and all pinnings $\tau_U$ we have $Z^{\tau_U}_G(\beta, \lambda) \neq 0$.
    
    Fix $\Delta \ge 3$ and let $G = (V,E) \in \mathcal{G}_{\Delta}$.
    Create a set of vertices $W = \{w_v : v \in V\}$ and edges $F = \{\{v, w_v\} : v \in V\}$, and define a graph $H(G) = (V \cup W, E \cup F) \in \mathcal{G}_{\Delta + 1}$.
    Observe that for every pinning $\tau_U$ for $U \subseteq V$ it holds that
    \[
        Z^{\tau_U, \pmb{+}_W}_{H(G)}(\beta, \lambda) = \lambda^{\size{W}} Z^{\tau_U}_{G}(\beta, e^{\beta} \lambda) ,
    \]
    where $\tau_U, \pmb{+}_W$ is the pinning on $H$ that agrees with $\tau_U$ on $U$ and assigns all vertices in $W$ to $+1$. 
    Hence, by our assumption, it holds that for all $\beta > \beta_u(\Delta + 1)$ and $\lambda \in (0, 1/\lambda_u(\Delta + 1), \beta)$ there is some $\delta = \delta(\Delta + 1, \beta,\lambda)$ such that the ferromagnetic Ising model on $\mathcal{G}_{\Delta}$ is $\delta$-absolutely zero-free at $e^{\beta} \lambda$.
    In particular, suppose we can choose $\Delta \ge 3$, $\beta > \beta_u(\Delta) \ge \beta_u(\Delta + 1)$ and $\lambda \in (0, 1/\lambda_u(\Delta + 1, \beta))$ such that $\lambda' \coloneqq e^{\beta} \lambda \in (1/\lambda_u(\Delta, \beta), \lambda_u(\Delta, \beta))$; then this would imply absolute zero-freeness (and hence rapid mixing of Glauber dynamics) for some $\lambda'$ outside the tree-uniqueness regime, which contradicts our slow mixing result \Cref{thm:GlauberRG}.
    To prove that such a choice of $\Delta$, $\beta$ and $\lambda$ is indeed possible, we reference the proof that weak spatial mixing does not imply strong spatial mixing for the ferromagnetic Ising model, found in Appendix 2 of \cite{sinclair2014approximation}. 
\end{proof}

\section{Rapid Mixing}\label{sec:fast}
\begin{figure}[!htbp]
\centering
\vspace{-1ex}
$$\xymatrix{
 & \txt{$\ell_{\infty}$-independence \\ for all pinnings\\ (grand canonical)} \ar[dr]|-{\text{Sec. \ref{sec:stability}, \ref{sec:spectral_independence}}} & && && \\
 \txt{Absolute \\ zero-freeness} \ar[ur] \ar[dr]|-{\text{Edgeworth expansion. Sec. } \ref{subsec:edgeworth}}  && \txt{$\ell_{\infty}$-independence \\ for ``small'' pinnings\\ (canonical)}\ar[rr]^{\text{Local-to-global.}}_{\text{Sec. } \ref{sec:loc_scheme}}  && \txt{$\Omega(1)$-spectral gap\\ of $(k, l)$-down-up walk\\ for ``small'' pinnings} 
 \ar[dd]|-{\txt{\scriptsize{Localization schemes. Sec.~\ref{sec:loc_scheme}}}}&& \\
&  \txt{Strong LCLT \\ for ``small'' pinnings} \ar[ur]|-{\text{Sec. \ref{sec:stability}, \ref{sec:spectral_independence}}}& && && \\
\txt{Contractive \\ coupling} \ar[r]_-{\txt{ \scriptsize{Sec. \ref{sec:rapid_large}}}} &&
\txt{\hspace{-25ex}$\Omega(1/k)$-spectral gap of \\ \hspace{-30ex}down-up walk for \\\hspace{-30ex} ``large'' pinnings} 
\ar[rr]^-{\txt{\scriptsize{Localization schemes.}}}_-{\text{Sec. } \ref{sec:loc_scheme}} &&\txt{$\Omega(1/k)$-spectral gap \\for down-up walk \\and Kawasaki dynamics}
} $$
\vspace{-0.5ex}
\caption{The structure of the rapid mixing proof }
\label{fig:structure}
 \end{figure}
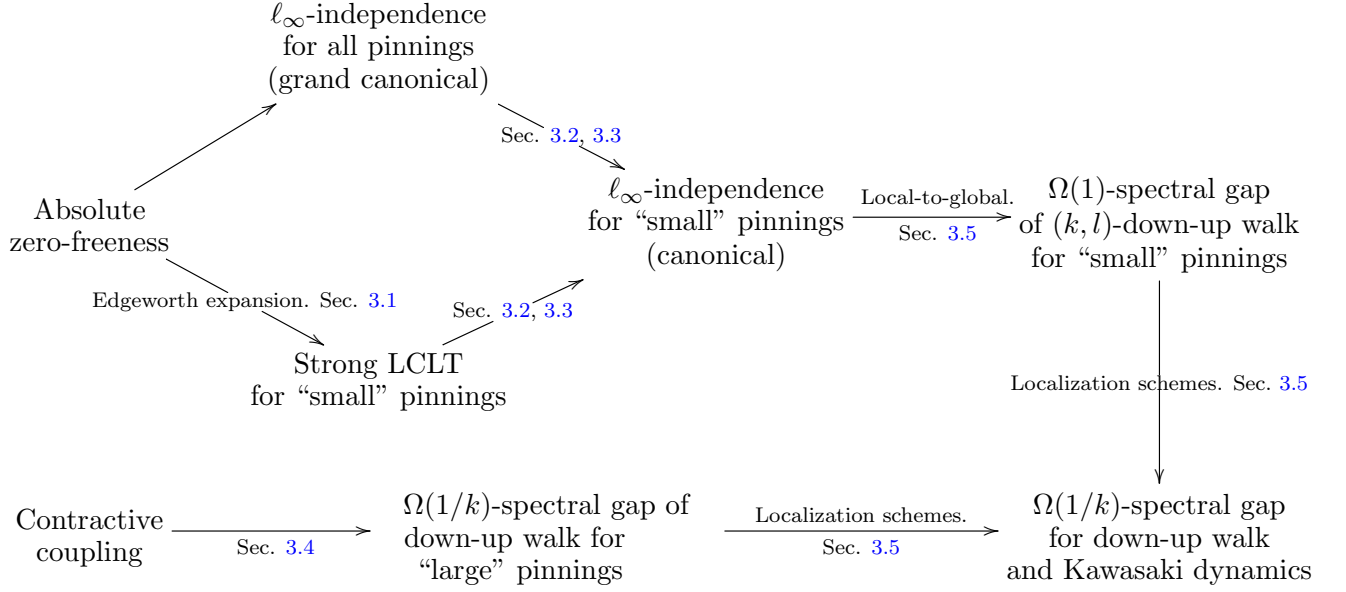

The main goal of this section is to prove conditions on $k$ such that the spectral gap of $\kawasaki_{\beta, k}$ is $\Omega(1/n)$.
Our setting continues to be an arbitrary $G \in \mathcal G_{\Delta}$. Our first step (\Cref{subsec:edgeworth}) is to derive a refined local central limit theorem for the number of $+1$ spins in the Ising model by using an Edgeworth expansion. 
Using this, we prove a stability result for the distribution of the number of $+1$ spins under adding vertices to the pinning (\Cref{sec:stability}).
We are then able to derive spectral independence for $\hat{\mu}^{U}_{\beta, k}$ for all $k - \size{U} \in \Omega(n)$ in \Cref{sec:spectral_independence}.
Until this point, we follow the template of~\cite{JMPV23} in their approach to fixed-size independent sets, but for smaller values of $k-|U|$, we require different techniques.
For this regime, we separately obtain an $\Omega(1/k)$ bound on the spectral gap of $\plusDownUp^{U}_{\beta, k}$ whenever $k - \size{U} \le \alpha n$ for some $\alpha > 0$, which we derive via a contracting coupling in \Cref{sec:rapid_large}. Putting everything together using a localization scheme and a local-to-global result gives the desired spectral gap for $\plusDownUp_{\beta, k}$ in \Cref{sec:loc_scheme}. By a comparison argument, this result carries over to $\kawasaki_{\beta, k}$. The structure of the entire proof is illustrated in Figure \ref{fig:structure}.

\subsection{Edgeworth Expansion} \label{subsec:edgeworth}
All results in this subsection should be understood in the context of the following assumptions.
\begin{condition}\label{cond:fast}
    \begin{enumerate}
        \item Let $\beta \ge 0$, and let $D \subset \R_{>0}$ be compact such that there is some $\delta > 0$ for which the Ising model is absolutely $\delta$-zero-free for all $\lambda \in D$. Further, let $\lambda \in D$.
        \item Let $\alpha \in [0, 1)$, let $U \subset V$ with $\size{U} \le \alpha n$ and let $\tau_U$ be a pinning of $U$.
        \item Let $\sigma \sim \mu^{\tau_U}_{\beta, \lambda}$ and let $X = \size{\sigma}^{+}$.
    \end{enumerate}
\end{condition}

Because many proofs in this subsection are identical or analogous to those appearing elsewhere \cite{davies2023approximately, JMPV23, JPSS21}, we either omit proofs or postpone them to \Cref{sec:proof edgeworth}.

Our first result is a strengthened version of a local central limit theorem for $X$. Recall that the cumulants of a random variable are defined by the {\em cumulant generating function} $K_X(t) = \log(\E[e^{tX}])$. For $j \in \N$, the {\em $j$th cumulant} of $X$ is the coefficient of $t^j/j!$ in the Maclaurin series of $K_{X}(t)$, which can be computed as 
$$\kappa_j(X) = \frac{d^j}{dt^j} K_X(t) \Big\rvert_{t = 0} .$$

\begin{theorem} \label{thm:edgeworth}
    Suppose Condition~\ref{cond:fast} holds.
    Let $d \in \N$ and let $\ell \in \R$ such that $\E[X] + \ell \in \Z_{\geq 0}.$
    Set $s = \sqrt{\Var(X)}$ and $\beta_j = \frac{\kappa_j(X)}{j! s^j}$ for all $j \in \N$, and write $H_k(\cdot)$ for the $k^{\text{th}}$ Hermite polynomial. 
    It holds that
    \[
        \mu_{\beta, \lambda}^{\tau_{U}}(X - \E[X] = \ell) = \frac{e^{-\frac{\ell^2}{2 s^2}}}{\sqrt{2 \pi} s}  \left(1 + \sum_{r \ge 3} H_{r}(\ell/s) \sum_{k_3, \dots, k_{2d+1}} \prod_{j=3}^{2d+1} \frac{\beta_j^{k_j}}{k_j !} \right) + O\left(n^{-d}\right) 
    \]
    where the inner sum is over tuples $k_3, \dots, k_{2d+1} \in \Z_{\geq 0}$ such that $\sum_j k_j \cdot j = r$ and $\sum_j k_j \cdot \frac{j-2}{2} \le d$, and the implied constants in the asymptotic notation depend only on $\Delta, \beta, \delta,  D$, $d$ and $\alpha$. 
\end{theorem}

To prove \Cref{thm:edgeworth}, we use an Edgeworth expansion similar to \cite[Proposition 10]{JMPV23}, which requires three ingredients.
We begin with a lemma that bounds the variance of $X$.
\begin{lemma} \label{lemma:variance} 
    Assume Condition~\ref{cond:fast} holds. Then $\Var(X) = \Theta(n)$ with implied constants depending only on $\Delta, \beta$, $\delta$, $D$ and $\alpha$.
\end{lemma}
The proof of Lemma~\ref{lemma:variance} follows  the proof of~\cite[Lemma 9]{davies2023approximately} and~\cite[Lemma 3.2]{JPSS21}. The lower bound is a general property of spin systems, while the upper bound relies on the zero-freeness of the partition function for all considered pinnings.

The second ingredient is the following bound on the characteristic function of $X$. Its proof is analogous to that of \cite[Lemma 3.5]{JPSS21}. 
\begin{lemma} \label{lemma:characteristic_function}
    Assume Condition~\ref{cond:fast} holds.
    There exists $c = c(\Delta, \beta, D, \alpha) > 0$ such that for all $t \in [- \pi, \pi]$ it holds that $|\E[e^{itX}]| \le e^{- c t^2 n}$.
\end{lemma}

The last ingredient that we need for deriving the Edgeworth expansion is an asymptotic bound on the cumulants of $X$, the proof of which is analogous to that of \cite[Lemma 11]{JMPV23}. 
\begin{lemma} \label{lemma:cumulant_bounds}
    Suppose Condition~\ref{cond:fast} holds.  
    There exists some $\varepsilon = \varepsilon(\Delta, \beta, D, \delta) > 0$ such that for all $j \in \N$, it holds that
    \[
        |\kappa_j(X)| = O(j! 2^{j} \varepsilon^{-j} n) ,
    \]
    where the implied constants depend on $\Delta$, $\beta$, $\delta$ and $D$.
    Moreover, for $t \in \C$ with $|t| \le \varepsilon/4$ and all $d \in \N$ it holds that 
    \[
        \left\lvert \ln \frac{Z^{\tau_U} (\beta, \lambda e^{t})}{Z^{\tau_U} (\beta, \lambda)} - \sum_{j=1}^{d} \kappa_{j}(X) \frac{t^j}{j!} \right\rvert = O(|2t/\varepsilon|^{d+1} n)
    \]
    with constants depending on $\Delta$, $\beta$, $\delta$ and $\lambda$.
\end{lemma}

We now have all the tools to prove \Cref{thm:edgeworth}. The main idea is to rewrite the probability mass function of $X$ via its Fourier transformation and compare it to a suitable scaled normal distribution. 
\Cref{lemma:variance,lemma:characteristic_function,lemma:cumulant_bounds} are then used to control the error terms. The details can be found in Appendix~\ref{proof:edgeworth}.

Throughout the rest of the section, we use the following corollary of \Cref{thm:edgeworth}.
\begin{corollary} \label{cor:edgeworth}
    Suppose in the setting of \Cref{thm:edgeworth} that $|\ell| \le L$ for some $L \in \Z_{\geq 0}$. Then
    \[
        \mu_{\beta, \lambda}^{\tau_{U}}(X - \E[X] = \ell) = \frac{e^{-\frac{\ell^2}{2 s^2}}}{\sqrt{2 \pi} s} + O\left(n^{-3/2}\right) 
    \]
    with an implied constant in the asymptotic notation that depends only on $\Delta, \beta, \delta, D$, $L$ and $\alpha$.
\end{corollary}
The result follows from bounding the higher-order terms in \Cref{thm:edgeworth}.

\subsection{Stability} \label{sec:stability}
We proceed under the assumptions of \Cref{cond:fast}.
Our main application of \Cref{cor:edgeworth} is to obtain the following stability result for the probability mass function of $X$ when adding vertices to a pinning. 

\begin{lemma} \label{lemma:stability}
    Suppose \Cref{cond:fast} holds and assume further that $\size{U} + 1 \le \alpha n$.
    Let $k \in \Z_{\geq 0}$ be such that $|\E[X] - k| \le L$ for some $L \in \R_{\ge 0}$.
    For all $v \in V \setminus U$ it holds that 
    \begin{gather}
        \mu_{\beta, \lambda}^{\tau_U} (X=k) = \Theta(n^{-1/2}), \label{lemma:stability:eq1} \\
        \left\lvert \mu_{\beta, \lambda}^{\tau_U} (X=k) - \mu_{\beta, \lambda}^{\tau_U} (X=k \mid \sigma(v) = +1) \right\rvert = O(n^{-3/2}) \label{lemma:stability:eq2} 
    \end{gather}
    with implied constants depending only on $\Delta, \beta$, $\delta$,  $D$, $L$ and $\alpha$.
\end{lemma}

While \eqref{lemma:stability:eq1} in \Cref{lemma:stability} follows immediately from \Cref{cor:edgeworth}, proving \eqref{lemma:stability:eq2} 
requires us to bound how much the mean and variance of $X$ change when a spin assignment for $v$ is added to $\tau_U$.
To obtain this, we derive the following more general stability result for the cumulants of $X$. 

\begin{lemma} \label{lemma:cumulant_stability}
    Suppose Condition~\ref{cond:fast} holds.
    Let $v \in V \setminus U$, and let $\tau_U, \pmb{+}_v$ denote the pinning on $U \cup \{v\}$ that maps $v$ to $+1$ and all other vertices $u \in U$ to $\tau_{U}(u)$.
    Let $X^{+} = \size{\sigma'}^{+}$ for $\sigma' \sim \mu^{\tau_U, \pmb{+}_v}_{\beta, \lambda}$.
    For all $j \in \N$ it holds that $|\kappa_j(X^+) - \kappa_j(X)| = O(1)$ with implied constants only depending on $\Delta$, $\beta$, $\delta$, $D$ and $j$.
\end{lemma}

The analog of \Cref{lemma:cumulant_stability} for the hard-core model was proven in \cite{JMPV23}.
However, their arguments are tailored to the hard-core model and do not apply in our setting.
Instead, we provide a more general argument that uses Montel's characterization of normal families of complex functions, and that is inspired by a similar application in \cite{regts2023absence}.

A family $\mathcal{F}$ of meromorphic functions from a connected open set $U \subseteq \C$ to the extended complex plane $\hat{\C} = \C \cup \{\infty\}$  is called \emph{normal} if every sequence in $\mathcal{F}$ has a subsequence that converges uniformly on every compact $K \subset U$ with respect to the spherical metric on $\hat{\C}$.
An immediate consequence is that if all functions in $\mathcal{F}$ are holomorphic, then every sequence in $\mathcal{F}$ has a subsequence that either converges uniformly (with respect to the standard metric on $\C$) to a holomorphic function on all compact subsets of $U$, or that converges uniformly to the constant $\infty$ function (with respect to spherical metric on $\hat{\C}$) on all compact subsets (see \cite{zalcman1998normal} for a detailed discussion).
A simple criterion for  normality is given by Montel's theorem.
\begin{theorem}[Montel] \label{thm:montel}
    Let $U \subseteq \C$ be a connected open set, and let $\mathcal{F}$ be a family of meromorphic functions from $U$ to $\hat{\C}$.
    Suppose there exist distinct points $a, b, c \in \hat{\C}$ such that $f(z) \notin \{a, b, c\}$ for all $f \in \mathcal{F}$. Then $\mathcal{F}$ is normal.
\end{theorem}

Normal families and Montel's theorem have previously been used for proving uniform bounds on the complex occupation probabilities in the hard-core model, as in \cite{regts2023absence}.
A general version of the argument, which is also applicable to our setting, is given in the following statement.
\begin{corollary} \label{cor:montel}
    Let $U \subseteq \C$ be a connected open set, and let $\mathcal{F}$ satisfy the assumptions of \Cref{thm:montel}.
    Suppose further that all functions in $\mathcal{F}$ are holomorphic, and let $K \subset U$ be a compact set such that for some $z \in K$, it holds that $\sup_{f \in \mathcal{F}}\{ |f(z)|\} < \infty$.
    Then $\mathcal{F}$ is uniformly bounded on $K$.
\end{corollary}

\begin{proof}
    Suppose for the sake of contradiction that $\mathcal{F}$ is not uniformly bounded on $K$. Then there is a sequence of functions $f_n \in \mathcal{F}$ and a sequence $z_n \in K$ such that $f_n(z_n) \ge n$ for all $n \in \N$.
    By \Cref{thm:montel}, we know that $\mathcal{F}$ is normal.
    Let $(f_{n_k})_{k \in \N}$ be the subsequence guaranteed by \Cref{thm:montel}, and let $g$ denote its limit.
    By the fact that $\sup_{f \in \mathcal{F}}\{|f(z)|\} < \infty$, we know that $g(z) < \infty$. Thus, $g$ is not constant $\infty$ on $K$.
    This implies that $f_{n_k} \to g$ uniformly with respect to the Euclidean distance on $\C$ and hence $g$ is holomorphic too. 
    In particular, as $g$ is continuous and $K$ is compact, this implies that the image of $K$ under $g$ is bounded.
    Writing $M \coloneqq \sup_{ y \in K} \{|g(y)|\} < \infty$ we have
    \[
        n_k \le f_{n_k} (z_{n_k}) \le \sup_{y \in K}\{|f_{n_k}(y) - g(y)|\} + M .
    \]
    However, this is a contradiction since $n_k \to \infty$ but $\sup_{y \in K} \{|f_{n_k}(y) - g(y)|\} \to 0$ by uniform convergence as $k \to \infty$, which shows that $\mathcal{F}$ must be uniformly bounded on $K$.
\end{proof}

We are now ready to prove \Cref{lemma:cumulant_stability} using \Cref{cor:montel}.
\begin{proof}[Proof of \Cref{lemma:cumulant_stability}]
    First observe that
    \begin{align*}
        |\kappa_j(X^+) - \kappa_j(X)| 
        &= \left\lvert \frac{\text{d}^j}{\text{d} t^j} \ln \frac{Z^{\tau_U, \pmb{+}_v}(\lambda e^{t})}{Z^{\tau_U, \pmb{+}_v}(\lambda)} \Big\rvert_{t = 0} - \frac{\text{d}^j}{\text{d} t^j} \ln \frac{Z^{\tau_U}(\lambda e^{t})}{Z^{\tau_U}(\lambda)} \Big\rvert_{t = 0} \right\rvert \\
        &= \left\lvert  \frac{\text{d}^j}{\text{d} t^j} \ln \frac{Z^{\tau_U, \pmb{+}_v}(\lambda e^{t})}{Z^{\tau_U}(\lambda e^{t})}\Big\rvert_{t = 0} \right\rvert , 
    \end{align*}
    given the involved derivatives are well-defined.
    Indeed, note that \Cref{cond:fast} provides that all involved partition functions are non-zero complex analytic functions in $t$ for $|t| < \varepsilon$ and $\varepsilon = \varepsilon(\Delta, \beta, D, \delta) > 0$ as in \Cref{lemma:cumulant_bounds}.
    In particular, $t \mapsto \ln \frac{Z^{\tau_U, \pmb{+}_v}(\lambda e^{t})}{Z^{\tau_U}(\lambda e^{t})}$ is analytic in the domain $\mathcal{N}(0, \varepsilon)$ for all $\lambda \in D$, $G \in \mathcal{G}_{\Delta}$ and all choices of $\tau_U$ and $v$.
    
    By applying Cauchy's integral formula, it thus suffices to show that
    \[
        \max_{|t| \le \varepsilon/2 } \left\lvert \ln \frac{Z^{\tau_U, \pmb{+}_v}(\lambda e^{t})}{Z^{\tau_U}(\lambda e^{t})} \right\rvert \le C
    \]
    for some $C$ that only depends on $\Delta$, $\beta$ $\delta$, and $D$.
    This, in turn, follows from
    \begin{equation}\label{eqn:partition-ratio-bd}
        1/C' \le \left\lvert \frac{Z^{\tau_U, \pmb{+}_v}(\lambda e^{t})}{Z^{\tau_U}(\lambda e^{t})} \right\rvert
        \le C'
    \end{equation}
    for $C'$ depending on $\Delta$, $\beta$, $\delta$ and $D$.
    
    We derive such a bound by using \Cref{cor:montel}.
    For the upper bound, consider the family of functions 
    \[
        \mathcal{F} = \left\{t \mapsto \frac{Z_{G}^{\tau_U, \pmb{+}_v}(\lambda e^{t})}{Z_{G}^{\tau_U}(\lambda e^{t})} \,:\, \lambda \in D, G \in \mathcal{G}_{\Delta}, U \subset V, \tau_U: U \to \{+1, -1\}, v \in V\setminus U \right\}
    \]
    on the domain $\mathcal{N}(0, \varepsilon)$, where $\Delta$, $\beta$ and $D$ are fixed and satisfy the assumptions of the lemma.
    We aim to use \Cref{cor:montel} to show that $\mathcal{F}$ is uniformly bounded on the compact set $|t| \le \varepsilon/2$.
    Since all involved partition functions are non-zero and analytic in $t$, all functions in $\mathcal{F}$ are holomorphic.
    Moreover, for $t = 0$, we have that
    \[
        \frac{Z^{\tau_U, \pmb{+}_v}(\lambda e^{t})}{Z^{\tau_U}(\lambda e^{t})} = \mu^{\tau_U}_{\beta, \lambda} (\sigma(v) = +1) \in (0, 1).
    \]
    Thus, it remains to show that $\mathcal{F}$, viewed as functions to $\hat{\C}$, avoids at least three points.
    Due to the zero-freeness of the partition functions, all functions in $\mathcal{F}$ must avoid $0$ and $\infty$.
    Further, note that 
    \[
        Z^{\tau_U, \pmb{+}_v}(\lambda e^{t}) 
        = Z^{\tau_U}(\lambda e^{t}) - Z^{\tau_U, \pmb{-}_v}(\lambda e^{t}).
    \]
    As $Z^{\tau_U, \pmb{-}_v}(\lambda e^{t}) \neq 0$, this means that $\mathcal{F}$ also avoids $1$, and \Cref{cor:montel} implies the desired uniform bound.

    To obtain the lower bound in \eqref{eqn:partition-ratio-bd}, we simply apply the same argument as for the upper bound to the inverse fraction $\frac{Z^{\tau_U}(\lambda e^{t})}{Z^{\tau_U, \pmb{+}_v}(\lambda e^{t})}$, arguing that the resulting family $\mathcal{F}$ again avoids $0, \infty$ and $1$, and using the fact that for $v \notin U$ and $\lambda \in D$, we have $\mu^{\tau_U}_{\beta, \lambda} (\sigma(v) = +1)$ bounded away from $0$ by a constant that only depends on $\Delta$, $\beta$ and $D$.
\end{proof}

The proof of \Cref{lemma:stability} then follows from \Cref{cor:edgeworth} and \Cref{lemma:cumulant_stability}. 
 
\begin{proof}[Proof of \Cref{lemma:stability}]\label{proof:stability}
    First, we note that \eqref{lemma:stability:eq1} follows immediately from \Cref{cor:edgeworth} and the fact that $s = \sqrt{\Var(X)} = \Theta(\sqrt{n})$ by \Cref{lemma:variance}.
    To prove \eqref{lemma:stability:eq2}, let $\sigma' \sim \mu_{\beta, \lambda}^{\tau_{U}, \pmb{+}_v}$ and $X' = \size{\sigma}^{+}$, where $\tau_{U}, \pmb{+}_v$ is the pinning that agrees with $\tau_U$ on $U$ and maps $v$ to $+1$.
    Note that $\mu_{\beta, \lambda}^{\tau_U} (X=k \mid \sigma(v) = +1) = \mu_{\beta, \lambda}^{\tau_{U}, \pmb{+}_v}(X' = k)$, and set $\ell = k - \E[X]$, $s = \sqrt{\Var(X)}$, $\ell' = k - \E[X']$ and $s' = \sqrt{\Var(X')}$.
    By \Cref{lemma:cumulant_stability} we know that
    \[
        |\E[X] - \E[X']| = |\kappa_1(X) - \kappa_1(X')| = O(1)
    \]
    with constants depending on $\Delta$, $\beta$, $\delta$ and $D$.
    In particular, we have $|\ell'| \le L + O(1)$ and, by \Cref{cor:edgeworth}, it holds that
    \[
        \absolute{\mu_{\lambda}^{\tau_{U}}(X = k) - \mu_{\lambda}^{\tau_{U}, \pmb{+}_v}(X' = k)} = \absolute{\frac{e^{-\frac{\ell^2}{2 s^2}}}{\sqrt{2 \pi} s} - \frac{e^{-\frac{\ell'^2}{2 s'^2}}}{\sqrt{2 \pi} s'}} + O\left(n^{-3/2}\right)  
    \]
    with constants depending on $\Delta$, $\beta$, $\delta$, $D$, $L$ and $\alpha$.
    We proceed by noting that by \Cref{lemma:variance} both $s = \Theta(\sqrt{n})$ and $s' = \Theta(\sqrt{n})$ with constant depending on $\Delta$, $\delta$, $\beta$, $D$ and $\alpha$.
    Further, by \Cref{lemma:cumulant_stability} we obtain 
    \[
        |s - s'| \le \frac{|s^2 - s'^2|}{s + s'} = \frac{|\kappa_2(X) - \kappa_2(X')|}{s + s'} = O(n^{-1/2}) 
    \]
    and thus
    \[
        \absolute{\frac{1}{s} - \frac{1}{s'}} = \frac{|s' - s|}{s \cdot s'} = O(n^{-3/2}) .
    \]
    Finally, by Taylor expansion we have
    \[
        \absolute{\frac{e^{-\frac{\ell^2}{2 s^2}}}{\sqrt{2 \pi} s} - \frac{e^{-\frac{\ell'^2}{2 s'^2}}}{\sqrt{2 \pi} s'}} = \absolute{\frac{1}{s} - \frac{1}{s'}} + O(n^{-3/2})
        = O(n^{-3/2}),
    \]
    which concludes the proof.
\end{proof}

\subsection{Spectral Independence}
\label{sec:spectral_independence} 
In this section, our goal is to show that for all $U \subset V$ sufficiently small compared to $k = \gamma n$, it holds that $\hat{\mu}^{U}_{\beta, k}$ satisfies $C$-spectral independence for some constant $C$ that only depends on $\Delta$, $\beta$ and $\gamma$.
From now on, it will suffice to focus exclusively on plus pinnings.
We show the following $\ell_{\infty}$-independence result.
\begin{theorem} \label{thm:spectral_independence_fm}
    Assume $0 \le \beta < \beta_u (\Delta)$ and $\gamma \in (0, 1/2]$, or $\beta \ge \beta_u (\Delta)$ and $\gamma \in (0, \frac{1 - \eta_a}{2})$ for $\eta_a = \eta_a(\Delta, \beta)$. 
    For all $k \coloneqq \gamma n \in \N$, all $\alpha \in [0, \gamma)$ and $U \subset V$ with $\size{U} \le \alpha n$ it holds that $\hat{\mu}^{U}_{\beta, k}$ is $C$-$\ell_{\infty}$-independent for a constant $C$ depending only on $\Delta, \beta$, $\gamma$ and $\alpha$.
\end{theorem}

The main idea for proving \Cref{thm:spectral_independence_fm} is to observe that, for all $\lambda > 0$, we may think of $\hat{\mu}^{U}_{k}$ as $\mu^{U}_{\lambda}$ conditioned on $X = k$ (where again $X = \size{\sigma}^{+}$ for $\sigma \sim \mu^{U}_{ \lambda}$).
Using this perspective, it is straightforward to see that $\hat{\mu}^{U}_{k}$ is $C$-$\ell_{\infty}$-independent if we can choose $\lambda$ such that $\mu^{U}_{\lambda}$ is $C'$-$\ell_{\infty}$-independent for some constant $C'$ and if we can show
\begin{align}
    \frac{\mu^{U}_{\lambda}(X = k \mid \sigma(v) = +1, \sigma(w) = +1)}{\mu^{U}_{\lambda}(X = k \mid \sigma(w) = i)} 
    = 1 + O(1/n) .
    \label{eq:use_stability}
\end{align}
For the latter, we aim to use our stability result from \Cref{lemma:stability}. 
This imposes the additional restrictions that $\lambda$ should be such that $E[X] \approx k$, and that $Z(\beta, \lam)$ is zero-free in a neighborhood of $\lambda$.
More precisely, since we want $C$ to be uniformly bounded over $G$ and $U$ as given in the theorem, we want that $C'$ and the implied constants in \eqref{eq:use_stability} only depend on $\beta$, $\Delta$, $\alpha$ and $\gamma$.
For this, it suffices if, for every $\beta, \Delta, \gamma$ and $\alpha$ as in \Cref{thm:spectral_independence_fm}, we can find a compact set $D \subset \R_{>0}$ such that the following hold.
\begin{enumerate}[(a)]
    \item For every $G \in \mathcal{G}_{\Delta}$ and every $U \subset V$ with $\size{U} \le \alpha n$ there is some $\lambda \in D$ such that $\E[X] = \gamma n$.\label{requirement:expectation}
    \item There is some $C' \ge 0$ such that the Ising model is $C'$-$l_{\infty}$-independent under any plus pinnings and for all $\lambda \in D$. \label{requirement:SI}
    \item There is some $\delta > 0$ such that the Ising model is absolutely $\delta$-zero-free for all $\lambda \in D$. \label{requirement:root}
\end{enumerate}
    
By \cite[Theorem 4.5]{shao2021contraction} and our definition of $\lambda_a$, part \ref{requirement:root} is satisfied if $\beta < \beta_u(\Delta)$ and $D$ is arbitrary, or if $\beta \ge \beta_u(\Delta)$ and $D \subset (0, 1/\lambda_a)$ for $\lambda_a = \lambda_a(\Delta, \beta)$.
By \Cref{thm:zero_free_SI}, part \ref{requirement:SI} holds for the same choice of $D$. 
The regime of $\gamma$ such that we can simultaneously satisfy \ref{requirement:expectation} is given by the following statement.
\begin{lemma} \label{lemma:translation_magnetization}
    Assume $0 \le \beta < \beta_u (\Delta)$ and $\gamma \in (0, 1/2]$, or $\beta \ge \beta_u (\Delta)$ and $\gamma \in (0, \frac{1 - \eta_a}{2})$ for $\eta_a = \eta_a(\Delta, \beta)$. 
    For all $\alpha \in [0, \gamma)$ there exists $\lambda_1 \coloneqq \lambda_1(\Delta, \beta, \gamma, \alpha) > 0$ and $\lambda_2 \coloneqq \lambda_2(\Delta, \beta, \gamma) \ge \lambda_1$ such that the following holds: 
    \begin{enumerate}
        \item If $\beta < \beta_u(\Delta)$ then $[\lambda_1, \lambda_2] \subset (0, 1]$, and if $\beta \ge \beta_u(\Delta)$ then $[\lambda_1, \lambda_2] \subset (0, 1/\lambda_a)$.
        \item For every $G \in \mathcal{G}_{\Delta}$ and $U \subset V$ with $\size{U} \le \alpha n$ there is some $\lambda \in [\lambda_1, \lambda_2]$ such that $\E[X] = \gamma n$ for $\sigma \sim \mu^{U}_{G, \beta, \lambda}$, $X = \size{\sigma}^{+}$.
    \end{enumerate}
\end{lemma}

\begin{proof}
We begin with some observations relating the magnetization to the external field. For $X = |\sig|^+$ and $U \subset V$, note that
  \[
        \E[X] = \sum_{v \in V} \mu_{\beta, \lambda}^{U} (\sigma(v) = +1) = \size{U} + \sum_{v \in V \setminus U} \mu_{\beta, \lambda}^{U} (\sigma(v) = +1) .
    \]
        Observing that for all $v \in V \setminus U$ it holds that
    \[
        \mu_{\beta, \lambda}^{U} (\sigma(v) = +1) \le \frac{\lambda e^{\beta \Delta}}{1 + \lambda e^{\beta \Delta}} \le \lambda e^{\beta \Delta}\,,
    \]
    we then get the general upper bound 
    \begin{equation}\label{eqn:magnetization-lb}
    \E[X] \le \size{U} + \lambda e^{\beta \Delta} \size{V \setminus U}\,.
    \end{equation}
    
Now suppose $\gamma \in (0, 1/2]$. We claim that there exists $\lam_1(\Delta, \beta, \gam, \alpha)$ with $0 < \lam_1 \leq 1$ such that $\E[X] = \gam n$ for some $\lam \in [\lam_1, 1]$. To see this, consider the map $f: \lambda \mapsto \E[X]$ and note that $f$ is a rational function in $\lambda$ and thus continuous.
    Thus, if $f(\lambda_1) \le \gamma n \le f(1)$ for some suitable $\lambda_1 = \lambda_1(\Delta, \beta, \gamma, \alpha)$, then the statement follows from applying the intermediate value theorem.
    By \eqref{eqn:magnetization-lb}, we have
    \[
        f(\lambda) \le \size{U} + \lambda e^{\beta \Delta} \size{V \setminus U} \le \left(\alpha + \lambda e^{\beta \Delta}\right) n .
    \]
    Thus, choosing $\lambda_1 = (\gamma - \alpha) e^{-\beta \Delta}$ yields the lower bound.
    For the upper bound, suppose $U = \emptyset$ and note that $f(1) = n/2$ by symmetry.
    It then follows that $f(1) \ge n/2 \ge \gamma n$ for all $U \subseteq V$ by the FKG inequality.

For the case that $\beta > \beta_u(\Delta)$, we further want to make sure that we only need to choose external fields with $\lambda \notin [1/\lambda_a(\Delta, \beta), \lambda_a(\Delta, \beta)]$. 
  We first observe that by symmetry of the Ising model, we can show $\mu_{\beta, \lambda} (\sigma) = \mu_{\beta, 1/\lambda} ( - \sigma)$ (where $-\sigma$ is the configuration obtained by reversing every spin).
    Hence, for $\sigma \sim \mu_{\beta, \lambda}$, $X = \size{\sigma}^{+}$ and $\sigma' \sim \mu_{\beta, 1/\lambda}$ and $X' = \size{\sigma'}^{+}$ it holds that
    \[
        \E[X] = n - \E[X'] = \frac12(n - \E[\sum_{v \in V} \sigma'(v)])
    \]
    Since we showed above that we may take $\lam < 1$, we have $\frac{1}{\lam} \geq 1$ and so we can apply Theorem~\ref{thm:extremalmag} to conclude that 
    \begin{equation}\label{eqn:extremal-magnetization}
    \E[X] \geq (1- \eta^+_{\Delta, \beta, 1/\lam})\frac{n}{2}
    \end{equation}
    We now take $\gam \in (0, \frac{1-\eta_a}{2})$. As before, we consider the map $f: \lambda \mapsto \E[X]$ and note that $f$ is continuous.
   Thus, it suffices to find $0 < \lambda_1 \le \lambda_2 < 1/\lambda_a$ depending only on $\Delta$, $\beta$, $\gamma$ and $\alpha$ such that $f(\lambda_1) \le \gamma n \le f(\lambda_2)$.
   The first part of the inequality follows from setting $\lambda_1 = (\gamma - \alpha) e^{-\beta \Delta}$ and applying \eqref{eqn:magnetization-lb}.
   For the second part, set $\lambda_2$ to be the unique solution to $(1 - \eta^{+}_{\Delta, \beta, 1/\lambda_2})/2 = \gamma$.
   By Proposition~\ref{propIsingTree}, $\lambda \mapsto \eta^{+}_{\Delta, \beta, \lambda}$ is strictly increasing, so $\gamma < \frac{1-\eta_a}{2}$ implies that $1/\lambda_2 > \lambda_a$.
   If $U = \emptyset$, then \eqref{eqn:extremal-magnetization} yields $f(\lambda_2) \ge \gamma n$. 
   For $U \neq \emptyset$, note that the event that an arbitrary set of vertices is mapped to $+1$ is a non-decreasing event. Hence, by linearity of expectation, the expected number of $+1$ vertices is non-decreasing under adding $+1$ pinnings. We can then apply the FKG inequality, which states that if $A, B$ are non-decreasing events, then $\mu(A \mid B) \ge \mu(B)$, to conclude.
\end{proof}

We are now ready to prove our spectral independence result.
\begin{proof}[Proof of \Cref{thm:spectral_independence_fm}]
Throughout, $\beta$ is fixed to satisfy the theorem assumptions, so we omit it from the notation for convenience.  We aim to prove that for $\tau \sim \hat{\mu}_{\beta, k}^{U} = \hat{\mu}_k^U$, it holds that
    \[
        \max_{w \in V \setminus U} \sum_{v \in V \setminus U} \absolute{\hat{\mu}_k^{U}\left(\tau(v) = +1 \mid \tau(w) = +1\right) - \hat{\mu}_k^{U}(\tau(v) = +1)} = O(1)
    \]
    with implied constants depending only on $\Delta, \beta$, $\gamma$ and $\alpha$.
    Note that $k \ge 1$ and $\size{U} < k$, and thus all involved probabilities are well-defined.
    
    By \Cref{lemma:translation_magnetization}
    we can find a compact set $D \subset \R_{> 0}$, only depending on $\Delta$, $\beta$, $\gamma$ and $\alpha$, and $\lambda \in D$ so that for $\sigma \sim \mu_{\lambda}^{U}$ and $X = \size{\sigma}^{+}$ it holds $\E[X]=k$. 
    If $\beta < \beta_u(\Delta)$ then $D \subset (0, 1]$, and if $\beta \ge \beta_u(\Delta)$ then $D \subset (0, 1/\lambda_a)$ for $\lambda_a = \lambda_a(\Delta, \beta)$. Using \cite[Theorem 4.5]{shao2021contraction} in the former case and our definition of $\lambda_a$ in the latter, we know that there is some $\delta > 0$, depending on $\Delta$, $\beta$, $\gamma$ and $\alpha$, such that the Ising model is absolutely $\delta$-zero-free for every $\lambda \in D$. 
    Moreover, by \Cref{thm:zero_free_SI} this implies that there is some $C' \geq 0$, only depending on $\Delta$, $\beta$, $\gamma$ and $\alpha$, such that $\mu^{U}_{\lambda}$ is $C'$-$\ell_{\infty}$-independent for every $\lambda \in D$.
    
    Now, fix $v, w \in V \setminus U$.
    We may assume $v \neq w$ since the case $v = w$ only contributes $O(1)$ to the sum.
    By definition, we have 
    \begin{align*}
        \hat{\mu}_{k}^{U}(\tau(v) = +1 \mid \tau(w) = +1)
        &= \mu_{\lambda}^{U}(\sigma(v) = +1 \mid \sigma(w) = +1, X = k) \\
        &= \mu_{\lambda}^{U}(\sigma(v) = +1 \mid \sigma(w) = +1) \cdot \frac{\mu_{\lambda}^{U\cup w}(X' = k \mid \sigma'(v) = +1)}{\mu_{\lambda}^{U\cup w}(X'=k)} ,
    \end{align*}
    where $\sigma' \sim \mu_{\lambda}^{U\cup w}$ and $X' = \size{\sigma'}^{+}$ for $\lambda$ as above.
    Next, observe that by \Cref{lemma:cumulant_stability} we have
    \[
        |\E[X] - \E[X']| = |\kappa_1(X) - \kappa_1(X')| = O(1)
    \]
    with constants depending on $\Delta$, $\beta$, $\gamma$ and $\alpha$.
    Thus, we have $|\E[X'] - k| = |\E[X] - k| + O(1)$ and applying \Cref{lemma:stability} to $\mu_{\lambda}^{U\cup w}$ we get
    \begin{align*}
        \hat{\mu}_{k}^{U}(\tau(v) = +1 \mid \tau(w) = +1) 
        &= \mu_{\lambda}^{U}(\sigma(v) = +1 \mid \sigma(w) = +1) \cdot \frac{\mu_{\lambda}^{U\cup w}(X'=k) + O(n^{-3/2})}{\mu_{\lambda}^{U\cup w}(X'=k)} \\
        &= \mu_{\lambda}^{U}(\sigma(v) = +1 \mid \sigma(w) = +1) \cdot \left( 1 + O(n^{-1})\right) .
    \end{align*}
    Analogously, we also have
    \[
        \hat{\mu}_{k}^{U}(\tau(v) = +1) 
        = \mu_{\beta, \lambda}^{U}(\sigma(v) = +1) \cdot \left( 1 + O(n^{-1})\right) .
    \]
    Summing over $v$---noting that the asymptotics are independent of $w$ and $v$---and applying the $\ell_{\infty}$-independence for the Ising model concludes the proof. 
\end{proof}

\subsection{Spectral Gap for Large Pinnings}\label{sec:rapid_large}

Our last required ingredient is a proof that the spectral gap of the pinned down-up walk $\plusDownUp^{U}_{\beta, k}$ is in $\Omega(1/n)$ whenever $k = \gamma n$ and $U \subset V$ are such that $k - \size{U}$ is small enough.
In the setting of fixed-size independent sets studied in \cite{JMPV23}, such a result was previously known (and in fact proved by Bubley and Dyer in the paper introducing the path coupling technique~\cite{bubley1997path}).
We prove the following statement for the fixed-magnetization Ising model.  Unlike in the case of the hard-core model, a straightforward application of path coupling with the Hamming metric does not work, and instead we introduce a modified metric on the state space.  
\begin{lemma} \label{lemma:low_magnetization} 
    Let $G \in \mathcal{G}_{\Delta}$ with $n$ sufficiently large. 
    There exists some $\alpha = \alpha(\Delta, \beta) > 0$ such that for all $0 < k \le n/2$ and all $U \subset V$ with $0 < k - \size{U} \le \alpha n$ it holds that $\plusDownUp_{\beta, k}^{U}$ has spectral gap $\Omega(1/k)$ with constants depending on $\beta$ and $\Delta$.
\end{lemma} 

We proceed with a proof of this statement. 
 Throughout this section, we switch back to studying Kawasaki dynamics as a proxy for the down-up walk.
In particular, we need the following notion of Kawasaki dynamics with plus pinnings:
\begin{definition}[Kawasaki dynamics with pinnings] \label{def:kawasaki-pinned}
    For $U \subseteq V$ with $\size{U} < k$ we define the Kawasaki dynamics on $\Omega_k^{U}$ as a Markov chain $\kawasaki_{\beta, k}^{U} = (X_t)_{t \geq 0}$, given by the following update rule:
    \begin{enumerate}
        \item Pick $v \in X_{t}^{-1} (+1) \setminus U$ and $w \in X_{t}^{-1} (-1) \setminus U$ uniformly at random, and set $X \in \Omega_{k}$ such that $X(v) = X_t(w), X(w) = X_t(v)$, and $X(u) = X_t(u)$ for $u \neq v, w$.
        \item Set $X_{t+1} = X$ with probability 
        $$\phi(X_t, X) := \min\left\{1, \frac{\hat{\mu}_{\beta, k}^{U}(X)}{\hat{\mu}_{\beta, k}^{U}(X_t)}\right\}$$ 
        and set $X_{t+1} = X_t$ otherwise.
    \end{enumerate} 
\end{definition}

Note that for $U = \emptyset$ this Markov chain is identical to $\kawasaki_{\beta, k}$.
\Cref{obs:ergodicity} then generalizes as follows.
\begin{observation} \label{obs:ergodicity_pinning}
    For all $U \subset V$ with $\size{U} < k$ it holds that $\kawasaki_{\beta, k}^{U}$ and $\plusDownUp_{\beta, k}^{U}$ are ergodic and reversible with respect to $\hat{\mu}^{U}_{\beta, k}$.
    Moreover, there is a constant $C \ge 1$ that only depends on $\Delta$ and $\beta$, such that for all $\sigma_1 \neq \sigma_2$
    \[
        \frac{1}{C} \cdot \plusDownUp_{\beta, k}^{U}(\sigma_1, \sigma_2)
        \le \kawasaki_{\beta, \Delta}^{U}(\sigma_1, \sigma_2)
        \le C \cdot \plusDownUp_{\beta, k}^{U}(\sigma_1, \sigma_2).
\]
\end{observation}
By \Cref{obs:poincare_comparison} and \Cref{obs:ergodicity_pinning}, it suffices to prove \Cref{lemma:low_magnetization} for the pinned Kawasaki dynamics instead of the pinned down-up walk. 

To prove \Cref{lemma:low_magnetization}, we will change our perspective on the Ising model slightly.
Instead of viewing Ising configurations as maps from $V$ to $\{+1, -1\}$, we view them as injections from a list of $k$ $+1$ spins to $V$. More precisely, we set 
\[
    \Omega_k^{U} \coloneqq \{\sigma: \{1, \dots, k - \size{U}\} \to V \setminus U \,:\, \sigma \text{ is injective}\} .
\] 
For every $\sigma \in \Omega_k^{U}$, let $V^+(\sig)$ denote the image of $\sig$ and $V^-(\sig) \coloneqq (V \setminus U) \setminus V^+(\sig)$. 
To link this to the previously-used definition of $\Omega_k^{U}$, we say that $\sigma \in \Omega_k^{U}$ assigns spin $+1$ to a vertex $v \in V$ if $v \in V^+(\sigma) \cup U$ and a $-1$ spin otherwise. 
With this connection, previously-introduced notions, such as the number of monochromatic edges $m_{G}(\sigma)$ and probability of a configuration under the Ising model $\hat{\mu}_{\beta, k}^{U}(\sigma)$, translate naturally to this new perspective. 

With this new definition of the state space $\Omega_k^{U}$, we can rephrase an update step of the Kawasaki dynamics for a configuration $X_t$ as follows: 
\begin{enumerate}
    \item Choose $i \in \{1, \dots, k - \size{U}\}$ and $v \in V^-(X_t)$ independently uniformly at random. \label{step:new_kawasaki:select}
    \item Let $X \in \Omega_k^{U}$ be such that $X(i) = v$ and $X(j) = X_t(j)$ for all $j \neq i$.
    \item Set $X_{t+1} = X$ with probability $\phi(X_t, X)$,
    and set $X_{t+1} = X_{t}$ otherwise. \label{step:new_kawasaki:accept}
\end{enumerate}

It is straightforward to check that this is indeed equivalent to \Cref{def:kawasaki-pinned} of the Kawasaki dynamics.
Throughout this section, $(X_t)_{t \geq 0}$ and $(Y_t)_{t \geq 0}$ will denote separate copies of the Kawasaki dynamics Markov chain. 
Our goal will be to couple these chains so that, for a suitable notion of distance, the coupled dynamics contracts in expectation.
The coupling we analyze is constructed as follows:
\begin{enumerate}
    \item Choose $i \in [k - \size{U}]$ and $v \in V^{-}(X_t)$ independently and uniformly at random, and let $X \in \Omega_k^{U}$ be such that $X(i) = v$ and $X(j) = X_t(j)$ for all $j \neq i$. \label{step:coupling:X}
    \item Construct $Y \in \Omega_k^{U}$ as follows: \label{step:coupling:target}
    \begin{enumerate}
        \item If $v \in V^{-}(X_t) \cap V^{-}(Y_t)$ let $Y \in \Omega_k$ be such that $Y(i) = v$ and $Y(j) = Y_t(j)$ for $j \neq i$. \label{step:coupling:same_v}
        \item Otherwise, if $v \in V^{-}(X_t) \setminus V^{-}(Y_t)$, pick $v' \in V^{-}(Y_t) \setminus V^{-}(X_t)$ uniformly at random, and let $Y \in \Omega_k$ be such that $Y(i) = v'$ and $Y(j) = Y_t(j)$ for $j \neq i$. \label{step:coupling:different_v}
    \end{enumerate} 
    \item Set $(X_{t+1}, Y_{t+1}) = (X, Y)$ with probability $\min(\phi(X_t, X), \phi(Y_t, Y))$, set $(X_{t+1}, Y_{t+1}) = (X_t, Y_t)$ with probability $\min(1 - \phi(X_t, X), 1 - \phi(Y_t, Y))$, and distribute the remaining probability mass to obtain a valid coupling. \label{step:coupling:accept}
\end{enumerate}

To state the notion of distance under which the above coupling is a contraction, we first need to introduce some new notation.
We denote by $D_t$ the set of disagreements between $X_t$ and $Y_t$, formally defined as
\[
    D_t \coloneqq \{j \in \{1, \dots, k - \size{U}\} \,:\, X_t(j) \neq Y_t(j)\}.
\]
Moreover, we write $B_t$ for the set of ``bad'' disagreements between $X_t$ and $Y_t$, which are disagreements that are adjacent to an agreement.
Formally, that is
\[
    B_t \coloneqq \{j \in D_t \,:\, \exists i \in D_t^c \text{ s.t. } X_t(i) \in N(X_t(j)) \cup N(Y_t(j))\}.
\] 
We aim to prove \Cref{lemma:low_magnetization} by showing that there exist $\varphi, c > 0$ such that for all $t \in \Z_{\geq 0}$
\[
    \E[\varphi \size{D_{t+1}} + \size{B_{t+1}} \mid X_t, Y_t ] \le (1 - c/k) \cdot (\varphi \size{D_{t}} + \size{B_{t}}) .
\]
To achieve this, we first prove two lemmas that recursively bound the expected number of disagreements and bad disagreements.
The first lemma states intuitively that, for $k'  := k - |U|$ sufficiently small compared to $n$, the number of disagreements decreases in expectation, except when $|B_t|/k'$ is too large compared to $|D_t|/k'$. 
We prove this by analyzing three types of transitions for our coupling:
\begin{enumerate}[(a)]
    \item We choose a disagreement $+$ and an agreeing $-$ vertex, thus decreasing the number of disagreements if both updates succeed.
    \item We choose an agreement $+$ that does not neighbor any other disagreements and change it into a disagreement, thus increasing the number of disagreements. Note that this occurs only if we either swap with a target vertex in $V^{-}(X_t) \setminus V^-(Y_t)$ or if the update succeeds in exactly one of the two chains.
    \item We choose an agreement $+$ that has a disagreement neighbor and change it into a disagreement, thus increasing the number of bad disagreements. 
\end{enumerate}
If $k'$ is small enough compared to $n$, type (a) transitions will be more likely than type (b) transitions, causing the expected number of disagreements to decrease by at least $\frac{b_1}{k'} |D_t|$ for some constant $b_1 > 0$.
On the other hand, type (c) transitions increase the expected number of disagreements by at most $\frac{b_2}{k'} |B_t|$.
We now proceed with the details.

\begin{lemma} \label{lemma:contraction:D}
    There exist $\alpha, b_1, b_2 > 0$, only depending on $\beta$ and $\Delta$, such that for $k':= k - \size{U} \le \alpha n$ and all $t \in \Z_{\geq 0}$ it holds that
    $$\E[|D_{t+1}|\,|\,X_t, Y_t] \leq \left(1 - \frac{b_1}{k'}\right)|D_t| + \frac{b_2}{k'}|B_t|$$
\end{lemma}

\begin{proof}
    We start by applying linearity of expectation.
    \begin{align*}
        \E[\size{D_{t+1}} \mid X_t, Y_t] 
        &= \size{D_t} - \E[\size{D_t \setminus D_{t+1}} \mid X_t, Y_t] + \E[\size{D_{t+1} \setminus D_t} \mid X_t, Y_t] \\
        &= \size{D_t} - \sum_{j \in D_t} \Prob[j \in D_{t+1}^c \mid X_t, Y_t] + \sum_{j \in D_t^c} \Prob[j \in D_{t+1} \mid X_t, Y_t] .
    \end{align*}
    To lower-bound the first sum, note that every $j \in D_t$ transitions to $D_{t+1}^c$ whenever we pick $i = j$ and $v \in V^{-}(X_t) \cap V^{-}(Y_t)$ in step (\ref{step:coupling:X}) of the coupling and both chains accept the move in step (\ref{step:coupling:accept}).
    Hence, for $k' \le \alpha n$ with $\alpha \le 1/8$, we have
    \begin{align}
        \sum_{j \in D_t} \Prob[j \in D_{t+1}^c \mid X_t, Y_t] 
        &\ge \frac{\size{D_t}}{k'} \cdot \frac{n - \size{U} - 2k'}{n - k} e^{- 2 \beta \Delta} \notag\\
        &\ge \frac{\size{D_t}}{k'} (1 - 4 \alpha) e^{- 2 \beta \Delta} \notag\\
        &\ge \frac{\size{D_t}}{k'} \cdot \frac{e^{-2\beta \Delta}}{2} \label{lemma:contraction:D:decrease}.
    \end{align}
    To upper-bound the second sum, define $A_t$ to be the set of agreements at time $t$ that have at least one disagreement in their neighborhood.
    Formally, that is
    \[
        A_t \coloneqq \{j \in D_t^c \,:\, \exists i \in D_t \text{ s.t. } X_t(j) \in N(X_t(i)) \cup N(Y_t(i))\} .
    \]
    Further, we set $C_t = D_t^c \cap A_t^c$ to be the set of agreements that do not have a disagreement in their neighborhood.
    Note that $D_t^c = A_t \cup C_t$  and thus
    \[
        \sum_{j \in D_t^c} \Prob[j \in D_{t+1} \mid X_t, Y_t] 
        =  \sum_{j \in C_t} \Prob[j \in D_{t+1} \mid X_t, Y_t] + \sum_{j \in A_t} \Prob[j \in D_{t+1} \mid X_t, Y_t] . 
    \]
    To move $j \in C_t$ to $D_{t+1}$ we have to choose $i=j$ in step (\ref{step:coupling:X}) of the coupling and one of the following events needs to occur:
    \begin{itemize}
        \item we choose $v \in V^{-}(X_t) \setminus V^{-}(Y_t)$ in step (\ref{step:coupling:X}), or
        \item we transition to $(X_{t+1}, Y_{t+1}) = (X, Y_t)$ or $(X_{t+1}, Y_{t+1}) = (X_{t}, Y)$ in step (\ref{step:coupling:accept}).
    \end{itemize}
    The latter can only happen if $\phi(X_t, X) \neq \phi(Y_t, Y)$, which requires $m_G(X_t) - m_G(X) \neq m_G(Y_t) - m_G(Y)$. 
    For $j \in C_t$ this is only the case if we choose $v \in N(X_t(j'))$ or $v \in N(Y_t(j'))$ for some $j' \in D_t$.
    Hence, for $k' \le \alpha n$ with $\alpha \le \frac{e^{-2 \beta \Delta}}{16 \Delta + 8}$, we have
    \begin{align}
        \sum_{j \in C_t} \Prob[j \in D_{t+1} \mid X_t, Y_t]
        &\le \frac{\size{C_t}}{k'} \cdot \left(\frac{2 \Delta \size{D_t}}{n - k} + \frac{\size{V^{-}(X_t) \setminus V^{-}(Y_t)}}{n - k} \right) \notag \\
        &\le \frac{\size{D_t}}{n} (4 \Delta + 2) \notag \\
        &\le \frac{\size{D_t}}{k'} \cdot \frac{e^{-2\beta \Delta}}{4}, \label{lemma:contraction:D:increase}
    \end{align}
    where the second inequality follows from bounding the first term with $\size{C_t} \le k' \le \alpha n$, the last term with $\size{V^{-}(X_t) \setminus V^{-}(Y_t)} \le \size{D_t}$,  and the denominators with $n - k \ge n/2$. In the third line, we apply our bound for $\alpha$.
    
    Finally, for the probability of moving $j \in A_t$ into $D_{t+1}$, we use the crude bound of $1/k'$ for choosing $i = j$ in step (\ref{step:coupling:X}) to get
    \[
        \sum_{j \in A_t} \Prob[j \in D_{t+1} \mid X_t, Y_t] 
        \le \frac{\size{A_t}}{k'} 
        \le \frac{2 \Delta \size{B_t}}{k'}.
    \]
    Combining this with \eqref{lemma:contraction:D:decrease} and \eqref{lemma:contraction:D:increase} yields
    \[
        \E[\size{D_{t+1}} \mid X_t, Y_t] \le \left(1 - \frac{e^{-2 \beta \Delta}}{4k'}\right) \size{D_t} + \frac{2 \Delta}{k'} \size{B_t} ,
    \]
    which proves the claim.
\end{proof}
The second lemma states that, for $k'  := k - |U|$ sufficiently small compared to $n$, the number of bad disagreements decreases in expectation, except when $|D_t|/(n-k)$ is too large compared to $|B_t|/k'$. 
As before, we consider the following types of transitions:
\begin{enumerate}[(a)]
    \item We choose a disagreement $+$ and an agreeing $-$ vertex, thus decreasing the number of disagreements if both updates succeed.
    \item We choose a non-bad disagreement from $B_t \setminus D_t$ and turn it into a bad disagreement. Roughly, this happens only if we either move the disagreement into the neighborhood of another disagreement or if we move another disagreement into its own neighborhood.
    \item We choose an agreement $+$ that does not neighbor any other disagreements and change it into a disagreement, thus increasing the number of disagreements. Note that this occurs only if we either swap with a target vertex in $V^{-}(X_t) \setminus V^-(Y_t)$ or if the update succeeds in exactly one of the two chains (analogous to type (b) transitions above).
    \item We choose an agreement $+$ that has a disagreement neighbor and change it into a bad disagreement, thus increasing the number of bad disagreements. Note that this occurs only if we either move the disagreement into the neighborhood of an agreement $+$ or attempt to move it to a target vertex in $V^-(X_t) \setminus V^-(Y_t)$, succeeding in $X_t$ but not in $Y_t$.
\end{enumerate}
If $k'$ is small enough compared to $n$, type (a) transitions decrease the expected number of bad disagreements by at least $\frac{a_1}{k'} |B_t|$ for some $a_1 > 0$.
On the other hand, type (b)-(d) transitions increase the expected number of bad disagreements by at most $\frac{a_2}{n-k} |D_t|$. We now carry out the detailed computations.

\begin{lemma} \label{lemma:contraction:B}
     There exist $\alpha, a_1, a_2 > 0$, only depending on $\beta$ and $\Delta$, such that for $k':= k - \size{U} \le \alpha n$ and all $t \in \Z_{\geq 0}$ it holds that
    \[
        \E[\size{B_{t+1}} \mid X_t, Y_t] \le \left(1 - \frac{a_1}{k'}\right) \size{B_t} + \frac{a_2}{n - k} \size{D_t} .
    \]
\end{lemma}

\begin{proof}
    As before, we start by noting that
    \begin{align*}
        \E[\size{B_{t+1}} \mid X_t, Y_t] 
        &= \size{B_t} - \E[\size{B_{t} \setminus B_{t+1}} \mid X_t, Y_t] + \E[\size{B_{t + 1} \setminus B_{t}} \mid X_t, Y_t]  \\
        &= \size{B_t} - \sum_{j \in B_t} \Prob[j \in B_{t+1}^c \mid X_t, Y_t] + \sum_{j \in B_t^c} \Prob[j \in B_{t+1} \mid X_t, Y_t]. 
    \end{align*}
    For a lower bound on the first sum, note that every $j \in B_t$ transitions to $D_{t+1}^c \subseteq B_{t+1}^c$ whenever we pick $i = j$ and $v \in V^{-}(X_t) \cap V^{-}(Y_t)$ in step (\ref{step:coupling:X}) of the coupling and both chains accept the move in step (\ref{step:coupling:accept}).
    Thus, for $k' \le \alpha n$ with $\alpha \le 1/8$, similar computations to those for \eqref{lemma:contraction:D:decrease} yield
    \begin{align}
        \sum_{j \in B_t} \Prob[j \in B_{t+1}^c \mid X_t, Y_t] 
        \ge \frac{\size{B_t}}{k'} \cdot \frac{e^{-2\beta \Delta}}{2} \label{lemma:contraction:B:decrease}.
    \end{align}
    
    For an upper bound on the second sum, we first use the fact that $D_t^c \subseteq B_t^c$ to split up the sum into
    \[
        \sum_{j \in B_t^c} \Prob[j \in B_{t+1} \mid X_t, Y_t]
        = \sum_{j \in D_t \setminus B_t} \Prob[j \in B_{t+1} \mid X_t, Y_t] + \sum_{j \in D_t^c} \Prob[j \in B_{t+1} \mid X_t, Y_t] .
    \]
    In order for $j \in D_t \setminus B_t$ to transition to $B_{t+1}$, one of three cases must occur.
    \begin{itemize}
        \item We choose some $i \neq j$ and $v \in N(X_t(j)) \cup N(Y_t(j))$ in step (\ref{step:coupling:X}) of the coupling, and both chains accept the move in step (\ref{step:coupling:accept}).
        This happens with probability at most
        \[
            \frac{k'-1}{k'}\cdot \frac{2 \Delta}{n - k} .
        \]
        \item We choose $i = j$ and $v \in N(X_t(j'))$ for some $j' \in D_t^c$ in step (\ref{step:coupling:X}), and only one of the chains accepts the move in step (\ref{step:coupling:accept}).
        This has probability at most
        \[
            \frac{1}{k'} \cdot \frac{\Delta \size{D_t^c}}{n - k} .
        \]
        \item We choose $i = j$ and $v \in V^{-}(X_t) \setminus V^{-}(Y_t)$ in step (\ref{step:coupling:X}), in which case we might choose $v'$ in neighborhood of some agreement $j' \in D_t^c$ in step (\ref{step:coupling:different_v}) of the coupling. 
        The probability of this is at most
        \[
            \frac{1}{k'} \cdot \frac{\size{V^{-}(X_t) \setminus V^{-}(Y_t)}}{n - k} .
        \]
    \end{itemize}
    Adding up the cases above, summing over all $j \in D_t \setminus B_t$, and using that $\size{D_t^c} \le k'$ and $\size{V^{-}(X_t) \setminus V^{-}(Y_t)} \le \size{D_t} \le k'$ gives
    \begin{align}
        \sum_{j \in D_t \setminus B_t} \Prob[j \in B_{t+1} \mid X_t, Y_t]
        \le \size{D_t \setminus B_t} \frac{3 \Delta + 1}{n - k} 
        \le \size{D_t} \frac{3 \Delta + 1}{n - k}. \label{lemma:contraction:B:increase:1}
    \end{align}

    For bounding the probability of transitioning from $j \in D_t^c$ to $j \in B_{t+1}$, we again look at two cases.
    To this end, define $A_t$ and $C_t$ as in the proof of \Cref{lemma:contraction:D} and recall that $D_t^c = A_t \cup C_t$ and thus
    \[
        \sum_{j \in D_t^c} \Prob[j \in B_{t+1} \mid X_t, Y_t] 
        = \sum_{j \in C_t} \Prob[j \in B_{t+1} \mid X_t, Y_t] + \sum_{j \in A_t} \Prob[j \in B_{t+1} \mid X_t, Y_t] .
    \]
    
    To bound the probability of moving $j \in C_t$ to $B_{t+1}$, we can use the fact that $B_{t+1} \subseteq D_{t+1}$ and the same arguments as for deriving \eqref{lemma:contraction:D:increase} to get
    \begin{align}
        \sum_{j \in C_t} \Prob[j \in B_{t+1} \mid X_t, Y_t]
        \le \frac{\size{C_t}}{k'} \cdot \frac{(2\Delta + 1) \size{D_t}}{n - k} 
        \le \size{D_t} \frac{2 \Delta + 1}{n - k} , \label{lemma:contraction:B:increase:2}
    \end{align}
    where the last inequality uses $\size{C_t} \le k - \size{U}$.
    
    Finally, for moving $j \in A_t$ to $B_{t+1}$, we need to choose $i = j$ in step (\ref{step:coupling:X}) to make it a disagreement and we have to move it in the neighborhood of an agreement $j' \in D_t^c$ or choose $v \in V^{-} (X_t) \setminus V^{-}(Y_t)$.
    Hence, by bounding $\size{A_t} \le 2 \Delta \size{B_t} \le 2 \Delta \size{D_t}$ as well as $\size{D_t^c} \le k'$ and $\size{V^{-}(X_t) \setminus V^{-}(Y_t)} \le k'$, we get
    \begin{align*}
        \sum_{j \in A_t} \Prob[j \in B_{t+1} \mid X_t, Y_t]
        &\le \size{A_t} \frac{1}{k'} \cdot \left(\frac{\Delta \size{D_t^c}}{n - k} + \frac{\size{V^{-}(X_t) \setminus V^{-}(Y_t)}}{n - k} \right) \\
        &\le \size{D_t} \frac{2 \Delta^2 + 2 \Delta}{n - k}. 
    \end{align*} 
    Combining this with \eqref{lemma:contraction:B:decrease}, \eqref{lemma:contraction:B:increase:1} and \eqref{lemma:contraction:B:increase:2} shows that
    \[
        \E[\size{B_{t+1}} \mid X_t, Y_t] \le \left( 1 - \frac{e^{- 2 \beta \Delta}}{2k'}\right) \size{B_t} + \frac{2 \Delta^2 + 7 \Delta + 2}{n - k} \size{D_t},
    \]
    which concludes the proof.
\end{proof}

We now turn towards proving the main result of the section.
The intuitive idea is that, if $k'  := k - |U|$ is small enough compared to $n-k$, we can combine \Cref{lemma:contraction:D} and \Cref{lemma:contraction:B} to find some weighted sum of the number of disagreements and bad disagreements that decreases in expectation, which results in a lower bound for the spectral gap by \Cref{thm:gap_from_contraction}.

\begin{proof}[Proof of \Cref{lemma:low_magnetization}]
    We prove our claim by applying \Cref{thm:gap_from_contraction}, meaning that we need to find a function $\rho: \Omega_k^{U} \times \Omega_k^{U} \to \R_{\ge 0}$ such that $\rho(\sigma_1, \sigma_2) = 0$ if and only if $\sigma_1 = \sigma_2$ and for all $t \in \Z_{\geq 0}$
    \[
        \E[\rho(X_{t+1}, Y_{t+1}) \mid X_t, Y_t] \le \left( 1 - \frac{c}{k} \right) \rho(X_t, Y_t)
    \]
    for some $c > 0$. 
    Let $a_1, a_2$ be as in \Cref{lemma:contraction:B}  and $b_1, b_2$ as in \Cref{lemma:contraction:D}.
    Set $\varphi = \frac{a_1}{2 b_2} > 0$, choose $\alpha > 0$ such that $ \alpha \le \frac{\varphi b_1}{4 a_2}$ and such that it is sufficiently small to satisfy \Cref{lemma:contraction:B,lemma:contraction:D}, and assume $k':= k - \size{U} \le \alpha n$.
    By \Cref{lemma:contraction:B,lemma:contraction:D} we have
    \begin{align*}
        \E[\varphi \size{D_{t+1}} + \size{B_{t+1}} \mid X_t, Y_t ] 
        &\le \left(1 - \frac{b_1}{k'}\right) \varphi \size{D_t} + \varphi \frac{b_2}{k'} \size{B_t} + \left(1 - \frac{a_1}{k'}\right) \size{B_t} + \frac{a_2}{n - k} \size{D_t} \\
        &= \left(1 - \frac{b_1}{k'}\right) \varphi \size{D_t}  + \left(1 - \frac{a_1}{2k'}\right) \size{B_t} + \frac{a_2}{n - k} \size{D_t} \\
        &\le \left(1 - \frac{b_1}{2k'}\right) \varphi \size{D_t}  + \left(1 - \frac{a_1}{2k'}\right) \size{B_t} ,
    \end{align*}
    where we use $\varphi \frac{b_2}{k'} = \frac{a_1}{2k'}$ to obtain the second line and  
    \[
        \frac{a_2}{n - k} \le \frac{2 \alpha \cdot a_2}{k'} \le \varphi \frac{b_1}{2k'} 
    \]
    to obtain the last line.
    Hence, setting $c = \frac{1}{2} \min\{a_1, b_1\} > 0$ and $\rho(X_t, Y_t) = \varphi \size{D_{t}} + \size{B_{t}}$, \Cref{thm:gap_from_contraction} proves that the spectral gap of $\kawasaki^{U}_{\beta, k}$ is at least $\Omega(1/k)$ with constants depending on $\beta$ and $\Delta$.
    Using \Cref{obs:ergodicity_pinning} and \Cref{obs:poincare_comparison} concludes the proof.
\end{proof}

\subsection{Bounding Spectral Gap}\label{sec:loc_scheme}
We conclude this section with our main goal, which is proving that the spectral gap of the Kawasaki dynamics is $\Omega(1/n)$ for all target magnetizations as given in part (\ref{thm:main_mixing:rapid}) of \Cref{thm:main-mixing}. 
This result is given in the following statement.
\begin{theorem} \label{thm:spectral_gap}
    Let $\gamma \in (0, 1)$ such that $\beta < \beta_u(\Delta)$ or $\gamma \notin \left[\frac{1 - \eta_a}{2}, \frac{1 + \eta_a}{2}\right]$ where $\eta_a = \eta_a(\Delta, \beta)$.
    For all $G \in \mathcal{G}_{\Delta}$ such that $k \coloneqq \gamma n \in \N$, the spectral gap of $\kawasaki_{\beta, k}$ on $G$ is in $\Omega(1/n)$ with implied constants only depending $\Delta$, $\beta$ and $\gamma$.
\end{theorem}
To prove this, we need to combine \Cref{lemma:low_magnetization} with some details related to localization schemes and simplicial complexes. As the proof is somewhat technical, we start with an overview of the key ideas.

We first note that the Kawasaki dynamics Markov chain is invariant under swapping all spins (i.e. mapping the configuration $\sigma$ to $-\sigma$).
Hence, we can focus on $k \le n/2$, or equivalently the magnetization regime $\eta \le 0$.
Moreover, by \Cref{obs:ergodicity} it suffices to prove the desired spectral gap for the down-up walk $\plusDownUp_{\beta, k}$ for the respective values of $k$.

We proceed by an argument similar to that in the proof of \cite[Theorem 20]{JMPV23}.
For any $\ell \le k-1$, we use a localization scheme to show that the spectral gap of $\plusDownUp_{\beta, k}$ is bounded below by the product of $\inf_{U \in \binom{V}{\ell}} \gap(\plusDownUp^{U}_{\beta, k})$ and the spectral gap of the $(k, \ell)$-down-up walk, which generalizes $\plusDownUp_{\beta, k}$ in that it resamples $k - \ell$ plus spins in each step.
By \Cref{lemma:low_magnetization}, we know that $\inf_{U \in \binom{V}{\ell}} \gap(\plusDownUp^{U}_{\beta, k}) \in \Omega(1/k)$ whenever $\ell$ is such that $k - \ell \le \alpha n$ for some $\alpha = \alpha(\Delta, \beta) > 0$.
For the $(k, \ell)$-down-up walk, we show a spectral gap of $\Omega(1)$ by applying a local-to-global theorem from \cite{chen2021optimal}.
For this, we use \Cref{thm:spectral_independence_fm} which says there exists some constant $C = C(\beta, \Delta, \gamma)$ such that $\hat{\mu}^{W}_{\beta, k}$ satisfies $C$-spectral independence for all $W \subset V$ with, say, $k - \size{W} \ge \frac{\alpha}{2} n$. This implies local spectral expansion for sufficiently many layers of the associated simplicial complex.

We now proceed with the full proof.

\begin{proof}
    First, note that the transition probabilities of $\kawasaki_{\beta, k}$ are invariant under mapping $\sigma \mapsto - \sigma$ for all $\sigma \in \Omega_k$, so we may assume $\gamma \le 1/2$.
    By \Cref{obs:ergodicity} the transition probabilities in $\kawasaki_{\beta, k}$ and $\plusDownUp_{\beta, k}$ for pairs of distinct states differ by at most some non-zero factor that only depends on $\beta$ and $\Delta$.
    Hence, by \Cref{obs:poincare_comparison}, it suffices to prove the statement for $\plusDownUp_{\beta, k}$.
    
    If $\gamma \le \alpha(\beta, \Delta)$ for $\alpha(\beta, \Delta)$ as in \Cref{lemma:low_magnetization}, then the claim follows immediately by applying the lemma with $U = \emptyset$. 
    So suppose that $\alpha(\beta, \Delta) < \gamma \le 1/2$.
    To prove the theorem in this case, we aim to use techniques from \cite{chen2022localization} and \cite{chen2021optimal}.
    To this end, it will be more suitable to consider the fixed-magnetization Ising model as a distribution supported on subsets of vertices of size $k$.
    
    More precisely, let $\Omega_k = \binom{V}{k} \coloneqq \{S \subseteq V\,:\, |S| = k\}$.
    Every $S \in \Omega_k$ is associated with the Ising configuration that maps exactly the vertices in $S$ to $+1$.
    The fixed-magnetization Ising model $\hat{\mu}_{\beta, k}$ translates naturally to a distribution on $\Omega_k$ as defined above.
    Moreover, for $U \subseteq V$ we write $\Omega_{k}^{U} \coloneqq \{S \in \Omega_k \mid U \subseteq S\}$ for the set of configurations containing $U$ (i.e., configurations that have the vertices in $U$ pinned to $+1$).
    Let $\mathcal{P}(\Omega_k)$ be the set of distributions on $\Omega_k$.
    For $\pi \in \mathcal{P}(\Omega_k)$ and $U \subseteq V$ we write $\pi(U)$ for the probability that a set drawn from $\pi$ contains $U$ (i.e., $\pi(\Omega_k^{U})$), and, for $\pi(U) > 0$, we write $\pi^U (\cdot)$ for the conditional distribution $\frac{\pi(~\cdot ~ \cap \Omega_k^{U})}{\pi(\Omega_k^{U})}$.

    We proceed to construct a localization scheme in the sense of \cite{chen2022localization} for distributions in $\mathcal{P}(\Omega_k)$.
    To this end, for $\pi \in \mathcal{P}(\Omega_k)$, we construct a localization process $(\pi_t)_{t \geq 0}$ recursively as follows:
    \begin{itemize}
        \item We set $U_0 = \emptyset$. 
        \item Given $U_t$, we define $\pi_t = \pi^{U_{t \wedge k}}$ (where $t \wedge k$ denotes the minimum). In particular, $\pi_0 = \pi$.
        \item For $1 \le t \le k - 1$, we construct $U_{t+1}$ by drawing $M \sim \pi_t$ and $v_t \in M \setminus U_t$ uniformly at random, and setting $U_{t+1} = U_t \cup \{v_t\}$. 
    \end{itemize}
    It can be checked that this is equivalent to the subset simplicial complex localization from \cite[Example 5]{chen2022localization} in the following sense: for every $\mathcal{P}(\Omega_k)$, the associated localization processes are identically distributed.
    
    As shown in \cite[Fact 8]{chen2022localization} the localization scheme above associates a probability distribution $\pi \in \mathcal{P}(\Omega_k)$ and an integer $0 \le \ell \le k-1$ with a Markov transition matrix on $\Omega_k$ via 
    \[
        Q_{\pi, \ell} (S_1, S_2) = \E\left[\frac{\pi_{\ell} (S_1) \pi_{\ell}(S_2)}{\pi(S_1)} \right] \text{ for every } S_1, S_2 \in \Omega_k .
    \]
    This transition matrix corresponds to the $(k, \ell)$-down-up walk $(X_t)_{t \geq 0}$ on $\Omega_k$ associated with $\pi$, which is given by the following update rule:
    \begin{enumerate}
        \item Sample $U \in \binom{X_t}{\ell}$ uniformly at random. \item Sample $X_{t+1} \sim \pi^{U}$.
    \end{enumerate}
    
    If $\pi$ is the fixed-magnetization Ising model $\hat{\mu}_{\beta, k}$, we write $\plusDownUp_{\ell, \beta, k}$ for the transition matrix $Q_{\pi, \ell}$.
    Further, if $\ell = k-1$, then the associated transition matrix corresponds to the down-up walk $\plusDownUp_{\beta, k}$ from \Cref{def:plusDownUp} (see \cite{chen2022localization} for details).
    
    From now on, fix $\pi = \hat{\mu}_{\beta, k}$. 
    Using Proposition 19 from \cite{chen2022localization}, we know that
    \[ 
        \gap(\plusDownUp_{\beta, k}) = \inf_{f: \Omega_k \to \R} \frac{\E[\Var_{\pi_{k-1}}(f)]}{\Var_{\pi}(f)} .
    \]
    Moreover, for every $0 \le \ell \le k-1$, we have
    \begin{align*}
        \frac{\E[\Var_{\pi_{k-1}}(f)]}{\Var_{\pi}(f)}
        = \E\left[
            \frac{\E[\Var_{\pi_{k-1}}(f) \mid \pi_{\ell}]}{\Var_{\pi_{\ell}}(f)} 
            \cdot \frac{\Var_{\pi_{\ell}}(f)}{\Var_{\pi}(f)}
            \right] .
    \end{align*}
    In the next step, we use the following claim.
    \begin{claim} \label{claim:pinned}
        In the setting above, we have
        \[
            \frac{\E[\Var_{\pi_{k-1}}(f) \mid \pi_{\ell}]}{\Var_{\pi_{\ell}}(f)} \ge \inf_{U \in \binom{V}{\ell}}\gap(\plusDownUp^{U}_{\beta, k}) \text{ a.s.},
        \]
        where $\plusDownUp^{U}_{\beta, k}$ is the pinned down-up walk as in \Cref{def:plusDownUp}.
    \end{claim}
    
    Using \Cref{claim:pinned}, we can apply Proposition 19 of \cite{chen2022localization} again to get
    \begin{align}
        \gap(\plusDownUp_{\beta, k}) \ge \inf_{U \in \binom{V}{\ell}}\gap(\plusDownUp^{U}_{\beta, k}) \cdot \inf_{f: \Omega_k \to \R} \frac{\E[\Var_{\pi_{\ell}}(f)]}{\Var_{\pi}(f)}
        = \inf_{U \in \binom{V}{\ell}}\gap(\plusDownUp^{U}_{\beta, k}) \cdot \gap(\plusDownUp_{\ell, \beta, k})
        \label{eq:gap_factorization}
    \end{align}

    Next, fix $\ell = k - \lfloor\frac{\alpha}{2} n \rfloor$ for $\alpha = \alpha(\Delta, \beta)$ as in \Cref{lemma:low_magnetization}.
    For this choice of $\ell$, it follows from \Cref{lemma:low_magnetization} that $\inf_{U \in \binom{V}{\ell}}\gap(\plusDownUp^{U}_{\beta, k}) \in \Omega(1/k)$ with constants depending only on $\beta$ and $\Delta$.
    Hence, it remains to lower-bound $\gap(\plusDownUp_{\ell, \beta, k})$ for the same value of $\ell$.
    To this end, we use our spectral independence result from \Cref{thm:spectral_independence_fm} and the variance contraction framework for simplicial complexes from \cite{chen2021optimal} to obtain the following claim.
    \begin{claim} \label{claim:k_l_walk}
        In the above setting, it holds that there is some constant $C > 0$, only depending on $\beta$, $\Delta$ and $\gamma$, such that
        \[
            \gap(\plusDownUp_{\ell, \beta, k}) \ge C .
        \]
    \end{claim}
    We defer the proofs of \Cref{claim:pinned} and \Cref{claim:k_l_walk} to \Cref{appendix:fast}. 
    Combining \Cref{lemma:low_magnetization} and \Cref{claim:k_l_walk} with \eqref{eq:gap_factorization} concludes the proof.
\end{proof}

\section{Metastability and Slow Mixing}\label{sec:slow}

In this section we prove slow-mixing results for both the Ising Glauber dynamics and fixed magnetization Kawasaki dynamics when $\beta>\beta_u(\Delta)$ and $|\log \lam| < \log \lam_u$ and $|\eta| < \eta_u$ respectively. The structure of the proof is illustrated below in Figure~\ref{fig:structure2}. 

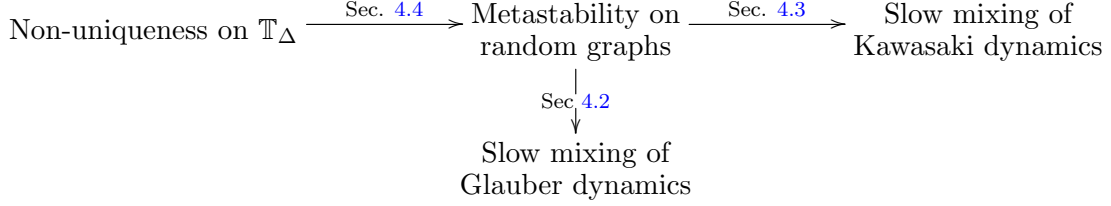
\begin{figure}[!htbp]
\centering
\vspace{-1ex}
$$\xymatrix{
\txt{Non-uniqueness on $\T_{\Delta}$}\ar[rr]^-{\text{Sec. }\ref{sec:metastable-proof}} && {\txt{Metastability on \\ random graphs}}\ar[rr]^-{\text{Sec. }\ref{secKawasakislow}}\ar[d]|-{\text{Sec }\ref{sec:Glauberslow}}&&\txt{Slow mixing of\\ Kawasaki dynamics} \\
&& \txt{Slow mixing of \\Glauber dynamics}&&
}$$
\vspace{-0.5ex}
\caption{The structure of the slow mixing proof }
\label{fig:structure2}
 \end{figure}

\noindent {\bf Note:} As in the previous section, both perspectives of fixed magnetization per vertex $\eta$ and fixed size $k$ will be useful in our arguments. We will use $Z_{G, \eta} (\beta, \lam)$ (where we sometimes drop the parameters $\beta$ and $\lam$ for convenience) to denote the contribution to the Ising model partition function $Z_G(\beta, \lam)$ from configurations of magnetization $\eta$. The notation $Z_{G, k}(\beta, \lam)$ will mean the contributions to $Z_G(\beta, \lam)$ from configurations of size $k$. When $k = \lfloor n \frac{\eta + 1}{2}\rfloor$, we have $Z_{G, \eta} = Z_{G,k}$ and will use the notations interchangeably.

\subsection{Tree recursions and random graph preliminaries}

Our goal is to understand how configurations of different magnetizations typically contribute to the partition function $Z_G(\beta,\lam)$ when $G$ is a  random $\Delta$-regular graph. To start, we shift to a slightly different model called the {\em configuration model}, which we will denote $\cG$. To generate a graph from this model for a given $\Delta$ and $n$, take $\Delta$ copies of $[n]$ and a uniformly random perfect matching on the $\Delta n$ vertices. Then identify the copies corresponding to the same vertex. This gives a random $\Delta$-regular multigraph, and it is well-known that properties holding with high probability for the configuration model also hold with high probability for the uniform random $\Delta$-regular graph when $\Delta$ is constant~\cite{JLR}.

We say the model has multiple {\em metastable states} if the function $\lim_{n \to \infty} \frac{1}{n} \E \log Z_{G,\eta}(\beta,\lam)$ has more than one local maximum as $\eta$ varies.   
A first attempt at understanding this phenomenon would be to look at the first moment, and understand the local maxima of the function
\begin{equation}\label{eqn:free-energy}
    f_{\Delta,\beta,\lam}(\eta) := \lim_{n \to \infty} \frac{1}{n} \log \E Z_{G,\eta}(\beta,\lam)
\end{equation} as a function of $\eta$ (with the crucial distinction between the two functions being the interchange of the expectation and logarithm). 

To derive an expression for $f_{\Delta, \beta, \lam}(\eta)$, we use computations similar to those found in~\cite{GSVY16, coja2016chromatic, Coja-Oghlan22}.
Suppose we have a configuration $\sigma\in \Omega$ on $G$ with $\frac{1+\eta}{2} n$ vertices assigned $+1$ and $\frac{1-\eta}{2}n$ assigned $-1$ (for convenience, we assume these two quantities are integers, as this will hold in the limit). 
Let $A = A^{\sig}$ 
be the vector $\left(\frac{1+\eta}{2}, \frac{1-\eta}{2}\right)$. In a slight abuse of notation, we will index the vector with $1$ and $-1$ so that $A(1) = \frac{1+\eta}{2}$ and $A(-1) = \frac{1-\eta}{2}$.
Let $B^{G, \sig}$ be a $2 \times 2$ matrix recording the edge statistics, again using the index $-1$ instead of $2$ so that $B(i,j)\Delta n$ is the number of edges with endpoints colored $i$ and $j$. Note that $B(1,1)+B(1,-1) = \frac{1+\eta}{2}$ and $B(-1,-1) + B(1,-1) = \frac{1-\eta}{2}$.
Observe that we may record the same information in a vector of 3 components $(b_+, b_0, b_-)$ where $b_+ = B(1, 1), b_- = B(-1, -1)$, and $b_0 = B(1, -1) = B(-1, 1)$.

Given $\sig \in \Omega$, if $B$ is a $2 \times 2$ symmetric matrix with positive entries whose top and bottom row sums are $A^{\sig}(1)$ and $A^{\sig}(-1)$, respectively, then we can compute the probability of the event $E_B(\cG, \sig) := \{B = B^{\cG, \sig}\}$ as follows:
there are 
${\frac{1+\eta}{2}\Delta n \choose b_0 \Delta n}{\frac{1-\eta}{2}\Delta n) \choose b_0 \Delta n}(b_0 \Delta n)!$ ways to choose the bichromatic edges. Using the fact that there are $(2m-1)!! := \frac{(2m)!}{2^m m!}$ perfect matchings between $2m$ vertices, we compute that there are $(b_+\Delta n-1)!!$ choices for the monochromatic $+$ edges and $(b_-\Delta n-1)!!$ choices for the monochromatic $-$ edges. Lastly, the state space has size $(\Delta n-1)!!$ so we have the following:
$$\Prob_{\cG}[E_B(\cG, \sig)] = \frac{1}{(\Delta n-1)!!} {\frac{1+\eta}{2} \Delta n \choose b_0\Delta n}{\frac{1-\eta}{2} \Delta n) \choose b_0\Delta n}(b_0\Delta n)!(b_+\Delta n-1)!!(b_-\Delta n-1)!!$$
Since ${n \choose \al n} \sim 2^{n H(\al) + o(1)}$ where $H(x) = -x \ln x - (1-x) \ln (1-x)$ is the binomial entropy function, we use the fact that $b_+ + b_0 = \frac{1+\eta}{2}$ and $b_- + b_0 = \frac{1-\eta}{2}$ along with Stirling's formula in the following.

\begin{align*}\Prob_{\cG}[E_{B}(\cG, \sig)] 
    &= \exp\left[\Delta n\left(\sum_{i,j} B(i,j) \ln \frac{A(i)}{B(i,j)}\right) + \frac{\Delta n}{2} \sum_{i,j} B(i,j) \ln B(i,j) + O\left(\frac{\ln n}{n}\right) \right]\\
    &= \exp\left[\frac{\Delta n}{2}\left(\sum_{i,j} (2 B(i,j) \ln \frac{A(i)}{B(i,j)} + B(i,j) \ln B(i,j))\right) + O\left(\frac{\ln n}{n}\right) \right]\\
    &= \exp\left[\frac{\Delta n}{2} \sum_{i,j} B(i,j) \ln \frac{A(i)^2}{B(i,j)} + O\left(\frac{\ln n}{n}\right)\right]
\end{align*}
We thus have
    $$\Prob_{\cG}[E_{B}(\cG, \sig)] = \exp\left[\frac{\Delta n}{2}\left(b_+ \ln \frac{\left(\frac{1+\eta}{2}\right)^2}{b_+} + b_0 \ln \frac{\left(\frac{1+\eta}{2}\right)^2 \left(\frac{1-\eta}{2}\right)^2}{b_0} + b_- \ln \frac{\left(\frac{1+\eta}{2}\right)^2}{b_-}\right) + O\left(\frac{\ln n}{n}\right)\right]$$
and so
$$\E[Z_{\cG, \eta}(\beta, \lam)] = {n \choose \frac{1+\eta}{2} n} \sum_{B} \Prob[E_B(\cG,  \sig)] \lam^{\frac{1+\eta}{2}n}e^{\beta(b_++b_-)\frac{\Delta n}{2}}$$
where the summation is over all $2 \times 2$ symmetric $B$ such that the top and bottom row sums are $\frac{1+\eta}{2}$ and $\frac{1-\eta}{2}$, respectively. 
We use this to rewrite \eqref{eqn:free-energy}
as
$$f_{\Delta, \beta, \lam}(\eta)
= \max_{B} g_{\Delta, \beta, \lam}(\eta,B)$$
where
$$g_{\Delta, \beta, \lam}(\eta,B) = (\Delta-1)\left(-H\left(\frac{1+\eta}{2}\right)\right) - \frac{\Delta}{2} \sum_{i,j} B(i,j) \log B(i,j) + \frac{\beta \Delta}{2} \sum_i B(i,i) + \frac{1+\eta}{2} \log \lam$$
and again the maximum is taken over $B$ as described above.
We note that for a fixed $\eta$, the maximizer of $g$ in $B$ is unique (as shown in~\cite{galanis2015inapproximability, GSVY16}).

\begin{lemma}
    Given $\eta \in (-1,1)$, there exists a unique $B$ maximizing the function $g_{\Delta, \beta, \lam}(\eta, B)$ with respect to the constraints $B(1,1) + B(1,-1) = \frac{1+\eta}{2}$ and $B(-1,-1)+B(1,-1) = \frac{1-\eta}{2}$.
\end{lemma}

The proof follows from a straightforward application of the Lagrange multiplier method. Thus, it suffices to understand the maxima of $f_{\Delta, \beta, \lam}(\eta)$ as a one-variable function with respect to $\eta$.

It was established in~\cite{GSVY16} (following~\cite{mossel2009hardness,galanis2014improved}) that the critical points of $f_{\Delta,\beta,\lam}(\eta)$ correspond exactly to fixed points of the \textit{tree recursion} for the Ising model on $\T_\Delta$. 
For a vertex $v$, let $R_v = \frac{\Prob(\sig(v) = +1)}{\Prob(\sig(v) = -1)}$. If $v$ is the root of a $(\Delta-1)$-ary tree with children $u_1, \dots, u_{\Delta-1}$, then we have the following {\em tree recursion} for $R_v$:
$$
R_v = \frac{\lam \prod_{i=1}^{\Delta-1} (\frac{R_{u_i}}{1+R_{u_i}}e^\beta + \frac{1}{1+R_{u_i}})}{\prod_{i=1}^{\Delta-1} (\frac{R_{u_i}}{1+R_{u_i}} + \frac{1}{1+R_{u_i}}e^\beta)}
$$
Fixed points are solutions to the equation
\begin{equation}
\label{eqTreeRecursion}
    R = \frac{\lam(Re^\beta + 1)^{\Delta-1}}{(R+e^\beta)^{\Delta-1}} \,.
\end{equation}
The fixed points have the following relationship to  $f_{\Delta,\beta,\lam}(\eta)$.
\begin{theorem}[{\hspace{1sp}\cite[Theorem 9, Lemma 11]{GSVY16}}]\label{thm:fixpt-correspondence}
    There is a 1-to-1 correspondence between the fixed points of the tree recursion given in~\eqref{eqTreeRecursion} and the critical points of $f_{\Delta, \beta, \lam}(\eta)$. Moreover, the stable fixed points of the tree recursion given in~\eqref{eqTreeRecursion} are in 1-to-1 correspondence with Hessian local maxima of $f_{\Delta,\beta,\lam}(\eta)$. 
\end{theorem}
Recall that a fixed point is {\em stable} if the absolute value of the derivative at that point is less than 1. A local maximum is a {\em Hessian local maximum} if the Hessian is negative definite at that point. In particular, as our functions are univariate (after fixing $\Delta, \beta, \lam$), this is simply saying that the second derivative is negative which implies the existence of a local maximum. 

For the above theorem to be useful, we need to understand the fixed points of the tree recursion. 

\begin{proposition}
    \label{thm:fixpt} 
    For $\beta > \beta_u$, the following hold:
    \begin{enumerate}
        \item If $|\log \lam| > \log \lam_u$, then $\eqref{eqTreeRecursion}$ has a unique fixed point. It is stable and hence corresponds to the global maximizer of $f_{\Delta,\beta,\lam}$.  This maximizer is $\eta^+_{\Delta,\beta,\lam} = \eta^-_{\Delta,\beta,\lam}$.
        \item If $|\log \lam| = \log \lam_u$, then \eqref{eqTreeRecursion} has two distinct fixed points, one of which is stable and corresponds to the global maximizer of $f_{\Delta, \beta, \lam}$. The other corresponds to an inflection point of $f_{\Delta, \beta, \lam}$. 
        \item If $|\log \lam| < \log \lam_u$, then $\eqref{eqTreeRecursion}$ has three distinct fixed points. The largest and the smallest are both stable, corresponding to the only two local maxima of $f_{\Delta,\beta,\lam}$.  When $\lam >1$, $\eta^+_{\Delta,\beta,\lam}$ is the unique global maximizer; when $\lam<1$, $\eta^-_{\Delta,\beta,\lam}$ is the unique global maximizer; when $\lam=1$ then $\eta^+_{\Delta,\beta,\lam},\eta^-_{\Delta,\beta,\lam}$ are both global maximizers.
    \end{enumerate}
\end{proposition}

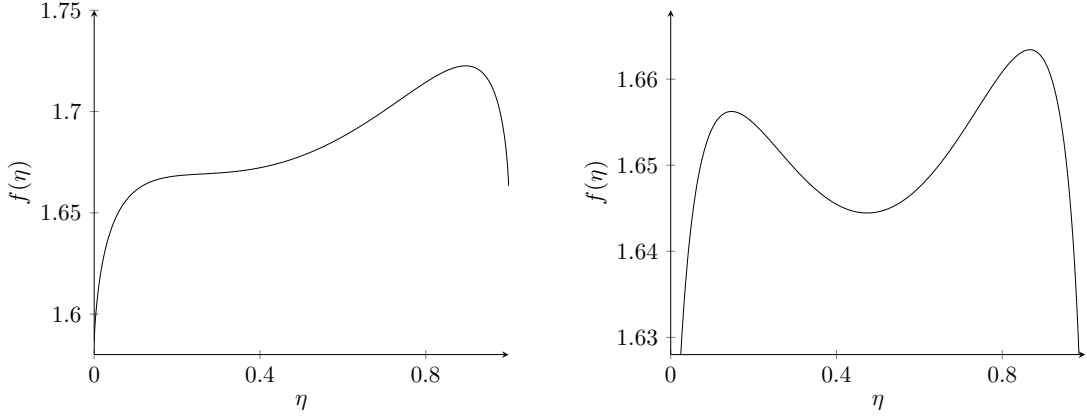
\begin{figure}[h]
\centering
\begin{center}
\begin{tikzpicture}[scale=0.8]
    \begin{axis}[axis lines = left, xlabel = \(\eta\), ylabel={\(f(\eta)\)}, xmin=0, xmax=1, ymin=1.58, ymax=1.75, xtick distance = 0.4]
    \addplot[name path=f] table {second-moment-data2.tex};
    \end{axis}
\end{tikzpicture}
\qquad
\begin{tikzpicture}[scale=0.8]
    \begin{axis}[axis lines = left, xlabel = \(\eta\), ylabel={\(f(\eta)\)}, xmin=0, xmax=1, ymin=1.628, ymax=1.668, xtick distance = 0.4, ytick distance=0.01]
    \addplot[name path=f] table {second-moment-data1.tex};
    \end{axis}
\end{tikzpicture}
\caption{Sketch of the function $f_{\Delta, \beta, \lam}(\eta)$ for $\Delta = 4$, $\beta = \ln(2)+0.1$, and \\(left) $\lam = 1.08$, (right) $\lam = 1.01$.}
\label{fig:second-moment}
\end{center}
\end{figure}

Portions of this statement have been shown in, for example, \cite{georgii2011gibbs, guolu2018, GSVY16}. For completeness, we prove the statement in Appendix~\ref{appendix:slow}. An illustration of $f_{\Delta, \beta, \lam}(\eta)$ is given in Figure~\ref{fig:second-moment}; the left plot appears for $\lam > \lam_u$ (Case 1 above) and the right plot appears for $1 < \lam < \lam_u$ (Case 3 above).

While the behavior described in part (3) suggests metastability, Proposition~\ref{thm:fixpt} is only about the expected partition function, and we will need to show that multiple local maxima exist with high probability over the random graph.  This will involve showing a lower bound on the partition function at the two local maxima and and upper bound everywhere else.  

Via Markov's inequality, the next statement gives a high probability approximate upper bound on $Z_{G,\eta}(\lam)$. 
\begin{lemma}
\label{lem:partitionUB}
Fix $\beta\ge 0, \lam > 0$. With probability $1-o(1)$ over the random $\Delta$-regular graph $G$ on $n$ vertices, it holds for every $\eta$ that
   $$
       Z_{G,\eta}(\lam) \le n^2 \cdot \E Z_{G,\eta}(\lam) \,.
   $$
\end{lemma}
\begin{proof}
    This follows from Markov's inequality and a union bound over all possible values of $\eta$. 
\end{proof}

We now prove  lower bounds on $Z_{G,\eta}$ for values of $\eta$ which are local maxima.  For a global maximum, this was proved in~\cite{GSVY16} via the second moment method.
\begin{theorem}[\hspace{1sp}{\cite[Theorem 8]{GSVY16}}]
\label{thm:partitionLB}
   Fix $ \lam > 0$ and suppose that $\eta$ is a global maximizer of $f_{\Delta,\beta,\lam}$. With probability $1-o(1)$ over the random $\Delta$-regular graph $G$ on $n$ vertices, 
   $$
       Z_{G,\eta}(\lam) \ge \frac{1}{n} \E Z_{G,\eta}(\lam) \,.
   $$
\end{theorem}
We next need a corresponding statement for the local maximizers.
\begin{proposition}\label{prop:localpartitionLB}
    Fix $\lam > 0$ and suppose that $\eta$ is a local maximizer of $f_{\Delta, \beta, \lam}$. For any $\zeta > 0$, with probability $1-o(1)$ over the random $\Delta$-regular graph $G$ on $n$ vertices, 
    $$Z_{G, \eta}(\lam) \geq e^{-\zeta n} \E[Z_{G, \eta}(\lam)]\,.$$
\end{proposition}
The proof of Proposition~\ref{prop:localpartitionLB} follows the template of Coja-Oghlan, Galanis, Goldberg, Ravelomanana, {\v{S}}tefankovi{\v{c}}, and Vigoda~\cite{Coja-Oghlan22} in proving metastability in the zero-field ferromagnetic Potts model (which in turn used ideas from~\cite{bapst2016harnessing,coja2022ising}).  We carry this out in Section~\ref{sec:metastable-proof}.

\subsection{Slow-Mixing of Glauber Dynamics}\label{sec:Glauberslow}

We now use the above ingredients to prove slow mixing of the Glauber dynamics in the external-field Ising model on the random regular graph.

\begin{proof}[Proof of Theorem~\ref{thm:GlauberRG}] Let $\beta > \beta_u(\Delta), \lam \in [1, \lam_u)$, and $G \sim \cG$. We will abbreviate $\mu_{G, \beta, \lam}$ as $\mu$.  Let $\eta = \eta^+_{\Delta,\beta,\lam}$, the mean magnetization of the root of $\T_{\Delta}$ under the $+$ boundary conditions with external field $\lam$, and let $\eta_- = \eta^-_{\Delta,\beta,\lam}$, the same but under the $-$ boundary conditions.

Recall that $\eta$ is the global maximizer of $f_{\Delta, \beta, \lam}$, and $\eta_-$ is a local maximizer of $f_{\Delta, \beta, \lam}$. 
This means that there exist $\eps > 0$ and $\delta > 0$ so that the following hold:
\begin{enumerate}
    \item $\E[Z_{G, \eta'}(\lam)] \leq e^{-\delta n} \E[Z_{G, \eta}(\lam)]$ for all $\eta'$ such that $|\eta' - \eta| > \eps$.\label{property:globalmax-glauber}
    \item $\E[Z_{G, \eta'}(\lam)] \leq e^{-\delta n}\E[Z_{G, \eta_-}(\lam)]$ for all $\eta'$ such that $|\eta' - \eta_-| \in (\eps, 2\eps)$.\label{property:localmax-glauber}
\end{enumerate}

Let $k$ be the number of $+$ spins in a configuration with magnetization $\eta$, and let $k_-$ be the number of $+$ spins in a configuration with magnetization $\eta_-$.  To lighten notation we drop the dependence on $\beta$ in $Z_{G}(\beta,\lam)$ and $\hat Z_{G,k}(\beta)$.

For $\eps>0$ as above,
define the following sets of configurations:
\begin{align*}
    S_1 &=  \Omega_k ,   \\
    S_2 &= \bigcup \left\{\Omega_j :  |j- k_-| \le \frac{\eps}{2} n\right\},  \\
    S_3 &= \bigcup \left\{\Omega_j : \frac{\eps}{2} n < |j- k_-| \le \eps n\right\}   \,.
\end{align*}
Now assume $G$ is such that the conclusions of Theorem~\ref{thm:partitionLB} and Lemma~\ref{lem:partitionUB} hold for $\lam$.  We will check the conditions of Lemma~\ref{lem:bottleneck} to deduce exponentially-slow mixing.

By the definition of the Glauber dynamics, the number of $+$ spins changes by at most $1$ in each step, so the Markov chain starting in $S_2$ must pass through $S_3$ to reach $S_1$.

Applying Theorem~\ref{thm:partitionLB}, we have 
\[ \mu(S_1  )  \ge  \frac{ \E Z_{G,\eta} (\lam)  }{  n Z_G(\lam)} \,.\]
On the other hand,
\[ \mu(S_2) \le n \max_{\eta' :  |\eta'- \eta_-| \le \eps} \frac{ Z_{G,\eta'}(\lam)}{Z_G(\lam)}\le n \max_{\eta' :  |\eta'- \eta_-| \le \eps} \frac{n^2 \E Z_{G,\eta'}(\lam)}{Z_G(\lam)} \le \frac{ n^3 e^{-\delta n} \E Z_{G,\eta} (\lam) }{  Z_G (\lam) }\]
where the second inequality follows from Lemma~\ref{lem:partitionUB}, and the last inequality utilizes Property~\ref{property:globalmax-glauber} above and the fact that $\absolute{\eta' - \eta_{-}} \le \epsilon$ implies $\absolute{\eta' - \eta} > \epsilon$ for $\epsilon$ small enough. Thus, $\mu(S_2) < \mu(S_1)$ for large enough $n$. 

Similarly, since $f_{\Delta,\beta,\lam}$ has a local maximum at $\eta_-$, by Proposition~\ref{thm:fixpt} we have
$$
 \mu(S_3) \le n \max_{\eta': |\eta' - \eta_-| \in [\eps,2 \eps]} \frac{ Z_{G,\eta'}(\lam)}{Z_G(\lam)}\le n \max_{\eta': |\eta' - \eta_-| \in [\eps,2 \eps]} \frac{n^2 \E Z_{G,\eta'}(\lam)}{Z_G(\lam)} \le \frac{ n^3 e^{-\delta n} \E Z_{G,\eta_-}(\lam) }{  Z_G (\lam) } \,.
 $$
Here, we use Property~\ref{property:localmax-glauber} of $\downeta$ for the last inequality. Finally, we have 
$$\mu(S_2) \geq \frac{Z_{G, \downeta}(\lam)}{Z_G(\lam)} \geq \frac{e^{-\zeta n}\E[Z_{G,\downeta}(\lam)]}{Z_G(\lam)}$$
by applying \Cref{prop:localpartitionLB}.
Taking $\zeta \ll \delta$ yields $\mu(S_3) \le e^{-\Omega(n)} \mu(S_2)$.
Applying Lemma~\ref{lem:bottleneck}, we conclude that the mixing time of Glauber dynamics on $G$ is $\exp (\Omega(n))$.
\end{proof}

\subsection{Slow Mixing of the Kawasaki Dynamics}
\label{secKawasakislow}

In this section, we prove that the Kawasaki dynamics are slow mixing at low temperatures when the magnetization satisfies $|\eta| < \eta_u$. 
Our setting will be a graph $H$ constructed as follows: let $G$ be sampled from the random $\Delta$-regular graph $\cG$, and let $H$ consist of $m$ copies of $G$ labelled $G_1, G_2, \dots G_m$, where $m$ will be determined based on $\eta$. 

The proof will proceed similarly to that of Theorem~\ref{thm:GlauberRG} in Section~\ref{sec:Glauberslow}: we will construct sets $S_1, S_2, S_3$ of configurations that satisfy the conditions of Lemma~\ref{lem:bottleneck}, then conclude exponentially slow mixing.
We introduce some notation to aid in describing our candidate sets. 
For a subgraph $G$, let $\sig_G$ denote $\sig$ restricted to $G$. Let $\vec{h} = (I_1, \dots, I_m)$ be such that $I_j \subset [0, 1, \dots, n]$. Then define $S(\vec{h})$ to be the set of all configurations on $H$ (with overall magnetization $\eta$) such that $|\sig_{G_j}|^+ \in I_j$ for all $j$.

\begin{proof}[Proof of Theorem~\ref{thm:main-mixing}]
Let $\beta > \beta_u(\Delta)$. We will separately consider the cases of $|\eta| \in (\eta_c, \eta_u)$ and $|\eta| \le \eta_c$. We start with the former and assume without loss of generality that $\eta > 0$ since the case of $\eta < 0$ will be symmetric.

By Proposition~\ref{propIsingTree}, there exists $\lam_{\eta} \in (1, \lam_u)$ such that $\eta = \eta^+_{\Delta, \beta, \lam_{\eta}}$.
Let $\lam_+ \in (\lam_\eta, \lam_u)$. Then set $\upeta = \eta^+_{\Delta, \beta, \lam_+}$ and $\downeta = \eta^-_{\Delta, \beta, \lam_+}$. 

Observe that by our choice of $\lam_\eta$ and $\lam_+$, we have that $\eta$ is the global maximizer of $f_{\Delta, \beta, \lam_{\eta}}$, $\upeta$ is the global maximizer of $f_{\Delta, \beta, \lam_+}$, and $\eta_-$ is a local maximizer of $f_{\Delta, \beta, \lam_+}$. 
As before, this means that there exist $\eps > 0$ and $\delta > 0$ so that the following properties hold:
\begin{enumerate}
    \item $\E[Z_{G, \eta'}(\lam_{\eta})] \leq e^{-\delta n} \E[Z_{G, \eta}(\lam_{\eta})]$ for all $\eta'$ such that $|\eta' - \eta| > \eps$.\label{property:globalmax}
    \item $\E[Z_{G, \eta'}(\lam_+)] \leq e^{-\delta n}\E[Z_{G, \eta_+}(\lam_+)]$ for all $\eta'$ such that $|\eta' - \eta_+| \in (\eps, 2\eps)$.\label{property:globalmaxnbhd}
    \item $\E[Z_{G, \eta'}(\lam_+)] \leq e^{-\delta n}\E[Z_{G, \eta_-}(\lam_+)]$ for all $\eta'$ such that $|\eta' - \eta_-| \in (\eps, 2\eps)$.\label{property:localmaxnbhd}
\end{enumerate}

For the remainder of this proof, we will use the perspective of size rather than magnetization. Let $k, k_+$, and $k_-$ be the sizes corresponding to $\eta, \upeta$, and $\downeta$, respectively (recall that $k = \lfloor \frac{(\eta+1)n}{2}\rfloor$). 

For $H$ as defined earlier, choose $m$ and $\ell$ such that 
$\frac{\ell}{m}\upeta + \frac{m-\ell}{m}\downeta = \eta$ (note that we may choose $\lam_+$ so that such $\ell, m \in \mathbb N$ exist). 
One can check that $|\sig_H|^+ = k$ when $|\sig_{G_i}|^+ = k_+$
for $1 \leq i \leq \ell$ and $|\sig_{G_j}|^+ = k_-$
for $\ell+1 \leq j \leq m$. 
We now define our configurations $S_1, S_2, S_3$ on the graph $H$.
We use the notation $[a \pm b]$ to mean the integers in the interval $[a-b, a+b]$. Let $\eps > 0$ be as specified above.
For $0 \leq i \leq \ell$ and $0 \leq j \leq m-\ell$, let $U_{i,j}$ be the set of vectors such that 
\begin{itemize}
    \item $i$ of the first $\ell$ components are equal to $[k_+ - 2\eps n, k_+-\eps n]$,
    \item the remainder of the first $\ell$ components are equal to $[
    k_+ \pm \eps n]$,
    \item $j$ of the last $m-\ell$ components are equal to $[k_- + \eps n, k_- + 2\eps n]$, and
    \item the remainder of the last $m-\ell$ components are equal to $[k_- \pm \eps n]$.
\end{itemize}

The three sets of configurations on $H$ that we will analyze are
\begin{align*} S_1 &= S([k]^m),\\
S_2 &= S(([k_+ \pm \eps n]^{\ell}, [k_- \pm \eps n]^{m-\ell})),\\
S_3 &= \bigcup_{(i,j) \neq (0,0)} \{S(\vec{u}) : u \in U_{i,j}\}.
\end{align*}
In words, $S_1$ contains the configurations that have size exactly $k$ on each $G_i$.
The set $S_2$ is defined similarly but with size $(k_+ \pm \eps n)$ on the first $\ell$ components and $(k_- \pm \eps n)$ on the remaining components, under the constraint that the total size is $mk$. The set $S_3$ consists of configurations that ``neighbor" those of $S_2$, where the configuration restricted to at least one component has an increased size (between $\eps n$ and $2\eps n$) compared to $S_2$ but the other components remain the same; again we require that the total size is $mk$.

It is clear that the first condition of Lemma~\ref{lem:bottleneck} holds: in order for the Kawasaki dynamics to pass from $S_2$ to $S_1$, either (1) the configuration on $G_i$ for some $1 \leq i \leq \ell$ must decrease from size $(k_+ \pm \eps n)$ to size $k$, so by appropriate choice of $k_+$ and $\eps$, this implies that at some time step, the configuration must have size in $[k_+ - 2\eps n, k_+ - \eps n]$, or (2) the configuration on $G_j$ for some $\ell+1 \leq j \leq m$ increases from size $(k_- \pm \eps n)$ to $k$ so at some time step, the configuration must have size in $[k_- + \eps n, k_- + 2\eps n]$.

Let $\hat{\mu}$ stand for $\hat{\mu}_{H, \beta, k}$, the fixed-magnetization Gibbs measure. To lighten notation, we drop the dependence on $\beta$ in $Z_{G, \eta}(\beta, \lam)$ and $\hat{Z}_{G, \eta}(\beta)$.
\begin{claim}\label{claim:large-eta}
     $\hat{\mu}(S_1) \geq \hat{\mu}(S_2)$ and $\hat{\mu}(S_3) \leq e^{-\Theta(n)} \hat{\mu}(S_2)$.
\end{claim}

\begin{poc}
    For the first bound, we have
    $$\hat{\mu}(S_1) \geq \frac{(Z_{G, k}(\lam_{\eta}))^m}{\lam_{\eta}^{mk}\hat{Z}_{H, mk}} \geq \frac{(\E Z_{G, k}(\lam_{\eta}))^m}{n^m\lam_{\eta}^{mk}\hat{Z}_{H, mk}}$$
    where the last inequality holds by Theorem~\ref{thm:partitionLB}.
    As there are $\poly(n)$ ways to choose a vector $\vec{v}$ such that $v_j \in I_j$, we also have that
    $$\hat{\mu}(S_2) \leq \poly(n) \frac{\max_{k_1, \dots, k_m} \prod_{i=1}^m Z_{G, k_i}(\lam_{\eta})}{\lam_{\eta}^{mk} \hat{Z}_{H, mk}}$$
    where the maximum is taken over all $k_1, \dots, k_m$ arising from configurations in $S_2$. By Lemma~\ref{lem:partitionUB} and Property~\ref{property:globalmax},
    $$Z_{G, k_i}(\lam_{\eta}) \leq n^2 \E Z_{G, k_i}(\lam_{\eta}) \leq n^2 e^{-\delta n} \E Z_{G, k}(\lam_{\eta})$$
    for some $\delta > 0$. Thus,
    $$\hat{\mu}(S_2) \leq \poly(n) \frac{(e^{-\delta n}\E Z_{G, k}(\lam_{\eta}))^m}{\lam_{\eta}^{mk} \hat{Z}_{H, mk}}$$ 
    and so
    $$\frac{\hat{\mu}(S_2)}{\hat{\mu}(S_1)} \leq e^{-\Theta(n)}\,.$$

    For the second condition, we bound $\hat{\mu}(S_2)$ and $\hat{\mu}(S_3)$ in a similar way but with $\lam = \lam_+$. On one hand,
     $$\hat{\mu}(S_3) \leq \poly(n) \frac{\max_{k_1, \dots, k_m}\prod_{i=1}^m Z_{G, k_i}(\lam_+)}{\lam_+^{mk} \hat{Z}_{H, mk}}$$
    where the maximum is taken over $k_1, \dots, k_m$ arising from configurations in $S_3$. By applying Lemma~\ref{lem:partitionUB} along with Properties~\ref{property:globalmaxnbhd} and \ref{property:localmaxnbhd}, we have
    $$\hat{\mu}(S_3) \leq  \poly(n)\frac{e^{-m\delta n}\E[Z_{G, k_+}(\lam_+)]^{\ell} \E[Z_{G, k_-}(\lam_+)]^{m-\ell}}{\lam_+^{mk} \hat{Z}_{H, mk}}\,.$$
    On the other hand, 
    $$\hat{\mu}(S_2) \geq \frac{(Z_{G, k_+}(\lam_+))^{\ell} (Z_{G, k_-}(\lam_+))^{m-\ell}}{\lam_+^{mk} \hat{Z}_{H, mk}}\,.$$
     We bound the first term in the numerator using Theorem~\ref{thm:partitionLB}:
    $$Z_{G, k_+} (\lam_+) \geq \frac{1}{n} \E[Z_{G, k_+}(\lam_+)]$$
    with high probability. For the second term in the numerator, we apply Proposition~\ref{prop:localpartitionLB}. For any $\zeta > 0$, we have
    $$Z_{G, k_-}(\lam_+) \geq e^{-\zeta n}\E[Z_{G, k_-}(\lam_+)]\,.$$
    Then
    $$\hat{\mu}(S_2) \geq \frac{e^{-\zeta(m-\ell)n}\E[Z_{G, k_+}(\lam_+)]^{\ell} \E[Z_{G,k_-}(\lam_+)]^{m-\ell}}{n^\ell \lam_+^{mk}\hat{Z}_{H,mk}}\,.$$
    Taking $\zeta \ll \delta$ finishes the proof.
    \end{poc}

    In the case that $|\eta| \leq \eta_c$ (again wlog we assume $\eta > 0$), we require a slightly different argument since we cannot apply Proposition~\ref{propIsingTree} to $\eta$. Observe that for any $\delta > 0$ and $\eta \in (0, \eta_c]$, we can find some $\upeta \in [\eta_c, \eta_c+\delta')$ such that $\eta$ can be written as a rational convex combination of $\upeta$ and $\downeta$ (where again $\downeta = \eta^-_{\Delta, \beta, \lam_{\upeta}}$). Indeed, we are done if $\frac{\eta_c}{\eta} \in \mathbb Q$ and in this case, we take $\upeta = \eta_c$ and $\downeta = -\eta_c$ and set $\lam_+ = 1$. Else, we know we can write $\eta$ as a real convex combination of $\eta_c$ and $-\eta_c$. By density of the rationals and continuity of $\eta^+_{\delta, \beta, \lam}$ when $\lam \geq 1$ (Proposition~\ref{propIsingTree}), we may perturb this convex combination to find $m, \ell \in \mathbb N$ along with $\upeta \in (\eta_c, \eta_c + \delta')$ for some $\delta' > 0$ such that $\ell \upeta + (m-\ell)\downeta = m\eta$.

    Define the sizes $k, k_+, k_-$ as before, corresponding to magnetizations $\eta, \upeta, \downeta$ respectively. In both cases, we then consider the following three sets of configurations:
    \begin{align*} S_1 &= S([k_+ \pm \eps n]^{(\ell)}, [k_- \pm \eps n]^{(m-\ell)}),\\ 
    S_2 &= S([k_- \pm \eps n]^{(m-\ell)}, [k_+ \pm \eps n]^{(\ell)}),\\
    S_3 &= \bigcup_{(i,j)\neq (0,0)} \{S(\vec{u}) : u \in U_{i,j}\}\,.
    \end{align*}
where $U_{i,j}$ is defined as in the first case but with our updated choices of $k_+$ and $k_-$.

By the same reasoning as before, we see that the Kawasaki dynamics must pass through $S_3$ in order to go from $S_2$ to $S_1$. We can also see that $\hat{\mu}(S_1) = \hat{\mu}(S_2)$. The proof that $\hat{\mu}(S_3) \leq e^{-\Theta(n)}\hat{\mu}(S_2)$ is identical to that of the previous case. 
\end{proof}

\subsection{Proof of Proposition~\ref{prop:localpartitionLB}}\label{sec:metastable-proof}
In this section, we prove a lower bound for the partition function at local maximizers of $f_{\Delta, \beta, \lam}$. 

For $\eps>0$, let $Z_{G, \eta}^\eps (\lam)$ denote the partition function $Z_{G}(\beta, \lam)$ restricted to configurations in the set $S_{\eta}(\eps) := \{\sig \in \Omega : \eta(\sig) \in (\eta \pm \eps)\}$. We require the following key lemma:
\begin{proposition}\label{lem:metastable-ising}
    Let $\lam \in (1, \lam_u)$. Suppose $\eta$ is a local maximizer of the function $f_{\Delta, \beta, \lam}$. 
    For every $\eps, \delta > 0$, with probability $1-o(1)$ over the random $\Delta$-regular graph $G$,
$$Z_{G, \eta}^\eps (\lam) \geq e^{-\delta n} \E[Z_{G, \eta}(\lam)] \,.$$
\end{proposition}

Proposition~\ref{lem:metastable-ising} follows from the methods used in~\cite{Coja-Oghlan22} to prove metastability of the Potts model on the random $\Delta$-regular graph. Because their proofs carry through once we have made the appropriate translation to our setting, we refrain from repeating their proofs and instead give a high-level overview.

The high probability lower bound on $Z_{G, \eta}^\eps (\lam)$ will follow from a first- and second-moment analysis of a modified random variable, 
$$
    Y_{G,\eta} = Z_{G, \eta}^\eps (\lam) \ind{\mathcal E(\eta)}
$$
where $\mathcal E(\eta)$ is the set of graphs $G$ for which two independent random samples from $\mu_{G, \beta, \lam}$ conditioned on $S_{\eta}(\eps)$ have approximately the same overlap as two independent and uniform configurations of magnetization $\eta$. 
 More precisely, we let $\bsig_{G, \eta} = \bsig_{G, \eta}(\eps)$ denote the distribution $\mu_{G, \beta, \lam}$ conditioned on $S_{\eta}(\eps)$. Let $\nu(\sig, \sig') = \frac{1}{n}(\sigma \cdot \sigma')$ (Observe that this quantity captures the information contained in the overlap matrix used in~\cite{Coja-Oghlan22} since here we have a two-spin model).
We then define $\mathcal E(\eta)$ to be the set of graphs $G$ for which $\E [\nu(\sig, \sig')] = \eta^2 + o(1)$ when $\sig, \sig'$ are independent samples from $\bsig_{G, \eta}$.

The indicator function $\ind{\mathcal E(\eta)}$ makes analyzing the second moment $\E [ Y_{G,\eta}^2  ]$ relatively simple.  Up to subexponential factors it matches the square of the first moment, $(\E[ Y_{G,\eta} ])^2$ (following the argument in~\cite[Proof of Corollary 3.10]{Coja-Oghlan22}).  The key to the proof is showing that not much is lost by including this indicator function; that is,
\begin{align}
\label{eqFMclose}
    \E [ Y_{G,\eta}] = (1+o(1)) \E [ Z_{G, \eta}^\eps (\lam) ] \, .
\end{align}
From this, the bound on $\E [ Y_{G,\eta}^2  ]$, and the fact that $Z_{G, \eta}^\eps (\lam) \ge Y_{G,\eta}$, Proposition~\ref{lem:metastable-ising} follows from the Paley-Zygmund inequality.

Showing~\eqref{eqFMclose} requires several ingredients, including the use of a \textit{planted model} and utilizing a correlation decay property (\textit{non-reconstruction}) of the plus and minus Ising measures on $\mathbb T_\Delta$.

The planted model at magnetization $\eta$ is the reweighted random graph model $\hat \cG_\eta$ defined by
$$
    \Prob[\hat \cG_\eta =G ] = \frac{ Z_{G,\eta}^\eps \Prob[\cG=G]  }{  \E_{\cG} [Z_{G,\eta}^\eps]  } \,.
$$
One can sample the random graph $\hat \cG_\eta$ from the planted model and then sample a configuration from the Gibbs measure conditioned on $S_\eta(\eps)$, which we can denote by $\bsig_{\hat \cG_{\eta},\eta}$.

On the other hand, the {\em Nishimori identities} (as used in \cite{CKPZ17}) allow us to reverse this experiment and first pick a configuration in $S_{\eta}(\eps)$ and then pick a random graph;  doing so will be useful in the proof.
To that end, we define a planted model $\hat{\cG}(\sig)$ with respect to a given configuration $\sig$ as
$$\Prob[\hat{\cG}_{\eta}(\sig) = G] = \frac{\Prob[\cG = G]e^{\beta m_G(\sig)}}{\E_{\cG}[e^{\beta m_G(\sig)}]}$$
and  define a random reweighted configuration $\hat{\bsig}_{\eta}(\eps)$ by
$$\Prob[\hat{\bsig}_{\eta} = \sig] = \frac{{\ind{\sig \in S_{\eta}(\eps)}}\E_{\cG}[e^{\beta m_G(\sig)}\lam^{|\sig|^+}]}{\E_{\cG}[Z^{\eps}_{G, \eta}(\lam)]}$$
Then the Nishimori identities tell us that 
$$(\hat{\cG}_{\eta}, \bsig_{\hat \cG_{\eta}, \eta}) \stackrel{d}{=} (\hat{\cG}(\hat{\bsig}_{\eta}), \hat{\bsig}_{\eta}) \,,$$
see~\cite[Proposition 3.1]{Coja-Oghlan22}.

We will  utilize the property of non-reconstruction for the planted model.
We first consider Gibbs measures on the infinite $\Delta$-regular tree, $\T_{\Delta}$.  As mentioned above, for a given set of Ising parameters $\Delta,\beta, \lam$ there may be a unique such measure or multiple measures. In this section we are in the regime in which multiple measures exist, and we focus on  two particular infinite-volume measures: the $+$ measure arising as the limit of finite-volume measures with all $+$ boundary conditions and the $-$ measure arising from all $-$ boundary conditions. Both of these measures can be constructed via a $2$-spin {\em broadcast process} (see, e.g., \cite{georgii2011gibbs}). Such  a broadcast process   arises from first sampling the spin of the root of $\T_{\Delta}$ and then sampling spins at each successive level conditional on the previous levels according to a $2 \times 2$ {\em broadcast matrix}.  Non-reconstruction holds for a broadcast process when the spins at level $\ell$ give no information about the spin at the root in the $\ell \to \infty$ limit (in terms of total variation distance).  
If this information decays exponentially in $\ell$, we say the process exhibits {\emph strong non-reconstruction} (see~\cite[Proposition 2.6]{Coja-Oghlan22}). 
Non-reconstruction (and indeed strong non-reconstruction) follows if the second eigenvalue $\theta$ of the broadcasting matrix satisfies $\theta^2 (\Delta-1) <1$ (the Kestum-Stigum bound~\cite{kesten1966additional}), but in general this bound is not tight, even for $2 \times 2$ non-symmetric matrices (see~\cite{mossel2003information} for background). Nevertheless, the broadcast process that attains the $+$ measure (and likewise the $-$ measure) satisfies strong non-reconstruction.

\begin{lemma}[\cite{martinelli2007fast}]\label{lem:nonreconstruction}
     For any $\beta \ge 0$ and $\lam \in \mathbb R$, the broadcast processes defining by the $+$ and $-$ measures on $\mathbb T_\Delta$ exhibit strong non-reconstruction.
\end{lemma}
This is the analog of~\cite[Proposition 2.6]{Coja-Oghlan22} in our setting. From this, we can show that an appropriate type of non-reconstruction persists even when we pass to the planted model $\hat{\cG}_{\eta}$. This is done by applying the Nishimori identities and constructing a coupling between the broadcasting process and the random configuration $\hat{\bsig}_{\eta}$ on neighborhoods of a fixed radius.

The last step is to show that we may pass from $Z^{\eps}_{G, \eta}(\lam)$ to $Y_{G, \eta}$ as described at the beginning of this subsection. 
\begin{lemma}
\label{lemOverlapPlanted}
    Let $\eta$ be a local maximizer of $f_{\Delta, \beta, \lam}$, and let $\sig, \sig'$ be independent samples from $\bsig_{\hat{\cG}_{\eta}, \eta}$. Then 
    $$\E[\nu(\sig, \sig') - \eta^2]=o(1)$$
\end{lemma}
This is proved exactly as in~\cite[Lemma 3.8]{Coja-Oghlan22}.  

With these ingredients, we can now prove~\eqref{eqFMclose}. We have
\begin{align*}
    \frac{\E [Y_{\cG,\eta} ]}{ \E[Z_{\cG,\eta}^\eps] } &= \frac{ \sum_{G \in \mathcal E(\eta)}  \Prob[\cG=G]  Z_{G,\eta}^\eps }  {  \E[Z_{\cG,\eta}^\eps]}  = \Prob[ \hat \cG_\eta \in \mathcal E(\eta)] = 1-o(1)  \,,
\end{align*}
with the last equality following from Lemma~\ref{lemOverlapPlanted}.  This completes the proof of Proposition~\ref{lem:metastable-ising}.

With Proposition~\ref{lem:metastable-ising} in hand, we now prove our lower bound for the  partition function restricted to the magnetization of a local maximizer.

\begin{proof}[Proof of \Cref{prop:localpartitionLB}]
    Let $\eta$ be a local maximizer of $f_{\Delta, \beta, \lam}$. To obtain our desired bound on $Z_{G, \eta}(\lam)$, we first consider how the partition function changes when increasing the size of the configuration one vertex at a time. Let $V^-(\sig) \subset V$ be the vertices assigned a $-1$ spin by $\sig$. 
    Given a configuration $\sig$ and $v \in V^-(\sig)$, 
    let $\sig \oplus v$ denote the configuration $\sig'$ where $\sig'(v) = +1$ and $\sig'(u) = \sig(u)$ for all $u \neq v$. Then for any $k \geq 0$, we have
    \begin{align*}
        \frac{Z_{G,k+1}(\lam)}{Z_{G,k}(\lam)}&= \frac{\frac{1}{k+1} \sum_{\sig \in \Omega_k} \sum_{v \in V^-(\sig)} \lam^{k+1}e^{\beta m_G(\sig \oplus v)}}{Z_{G, k}(\lam)}\\
        &= \frac{1}{k+1}\sum_{\sig \in \Omega_k} \frac{\lam^k e^{\beta m_G(\sig)}}{Z_{G, k}(\lam)} \left(\sum_{v \in V^-(\sig)} \frac{\lam^{k+1}e^{\beta m_G(\sig\oplus v)}}{\lam^k e^{\beta m_G(\sig)}}\right)\\
        &= \frac{1}{k+1} \E_{\sig \sim \hat{\mu}_k} \left[\sum_{v \in V^-(\sig)} \frac{\lam^{k+1}e^{\beta m_G(\sig\oplus v)}}{\lam^k e^{\beta m_G(\sig)}}\right]\\
        &\geq \frac{n-k}{k+1} \lam e^{-2 \Delta \beta}\,.
    \end{align*}
    Thus, for $t \geq 1$, we have
    $$Z_{G, k+t}(\lam) 
    \geq \left(\frac{n-(k+t)}{k+t} \lam e^{-2\Delta\beta}\right)^t Z_{G, k}(\lam)\,.$$
    
    In particular, since $\frac{n-(k+t)}{k+t}$ is minimized for $k + t = \Theta(n)$, we have that
    \begin{equation}\label{eqn:ratio-partition}
    Z_{G, k+t}(\lam) \geq e^{c_1t}Z_{G, k}(\lam)
    \end{equation}
    where $c_1 = c_1(\Delta, \beta, \eta, \lam) \geq \log \lam - 2\Delta \beta$.
    
    Similarly,
    $$Z_{G, k}(\lam) \geq \left(\frac{k}{n-k}\lam^{-1}e^{2\Delta\beta}\right)^t Z_{G, k+t}(\lam)\,.$$
    Recalling that $k = \lfloor n\frac{\eta+1}{2}\rfloor$, we then know that $\frac{k}{n-k} \geq \frac{\eta+1}{2} - \frac1n$ and thus
    \begin{equation}\label{eqn:ratio-partition2}
        Z_{G, k}(\lam) \geq e^{c_2t} Z_{G, k+t}(\lam)
    \end{equation}
    where $c_2 = c_2(\Delta, \beta, \eta, \lam)$. Let $c = \min\{c_1, c_2\}$. 
    
    We now combine the above computations with our previous lemmas. Fix $\eta$ as a local maximizer of $f_{\Delta, \beta, \lam}$ and let $\zeta > 0$. Set $\delta = \frac{\zeta}{4}$. If $c = c(\Delta, \beta, \eta, \lam) > 0$, set $\eps = \frac{\zeta}{c}$, and  if $c < 0$, set $\eps = \frac{\zeta}{-c}$. 
    Suppose first that $c < 0$. 
    By \Cref{lem:metastable-ising} we have with high probability,
    $$Z_{G, \eta}^{-\zeta/c} \geq e^{-(\zeta/4)n} \E[Z_{G, \eta}]\,.$$
    Thus, there exists some $\eta'$ where $|\eta' - \eta| < \frac{\zeta}{c}$ such that $Z_{G, \eta'} \geq \frac{1}{n}e^{-(\zeta/4)n} \E[Z_{G, \eta}] \geq e^{-(\zeta/2)n}\E[Z_{G,\eta}]$ 
    for $n$ large enough. 
    Let $k', k$ correspond to $\eta', \eta$, respectively. Then $|k' - k| < \frac{\zeta n}{-2c}$. Applying \eqref{eqn:ratio-partition} or \eqref{eqn:ratio-partition2} in the case that $\eta' \leq \eta$ or $\eta < \eta'$, respectively, with $t \leq \frac{\zeta n}{-2c}$ gives us
    $$Z_{G, \eta} \geq e^{ct}Z_{G, \eta'} \geq e^{-(\zeta/2)n}Z_{G, \eta'} \geq \frac1n e^{-\zeta n} \E[Z_{G, \eta}]\,.$$
    If $c > 0$, we perform similar computations but conclude using the bound $t \geq 1$ to obtain
    \begin{equation*}
        Z_{G, \eta} \geq e^cZ_{G, \eta'}  \geq e^{-\zeta n} \E[Z_{G, \eta}] \,. \qedhere
        \end{equation*}
\end{proof}

\section{Acknowledgments}

The authors thank Zongchen Chen and Marcus Michelen for very helpful discussions. WP supported in part by  NSF grant CCF-2309708. CY supported in part by NIH grant R01GM126554.

\bibliographystyle{abbrv}
\bibliography{references}

\newpage

\appendix

\section{Proof that Zero-Freeness Implies $\ell_{\infty}$-Independence} \label{appendix:zero_freeness_SI}
Our goal is to prove the following theorem. 
\begin{customthm}{\bf \ref{thm:zero_free_SI}}
    Fix $\beta \ge 0$ and $\Delta \in \N$. 
    Let $D \subset \R_{>0}$ be compact and assume there is some $\delta > 0$ such that the Ising model is absolutely $\delta$-zero-free at every $\lambda \in D$. 
    Then there is some constant $C > 0$, only depending on $D$, $\lambda$, $\beta$ and $\Delta$, such that for all $\lambda \in D$, $G \in \mathcal{G}_{\Delta}$ and all pinnings $\tau_U$ it holds that $\hat{\mu}^{\tau_U}_{G, \beta, \lambda}$ is $C$-$\ell_{\infty}$-independent.
\end{customthm}
To simplify notation, we write $M_{\tau_U}$ for the pairwise influence matrix $M_{\mu_{G, \beta, \lambda}^{\tau_U}}$ for any pinning $\tau_U$ whenever $\beta$, $\lambda$ and $G$ are clear from the context.
Further, it will be useful to consider a multivariate version of the (pinned) Ising partition function, defined by
\[
    Z_{G}^{\tau_U}(\beta, \pmb{\lambda}) \coloneqq \sum_{\sigma \in \Omega^{\tau_U}} e^{\beta m_{G}(\sigma)} \prod_{v \in V: \sigma(v) = +1} \lambda_{v}
\]
for any multivariate external field $\pmb{\lambda} = (\lambda_v)_{v \in V}$.
The following lemma expresses pairwise influence as derivatives of multivariate Ising partition functions.

\begin{lemma} \label{lemma:pairwise_influence_derivative}
    For every $\beta$, $G$, $\lambda$, every pinning $\tau_U$ and all $u, v \in V$ it holds that
    \[
        M_{\tau_U}[u, v] = \lambda \cdot \frac{\partial \ln( Z^{\tau_U, \pmb{+}_{u}}(\beta, \pmb{\lambda}) / Z^{\tau_U}(\beta, \pmb{\lambda}))}{\partial \lambda_v} \Big\rvert_{\pmb{\lambda} = \lambda \cdot \pmb{1}}
    \]
    where $\pmb{\lambda} = (\lambda_v)_{v \in V}$ and $\lambda \cdot \pmb{1}$ is the function that is constant $\lambda$ on $V$.
\end{lemma}

\begin{proof}
    Note that
    \[
        \lambda \cdot \frac{\partial \ln (Z^{\tau_U, \pmb{+}_{u}}(\beta, \pmb{\lambda}))}{\partial \lambda_v}\Big\rvert_{\pmb{\lambda} = \lambda \cdot \pmb{1}}
        = \frac{\lambda}{Z^{\tau_U, \pmb{+}_{u}}(\beta, \lambda)} \cdot \frac{\partial Z^{\tau_U, \pmb{+}_{u}}(\beta, \pmb{\lambda})}{\partial \lambda_v}\Big\rvert_{\pmb{\lambda} = \lambda \cdot \pmb{1}}
    \]
    and 
    \[
        \frac{\partial Z^{\tau_U, \pmb{+}_{u}}(\beta, \pmb{\lambda})}{\partial \lambda_v} = \sum_{\sigma \in \Omega^{\tau_U, \pmb{+}_{u, v}}} e^{\beta m_{G}(\sigma)} \prod_{w \in V \setminus v\,:\, \sigma(w) = +1} \{\lambda_w\,|\,w\in V\setminus v, \sig(w) = +1\}.
    \]
    Hence, we get
    \[
        \lambda \cdot \frac{\partial \ln (Z^{\tau_U, \pmb{+}_{u}}(\beta, \pmb{\lambda}))}{\partial \lambda_v} \Big\rvert_{\pmb{\lambda} = \lambda \cdot \pmb{1}} = \frac{Z^{\tau_U, \pmb{+}_{u, v}}(\beta, \lambda)}{Z^{\tau_U, \pmb{+}_u}(\beta, \lambda)} = \mu^{\tau_U}(\sigma(v) = +1 \mid \sigma(u) = +1) .
    \]
    Similarly, it holds that
    \[
        \lambda \cdot \frac{\partial \ln (Z^{\tau_U}(\beta, \pmb{\lambda}))}{\partial \lambda_v} \Big\rvert_{\pmb{\lambda} = \lambda \cdot \pmb{1}} = \frac{Z^{\tau_U, \pmb{+}_{v}}(\beta, \lambda)}{Z^{\tau_U}(\beta, \lambda)} = \mu^{\tau_U}(\sigma(v) = +1),
    \]
    and subtracting both proves the claim.
\end{proof}

The following lemma results from \Cref{lemma:pairwise_influence_derivative} by applying a multivariate chain rule and the FKG inequality; we obtain the following bound on the $\ell_{\infty}$-norm of the pairwise influence matrix.

\begin{lemma} \label{lemma:chain_rule_SI}
    For every $\beta$, $G$, $\lambda > 0$, every pinning $\tau_U$ and all $u \in V$ it holds that
    \[
        \sum_{v \in V} |M_{\tau_U}[u, v]| = f'(0)
    \]
    where $f: (- \lambda, \infty) \to \R, \zeta \mapsto \lambda \cdot \ln\left( \frac{Z^{\tau_U, \pmb{+}_{u}}(\beta, \lambda + \zeta)}{Z^{\tau_U}(\beta, \lambda + \zeta)} \right)$.
\end{lemma}

\begin{proof}
    For $v \in V$ define $g_v(\zeta) = \lambda + \zeta$ and note that 
    \[
        f(\zeta) = \lambda \cdot \ln\left( \frac{Z^{\tau_U, \pmb{+}_{u}}(\beta, (g_v(\zeta)_{v \in V})}{Z^{\tau_U}(\beta, (g_v(\zeta)_{v \in V}))} \right) .
    \]
    Hence, by the chain rule, \Cref{lemma:pairwise_influence_derivative} and the fact that $g'_v(0) = 1$, we have
    \[
        f'(0) = \sum_{v \in V} \lambda \cdot \frac{\partial \ln( Z^{\tau_U, \pmb{+}_{u}}(\beta, \pmb{\lambda}) / Z^{\tau_U}(\beta, \pmb{\lambda}))}{\partial \lambda_v} \Big\rvert_{\pmb{\lambda} = \lambda \cdot \pmb{1}} \cdot g'_v(0)
        = \sum_{v \in V} M_{\tau_U}[u, v].
    \]
    Observing that by the FKG inequality all entries of $M_{\tau_U}$ are non-negative concludes the proof.
\end{proof}

The last ingredient we need is the following bound.
\begin{lemma} \label{lemma:montel_SI}
    Let $\beta$, $\Delta$, $D$ and $\delta$ be as in \Cref{thm:zero_free_SI}.
    There exists a constant $C$, only depending on $\delta$, $\beta$, $\lambda$ and $\Delta$, such that for all $G \in \mathcal{G}_{\Delta}$, all $u \in V$, all $\lambda \in D$ and all pinnings $\tau_U$ it holds that the complex function $\left\vert \lambda \cdot \ln\left( \frac{Z^{\tau_U, \pmb{+}_{u}}(\beta, \lambda + \zeta)}{Z^{\tau_U}(\beta, \lambda + \zeta)} \right)\right\rvert \le C$ for all $\zeta \in \C$ with $|\zeta| \le \delta/2$.
\end{lemma}

\begin{proof}
    The proof follows from an application of Montel's theorem. 
    More precisely, we apply \Cref{cor:montel} using the zero-freeness and the fact that the Ising model on bounded degree graphs has vertex marginals bounded way from $0$ and $1$ similar to \Cref{lemma:cumulant_stability}.
\end{proof}

We are now ready to prove \Cref{thm:zero_free_SI}.
\begin{proof}[Proof of \Cref{thm:zero_free_SI}]
    For any $G \in \mathcal{G}_{\Delta}$, $u \in V$ and pinning $\tau_U$ consider $f$ as in \Cref{lemma:pairwise_influence_derivative} and and define $h: \mathcal{N}(0, \delta) \to \hat{\C}, \zeta \mapsto \lambda \cdot \ln\left( \frac{Z^{\tau_U, \pmb{+}_{u}}(\beta, \lambda + \zeta)}{Z^{\tau_U}(\beta, \lambda + \zeta)} \right)$.
    Note that $h$ is analytic on its domain due to the zero-freeness.
    Further, $f$ and $h$ agree on the real interval $(- \min(\delta, \lambda), \delta)$, and thus $f'(0) = h'(0)$.
    By \Cref{lemma:chain_rule_SI} it now suffices to show that $h'(0)$ is bounded by some constant that only depends on $D$, $\delta$, $\Delta$ and $\beta$, which follows immediately from \Cref{lemma:montel_SI} and the Cauchy integral formula.
\end{proof}

\section{Proofs for Fast Mixing}\label{appendix:fast}

\subsection{Proofs: Edgeworth Expansion and Stability}\label{sec:proof edgeworth} 

In this section, we prove several of the technical lemmas required for our fast mixing result. All of them have the following assumptions:
\begin{customcond}{\bf \ref{cond:fast}}
    \begin{enumerate}
        \item Let $\beta \ge 0$, and let $D \subset \R_{>0}$ be compact such that there is some $\delta > 0$ for which the Ising model is absolutely $\delta$-zero-free for all $\lambda \in D$. Further, let $\lambda \in D$.
        \item Let $\alpha \in [0, 1)$, let $U \subset V$ with $\size{U} \le \alpha n$ and let $\tau_U$ be a pinning of $U$.
        \item Let $\sigma \sim \mu^{\tau_U}_{\beta, \lambda}$ and let $X = \size{\sigma}^{+}$.
    \end{enumerate}
\end{customcond}

\begin{customlem}{\bf \ref{lemma:characteristic_function}}
    Assume Condition~\ref{cond:fast} holds.
    There exists $c = c(\Delta, \beta, D, \alpha) > 0$ such that for all $t \in [- \pi, \pi]$ it holds that $|\E[e^{itX}]| \le e^{- c t^2 n}$.
\end{customlem}

\begin{proof} \label{proof:characteristic_function}
    We start with the same construction as in the proof of \Cref{lemma:variance}.
    Set $W = V \setminus U$ and let $J$ be an independent set in $G[W]$ of size at least $\frac{1-\alpha}{\Delta + 1} n$.
    Using the law of total expectation and the spatial Markov property yields
    \begin{align*}
        |\E[e^{itX}]|  
        &= |\E[\E[e^{itX} \mid \sigma(V \setminus J)]]| \\
        &\le \max_{\xi \in \Omega^{\tau_U}} |\E[e^{itX} \mid \sigma(V \setminus J) = \xi(V \setminus J)]| \\
        &= \max_{\xi \in \Omega^{\tau_U}} \prod_{v \in J} |\E[e^{it \cdot \ind{\sigma(v) = +1}} \mid \sigma(V \setminus J) = \xi(V \setminus J)]| \\
        &\le \prod_{v \in J} \max_{\xi \in \Omega^{\tau_U}} |\E[e^{it \cdot \ind{\sigma(v) = +1}} \mid \sigma(V \setminus J) = \xi(V \setminus J)]|.
    \end{align*}
    Next, fix $v \in J$ and $\xi \in \Omega^{\tau_U}$ and let $p = \Prob(\sigma(v) = +1\ |\ \sigma(V \setminus J) = \xi(V \setminus J))$.
    It holds that 
    \begin{align*}
        |\E[e^{it \cdot \ind{\sigma(v) = +1}} \mid \sigma(V \setminus J) = \xi(V \setminus J)]|^2 
        &= p^2 + (1-p)^2 + 2 p (1-p) (e^{it} + e^{-it}) \\
        &= 1 - 2 p (1-p) (1 - \cos(t)) \\
        &\le 1 - 2 p (1-p) t^2 \\
        &\le e^{-2 p (1-p) t^2} .
    \end{align*}
    Note that $p (1-p) \ge \frac{\lambda}{(1 + \lambda)^2} e^{- \beta \Delta}$. 
    Thus, setting $c = \frac{1-\alpha}{\Delta + 1} e^{-\beta \Delta} \cdot \min_{\lambda \in D} \frac{\lambda}{(1+ \lambda)^2}  > 0$ concludes the proof.
\end{proof}

\begin{customlem} {\bf \ref{lemma:cumulant_bounds}}
    Suppose Condition~\ref{cond:fast} holds.  
    There exists some $\varepsilon = \varepsilon(\Delta, \beta, D, \delta) > 0$ such that for all $j \in \N$, it holds that
    \[
        |\kappa_j(X)| = O(j! 2^{j} \varepsilon^{-j} n) ,
    \]
    where the implied constants depend on $\Delta$, $\beta$, $\delta$ and $D$.
    Moreover, for $t \in \C$ with $|t| \le \varepsilon/4$ and all $d \in \N$ it holds that 
    \[
        \left\lvert \ln \frac{Z^{\tau_U} (\beta, \lambda e^{t})}{Z^{\tau_U} (\beta, \lambda)} - \sum_{j=1}^{d} \kappa_{j}(X) \frac{t^j}{j!} \right\rvert = O(|2t/\varepsilon|^{d+1} n)
    \]
    with constants depending on $\Delta$, $\beta$, $\delta$ and $\lambda$.
\end{customlem}

\begin{proof}\label{proof:cumulant_bounds}
    Observe that 
    \[
        \kappa_{j}(X) 
        = \frac{\text{d}^j}{\text{d} t^j} \ln \frac{Z^{\tau_U}(\beta, \lambda e^{t})}{Z^{\tau_U}(\beta, \lambda)} \Big\rvert_{t = 0} 
    \]
    provided the expectation exists in a neighborhood of $t=0$.
    
    We proceed by considering the function $t \mapsto \ln \frac{Z^{\tau_U}(\beta, \lambda e^{t})}{Z^{\tau_U}(\beta, \lambda)}$ in a complex neighborhood of $t=0$.
    Since $\lambda \in D$, we know that $Z^{\tau_U}(\beta, z) \neq 0$ for all $z \in \mathcal{N}(\lambda, \delta)$ for some $\delta > 0$ by \Cref{cond:fast}.
    Thus, setting $\varepsilon = \min_{\lambda \in D} \frac{2 \delta}{3 \lambda} > 0$ we have that $Z^{\tau_U}(\beta, \lambda e^{t}) \neq 0$ and $t \mapsto \ln \frac{Z^{\tau_U}(\beta, \lambda e^{t})}{Z^{\tau_U}(\beta, \lambda)}$ is analytic for all $|t| < \varepsilon$.

    Next, note that our desired bound on $|\kappa_j(X)|$ follows from Cauchy's integral formula once we show that 
    \[
        \max_{|t| = \varepsilon/2} \left\lvert \ln \frac{Z^{\tau_U}(\beta, \lambda e^{t})}{Z^{\tau_U}(\beta, \lambda)} \right\rvert = O(n)
    \]
    with constants depending only on $\Delta$, $\beta$, $\delta$ and $D$.
    To this end, note that $z \mapsto Z^{\tau_U}(\beta, z)$ is a polynomial; let its roots be $\{\xi_k\ |\ 1 \le k \le n\}$.
    Thus, we have
    \[
        \max_{|t| = \varepsilon/2} \left\lvert \ln \frac{Z^{\tau_U}(\beta, \lambda e^{t})}{Z^{\tau_U}(\beta, \lambda)} \right\rvert 
        = \max_{|t| = \varepsilon/2} \left\lvert \ln \left( \prod_{k = 1}^{n} \frac{\lambda e^{t} - \xi_{k}}{\lambda - \xi_k} \right) \right\rvert .
    \]
    Since $\lambda \in D$, we know that for all $k$, both $|\lambda - \xi_k|$ and $|\lambda e^{t} - \xi_k|$ are bounded away from $0$ by a constant only depending on $\Delta$, $\beta$, $\delta$ and $D$, which proves the desired bound and thus the first part of our claim.

    For the second part of the claim, considering the Taylor expansion of $t \mapsto \ln \frac{Z^{\tau_U}(\beta, \lambda e^{t})}{Z^{\tau_U}(\beta, \lambda)}$ around $0$ yields
    \[
        \left\lvert \ln \frac{Z^{\tau_U} (\beta, \lambda e^{t})}{Z^{\tau_U} (\beta, \lambda)} - \sum_{j=1}^{d} \kappa_{j}(X) \frac{t^j}{j!} \right\rvert 
        \le \sum_{j > d} |\kappa_j(X)| \frac{|t|^j}{j!}.
    \]
    Applying our bound on $|\kappa_j(X)|$ then gives us the desired result.
\end{proof}

\begin{customthm}{\bf \ref{thm:edgeworth}}
Suppose Condition~\ref{cond:fast} holds.
    Let $d \in \N$ and let $\ell \in \R$ such that $\E[X] + \ell \in \Z_{\geq 0}.$
    Set $s = \sqrt{\Var(X)}$ and $\beta_j = \frac{\kappa_j(X)}{j! s^j}$ for all $j \in \N$, and write $H_k(\cdot)$ for the $k^{\text{th}}$ Hermite polynomial. 
    It holds that
    \[
        \mu_{\beta, \lambda}^{\tau_{U}}(X - \E[X] = \ell) = \frac{e^{-\frac{\ell^2}{2 s^2}}}{\sqrt{2 \pi} s} \cdot  \left(1 + \sum_{r \ge 3} H_{r}(\ell/s) \sum_{k_3, \dots, k_{2d+1}} \prod_{j=3}^{2d+1} \frac{\beta_j^{k_j}}{k_j !} \right) + O\left(n^{-d}\right) 
    \]
    where the inner sum is over tuples $k_3, \dots, k_{2d+1} \in \Z_{\geq 0}$ such that $\sum_j k_j \cdot j = r$ and $\sum_j k_j \cdot \frac{j-2}{2} \le d$, and the implied constants in the asymptotic notation depend only on $\Delta, \beta, \delta,  D$, $d$ and $\alpha$.
\end{customthm}

\begin{proof}\label{proof:edgeworth}
    Using Fourier inversion for random variables on a lattice, we have
    \[
        \mu_{\beta, \lambda}^{\tau_{U}}(X - \E[X] = \ell) = \frac{1}{2 \pi s} \int_{|t| \le \pi s} e^{-it\ell/s} \E[e^{it(X - \E X)/s}] \text{d} t .
    \]
    We proceed by splitting up the integral into $|t| \le C \log(n)$ and $|t| > C \log(n)$ for a constant $C$ to be chosen later.
    By \Cref{lemma:characteristic_function} we have $|\E[e^{it(X - \E X)/s}]| = |\E[e^{itX/s}]| \le e^{-c t^2 n/s^2}$ for some $c = c(\Delta, \beta, D, \alpha) > 0$.
    Thus, we have
    \[
         \left\lvert \frac{1}{2 \pi s} \int_{C \log(n) < |t| \le \pi s} e^{-it\ell/s} \E[e^{it(X - \E X)/s}] \text{d} t \right\rvert
         \le \frac{1}{2 \pi s} \int_{|t| > C \log(n)} e^{-c t^2 n/s^2} \text{d} t
         \le \frac{e^{- c C n \log(n)/{s^2}}}{c \pi n/s} .
    \]
    Since $s^2 \in \Theta(n)$ by \Cref{lemma:variance}, we can choose $C = C(\Delta, \beta, \delta, D, \alpha, d)$ sufficiently large to ensure that this is in $O(n^{-d})$.
    For such $C$, we have
    \begin{align}
        \mu_{\beta, \lambda}^{\tau_{U}}(X - \E[X] = \ell) = \frac{1}{2 \pi s} \int_{|t| \le C \log(n)} e^{-it\ell/s} \E[e^{it(X - \E X)/s}] \text{d} t + O(n^{-d}) . \label{thm:edgewort:eq1}
    \end{align}

    We proceed with rewriting the integral for $|t| \le C \log(n)$ and $C$ as above.
    By \Cref{lemma:variance}, we know that $|t/s| \to 0$ as $n \to \infty$ for all such $t$.
    Thus, using \Cref{lemma:cumulant_bounds}, we obtain
    \begin{align*}
        \E[e^{it(X - \E X)/s}] 
        &= e^{-i t \E X/s} \exp\left[\log \frac{Z^{\tau_U}(\beta, \lambda e^{i t / s})}{Z^{\tau_U}(\beta, \lambda)}\right] \\
        &= e^{-i t \E X/s} \exp\left[\sum_{j = 1}^{2d + 1} \kappa_j(X) \frac{(it/s)^j}{j!} + O(n^{-d} \log(n)^{2d+2})\right] \\
        &= e^{-t^2 / 2} \exp\left[\sum_{j = 3}^{2d + 1} \beta_j (it)^j \right] \exp\left[O(n^{-d} \log(n)^{2d+2})\right] ,
    \end{align*}
    where the last step uses the definition of $\beta_j$ and the fact that $\kappa_1(X) = \E[X]$ and $\kappa_2(X) = s^2$.
    Taylor expansion of the exponential function and grouping terms appropriately yields
    \[
         \exp\left[\sum_{j = 3}^{2d + 1} \beta_j (it)^j \right] = 1 + \sum_{r \ge 3} (it)^r \sum_{\substack{k_3, \dots, k_{2d+1} \in \N_{0}: \\ \sum_{j} k_j \cdot j = r}} \prod_{j=3}^{2d+1} \frac{\beta_j^{k_j}}{k_j!} .
    \]
    We proceed with restricting the integer tuples $k_3, \dots, k_{2d+1}$ in the above sum.
    To this end, note that $|\beta_j| = O(n^{-(j-2)/2})$ and thus $\prod_{j=3}^{2d+1} \beta_j^{k_j} = O(n^{- \sum_{j} k_j \cdot \frac{j-2}{2}})$ with implicit constants depending only on $\Delta$, $\beta$, $\delta$, $D$, $\alpha$ and $d$.
    Moreover, for all integer sequences with $\sum_{j=3}^{2d+1} k_j \cdot j = r$ it holds that $\sum_{j=3}^{2d+1} k_j \cdot \frac{j-2}{2} \ge r/6$. 
    Define sets $S_d = \{(k_3, \dots, k_{2d+1}) : k_j \in \mathbb Z_{\geq 0}, \sum_j j \cdot k_j = r, \sum_j \frac{j-2}{2} k_j > d\}$ and $T_{d, m} = \{(k_3, \dots, k_{2d+1}) : k_j \in \mathbb Z_{\geq 0}, \sum_j j \cdot k_j = r, \sum_j \frac{j-2}{2} k_j = m\}$.
    Thus, we have
    \begin{align*}
        \left\lvert \sum_{r \ge 3} (it)^r \sum_{S_d} \prod_{j=3}^{2d+1} \frac{\beta_j^{k_j}}{k_j!} \right\rvert
        \le 
        \sum_{d' > d} O(n^{-d'}) \sum_{r \ge 3}^{6d'} |t|^r \sum_{T_{d, d'}} \prod_{j=3}^{2d+1} \frac{1}{k_j!} .
    \end{align*}
    Using $|t| \le C \log(n)$ and 
    \[
         \sum_{T_{d, d'}} \prod_{j=3}^{2d+1} \frac{1}{k_j!} \le e^{2d-2} 
    \]
    we obtain
    \begin{align*}
        \left\lvert \sum_{r \ge 3} (it)^r \sum_{S_d} \prod_{j=3}^{2d+1} \frac{\beta_j^{k_j}}{k_j!} \right\rvert
        = \sum_{d' > d} O(n^{-d'} \log(n)^6d')
        = O(n^{-d} \log(n)^{6d})
    \end{align*}
    with implied constants depending on $\Delta$, $\beta$, $\delta$, $D$, $\alpha$ and $d$.

    Substituting everything back into \eqref{thm:edgewort:eq1} gives
    \begin{align}
        \mu_{\beta, \lambda}^{\tau_{U}}(X - \E[X] = \ell) = \frac{1}{2 \pi s} \int_{|t| \le C \log(n)} e^{-it\ell/s} e^{-t^2 / 2} \left(1 + \sum_{r \ge 3} (it)^r \sum_{k_3, \dots, k_{2d+1}} \prod_{j=3}^{2d+1} \frac{\beta_j^{k_j}}{k_j!}\right) \text{d} t + O(n^{-d}) , \label{thm:edgeworth:eq2}
    \end{align}
    where the inner sum is over tuples $k_3, \dots, k_{2d+1} \in \N_{0}$ such that $\sum_{j=3}^{2d+1} k_j \cdot j = r$ and $\sum_{j=3}^{2d+1} k_j \cdot \frac{j-2}{2} \le d$, and where we used that $s = \Theta(\sqrt{n})$ to drop the poly-logarithmic factors from the error term.

    To get from \eqref{thm:edgeworth:eq2} to the desired expression, we first exchange integration and summation. 
    First, observe that by the inverse Fourier transform of the normal distribution, it holds that
    \begin{align*}
        \left\lvert e^{-\frac{\ell^2}{2 s^2}} - \frac{1}{\sqrt{2 \pi}} \int_{|t| \le C \log(n)} e^{-it\ell/s} e^{-t^2 / 2}  \text{d} t \right\rvert
        &\le \left\lvert\frac{1}{\sqrt{2 \pi}} \int_{|t| > C \log(n)} e^{-it\ell/s} e^{-t^2 / 2}  \text{d} t \right\rvert \\
        &\le \int_{|t| > C \log(n)} e^{-t^2 / 2}  \text{d} t \\
        &= o(n^{-d}) .
    \end{align*}
    Thus, it remains to bound 
    \[
        \sum_{r \ge 3} \left\lvert \frac{1}{\sqrt{2\pi}} \int_{|t| \le C \log(n)} e^{-it\ell/s} e^{-t^2 / 2} (it)^r \text{d} t - e^{-\frac{\ell^2}{2 s^2}} H_r(\ell/s)\right\rvert \sum_{k_3, \dots, k_{2d+1}} \prod_{j=3}^{2d+1} \frac{\beta_j^{k_j}}{k_j!} 
    \]
    where the inner sum is over tuples $k_3, \dots, k_{2d+1}$ as in \eqref{thm:edgeworth:eq2}.
    First, note that for $r > 6d$ there are no tuples $k_3, \dots, k_{2d+1} \in \Z_{\geq 0}$ that satisfy the requirements $\sum_{j=3}^{2d+1} k_j \cdot j = r$ and $\sum_{j=3}^{2d+1} k_j \cdot \frac{j-2}{2} \le d$, allowing us to restrict the outer sum to $3 \le r \le 6d$.
    Using the identity
    \[
         e^{-\frac{x^2}{2}} H_r(x) = \int_{\R} e^{-itx} e^{-t^2 / 2} (it)^r \text{d} t,
    \]
    and the fact that $|\beta_j| = O(n^{-(j-2)/2})$ yields
    \begin{align*}
        &\sum_{r \ge 3} \left\lvert \frac{1}{\sqrt{2\pi}} \int_{|t| \le C \log(n)} e^{-it\ell/s} e^{-t^2 / 2} (it)^r \text{d} t - e^{-\frac{\ell^2}{2 s^2}} H_r(\ell/s)\right\rvert \sum_{k_3, \dots, k_{2d+1}} \prod_{j=3}^{2d+1} \frac{\beta_j^{k_j}}{k_j!} \\
        &\hspace{2em} \le  O(1) \cdot \sum_{r = 3}^{6d} \frac{1}{\sqrt{2\pi}} \int_{|t| > C \log(n)} e^{-t^2 / 2} |t|^r \text{d} t \\
        &\hspace{2em} = o(n^{-d}) ,
    \end{align*}
    concluding the proof.
\end{proof}

\begin{customcor}{\bf \ref{cor:edgeworth}}
    Suppose in the setting of \Cref{thm:edgeworth} that $|\ell| \le L$ for some $L \in \Z_{\geq 0}$. Then
    \[
        \mu_{\beta, \lambda}^{\tau_{U}}(X - \E[X] = \ell) = \frac{e^{-\frac{\ell^2}{2 s^2}}}{\sqrt{2 \pi} s} + O\left(n^{-3/2}\right) 
    \]
    with implied constants in the asymptotic notation depend only on $\Delta, \beta, \delta, D$, $L$ and $\alpha$.
\end{customcor}

\begin{proof}\label{proof:cor_edgeworth}
    We prove this statement by applying \Cref{thm:edgeworth} for $d=2$, implying that we only need to consider $3 \le r \le 12$.
    Thus, using $|\ell| \le L$ and $s \in \Theta(n)$ by \Cref{lemma:variance}, we can upper bound $H_r(\ell / s)$ for all such $r$ by some constant that only depends on $\Delta, \beta, \delta, D$, $L$ and $\alpha$.
    Next, recall that $|\beta_j| = O(n^{-(j-2)/2})$.
    Thus, for all valid tuples $(k_3, k_4, k_5)$ accept $(1, 0, 0)$, $(0, 1, 0)$ and  $(2, 0, 0)$ it follows that $|\prod_{j=3}^{5} \beta_j^{k_j}| = O(n^{-3/2})$.
    Therefore, we obtain
    \[
        \mu_{\lambda}^{\tau_{U}}(X - \E[X] = \ell) = \frac{e^{-\frac{\ell^2}{2 s^2}}}{\sqrt{2 \pi} s} \cdot \left( 1 + H_3(\ell/s) \beta_3 + H_{4}(\ell/s) \beta_4 + H_6(\ell/s) \frac{\beta_3^2}{2}\right) + O\left(n^{-3/2}\right),
    \]
    and it remains to argue that
    \[
        \frac{e^{-\frac{\ell^2}{2 s^2}}}{\sqrt{2 \pi} s} \cdot \left\lvert H_3(\ell/s) \beta_3 + H_{4}(\ell/s) \beta_4 + H_6(\ell/s) \frac{\beta_3^2}{2}\right\rvert = O(n^{-3/2}) .
    \]
    To see this, recall that $s = \Theta(\sqrt{n})$, and note that $|\beta_3| = O(n^{-1/2})$, $|\beta_4| = O(n^{-1})$ and $H_3(x) = x^{3} -  3x$. 
\end{proof}

\subsection{Proof: Spectral Gap}\label{sec:spectral_gap}
In this section, we give the proofs of two claims required for the proof of \Cref{thm:spectral_gap}.

\begin{customclaim}{\bf \ref{claim:pinned}}
        In the setting of the proof of \Cref{thm:spectral_gap}, we have
        \[
            \frac{\E[\Var_{\pi_{k-1}}(f) \mid \pi_{\ell}]}{\Var_{\pi_{\ell}}(f)} \ge \inf_{U \in \binom{V}{\ell}}\gap(\plusDownUp^{U}_{\beta, k}) \text{ a.s.},
        \]
        where $\plusDownUp^{U}_{\beta, k}$ is the pinned down-up walk as in \Cref{def:plusDownUp}.
\end{customclaim}

\begin{proof}[Proof of \Cref{claim:pinned}]
    Recall that $\pi = \hat{\mu}_{\beta, k}$ and note that 
    \[
        \frac{\E[\Var_{\pi_{k-1}}(f) \mid \pi_{\ell}]}{\Var_{\pi_{\ell}}(f)} = \frac{\E[\Var_{\pi_{k-1}}(f) \mid U_{\ell}]}{\Var_{\pi_{\ell}}(f)} ,
    \]
    since the random probability measure $\pi_{\ell} = \hat{\mu}_{\beta, k}^{U_{\ell}}$ only depends on the random set $U_{\ell}$.
    Hence, it suffices to show that for all $U \in \binom{V}{\ell}$ we have
    \[
        \frac{\E[\Var_{\pi_{k-1}}(f) \mid U_{\ell} = U]}{\Var_{\hat{\mu}_{\beta, k}^{U}}(f)} \ge \gap(\plusDownUp_{\beta, k}^{U}) .
    \]
    To this end, note that for every $S_1, S_2 \in \Omega_k^{U}$ with $S_1 \neq S_2$ it holds that
    \begin{align*}
        \frac{\E[\pi_{k-1}(S_1) \pi_{k-1}(S_2) \mid U_{\ell} = U]}{\hat{\mu}_{\beta, k}^{U} (S_1)}
        &= \sum_{W \in \binom{V}{k-1}} \frac{\hat{\mu}_{\beta, k}^{W} (S_1) \cdot \hat{\mu}_{\beta, k}^{W} (S_2)}{\hat{\mu}_{\beta, k}^{U} (S_1)} \cdot \Prob[U_{k - 1} = W \mid U_{\ell} = U] \\
        &= \frac{\hat{\mu}_{\beta, k}^{S_1 \cap S_2} (S_1) \cdot \hat{\mu}_{\beta, k}^{S_1 \cap S_2} (S_2)}{\hat{\mu}_{\beta, k}^{U} (S_1)} \cdot \frac{1}{k - \ell} \cdot \hat{\mu}_{\beta, k}^{U}(S_1 \cap S_2) \\
        &= \frac{1}{k - \ell} \cdot \hat{\mu}_{\beta, k}^{S_1 \cap S_2} (S_2) \\
        &= \plusDownUp_{\beta, k}^{U}(S_1, S_2) .
    \end{align*}
    Similarly, we have $\frac{\E[\pi_{k-1}(S_1) \pi_{k-1}(S_1) \mid U_{\ell} = U]}{\hat{\mu}_{\beta, k}^{U} (S_1)} = \plusDownUp_{\beta, k}^{U}(S_1, S_1)$, and it is easily checked that $\plusDownUp_{\beta, k}^{U}$ is reversible with respect to $\hat{\mu}_{\beta, k}^{U}$.
    Applying the same computations as in the proof of Proposition 19 in \cite{chen2022localization} yields
    \[
        \gap(\plusDownUp_{\beta, k}^{U}) = \inf_{g: \Omega_k \to \R} \frac{\E[\Var_{\pi_{k-1}}(g) \mid U_{\ell} = U]}{\Var_{\hat{\mu}_{\beta, k}^{U}}(g)} 
        \le \frac{\E[\Var_{\pi_{k-1}}(f) \mid U_{\ell} = U]}{\Var_{\hat{\mu}_{\beta, k}^{U}}(f)},  
    \]
    which proves the claim.
\end{proof}

\subsection{Simplicial Complexes and Proof of \Cref{claim:k_l_walk}}
For proving \Cref{claim:k_l_walk}, we first need to introduce some background on simplicial complexes.
We first present some useful definitions and lemmas about simplicial complexes.

Recall the setting of given in the proof of \Cref{thm:spectral_gap}.
For the rest of the section, we fix some $2 \le k \le n$ and $\pi \in \mathcal{P}(\Omega_k)$. 
To avoid trivialities, assume that $\pi$ is supported on all of $\Omega_k$.
If the latter does not apply, we can restrict $\Omega_m$ for $0 \le m \le k$ to such sets $U \in \binom{V}{m}$ that satisfy $\pi(U)>0$. 
For a given integer $\ell < k$, we are interested in the spectral gap of $Q_{\pi, \ell}$, the natural $(k, \ell)$-down-up walk on $\Omega_k$ associated with $\pi$. 
It is easily checked that $Q_{\pi, \ell}$ is reversible with respect to $\pi$.

A frequent way to analyze the spectral gap of $Q_{\pi, \ell}$ is by studying it as a walk on the maximum faces of a suitable \emph{weighted simplicial complex} $\complex$ on the ground set $V$. We define the simplicial complex by $\complex = \bigcup_{m = 0}^{k} \Omega_m$.
The maximum faces are given by the sets in $\Omega_k$ and are weighted according to $\pi$.
Then $Q_{\pi, \ell}$ is exactly the $(k, \ell)$-down-up walk on $\complex$, as studied in, for example, \cite{alev2020improved,chen2021optimal}.

The main advantage of studying the $(k, \ell)$-down-up walk in the context of simplicial complexes is that we can use various \emph{local-to-global theorems} that relate its spectral gap with the second largest eigenvalues of a family of local walks.
Given a set $U \in \Omega_m$ for some $0 \le m \le k-2$, the \emph{local walk} at $U$ is defined as the random walk on $V \setminus U$ with transition matrix
\[
    Q^{U}_{\pi}(u, v) = \begin{cases}
        0 &\text{ if } u = v \\
        \frac{1}{k - \size{U} - 1} \cdot \pi^{U}(v \mid u) &\text{ otherwise }
    \end{cases} .
\]
We have the following notion of local spectral expansion.
\begin{definition} \label{def:local_expansion}
    We say that $\pi$ satisfies $(\zeta_0, \dots, \zeta_{k-2})$\emph{-local spectral expansion} if for all $0 \le m \le k-2$ and all $U \in \Omega_m$ the second-largest eigenvalue of the local walk $Q^{U}_{\pi}$ is at most $\zeta_m$.
\end{definition}

Deriving the spectral gap of the of the $(k, \ell)$-down-up walk from local spectral expansion can for example be done by the following theorem.

\begin{theorem}[\hspace{1sp}{\cite[Fact A.8, Theorem A.9]{chen2021optimal}}] \label{thm:gap_from_expansion}
    Suppose $\pi$ satisfies $(\zeta_0, \dots, \zeta_{k-2})$-local spectral expansion.
    For every $0 \le m \le k-1$ define $\Gamma_m = \prod_{j=0}^{m-1} \frac{1 - \zeta_j}{1 + \zeta_j}$ (in particular, $\Gamma_0 = 1$).
    For all $0 \le \ell \le k-1$ it holds that
    \[
        \gap(Q_{\pi, \ell}) \ge \frac{\sum_{m = \ell}^{k-1} \Gamma_m}{\sum_{m = 0}^{k-1} \Gamma_m} .
    \]
\end{theorem}

For various types of distributions, it has been shown that local spectral expansion is closely related to spectral independence.
This relationship is particularly well-studied for grand-canonical ensembles, in which case a different way of constructing the simplicial complex is required (see for example \cite{chen2021optimal,chen2023rapid,anari2021spectral}).
The following statement can be seen as a canonical ensemble version of statement of Theorem 1.5 in \cite{anari2021spectral} and is proven similarly.
\begin{lemma} \label{lemma:SE_from_SI}
    Let $U \in \Omega_m$ for some $0 \le m \le k-2$. If $\pi^U$ satisfies $C$-spectral independence for some constant $C$, then the second-largest eigenvalue of $Q^{U}_{\pi}$ is at most $\frac{C - 1}{k - \size{U} - 1}$.
\end{lemma}

\begin{proof}
    We start by noting that $Q^{U}_{\pi}$ is reversible with respect to $\hat{\pi}(i) = \frac{1}{k - \size{U}} \pi^{U}(i)$ for $i \in V \setminus U$.
    Hence, we know that the right eigenvectors of $Q^{U}_{\pi}$ are pairwise orthogonal with respect to the inner product
    \[
        \langle f, g \rangle_{\hat{\pi}} = \sum_{v \in V \setminus U} f(v) g(v) \hat{\pi}(v) \text{ for } f,g: V \setminus U \to \R .
    \]
    Next, we consider the linear map 
    \[
        \hat{Q} f = \frac{k - \size{U} - 1}{k - \size{U}} Q^{U}_{\pi} f - \langle f, \pmb{1} \rangle_{\hat{\pi}} \text{ for } f: V \setminus U \to \R.
    \]
    where $\pmb{1}$ denotes the all 1's vector.
    Since $Q^{U}_{\pi}$ is stochastic, we know that $\pmb{1}$ is an eigenvector of $Q^{U}_{\pi}$ associated with the eigenvalue of $1$.
    Now, suppose $f \neq \pmb{1}$ is an eigenvector of $Q^{U}_{\pi}$.
    Since $\langle f, \pmb{1} \rangle_{\hat{\pi}} = 0$, we have $\hat{Q} f = \frac{k - \size{U} - 1}{k - \size{U}} \lambda f$.
    Next, observe that 
    \[
        \hat{Q} f (u) = \frac{1}{k - \size{U}} \sum_{v \in V \setminus U} f(v) \pi^U(v \mid u) \ind{v \neq u} - \frac{1}{k - \size{U}} \sum_{v \in V \setminus U} f(v) \pi^U(v) = \frac{1}{k - \size{U}} \cdot (M_{\pi^U} f - f)(u), 
    \]
    where $M_{\pi^U}$ is the pairwise influence matrix of $\pi^U$ as in \Cref{def:influence}.
    Hence, $f$ must also be an eigenvector of $M_{\pi^U}$ with eigenvalue $(k - \size{U} - 1) \lambda + 1$. 
    Assuming $\pi^U$ satisfies $C$-spectral independence now yields $\lambda \le \frac{C - 1}{k - \size{U} - 1}$, which concludes the proof. 
\end{proof}

We now prove \Cref{claim:k_l_walk}. We recall the statement here.

    \begin{customclaim} {\bf \ref{claim:k_l_walk}}
        In the setting of the proof of \Cref{thm:spectral_gap}, it holds that there is some constant $C > 0$, only depending on $\beta$, $\Delta$ and $\gamma$, such that
        \[
            \gap(\plusDownUp_{\ell, \beta, k}) \ge C .
        \]
    \end{customclaim}

\begin{proof}[Proof of \Cref{claim:k_l_walk}]
    Recall that $\plusDownUp_{\ell, \beta, k}$ corresponds to the $(k, \ell)$-down-up walk $Q_{\pi, \ell}$ on $\Omega_k$ for $\pi = \hat{\mu}_{\beta, k}$.
    Further, recall that we assume $k \coloneqq \gamma n > \alpha n$ and $\ell = k - \lfloor \alpha n \rfloor$ for some $\alpha = \alpha(\Delta, \beta) > 0$.
    We aim for applying \Cref{thm:gap_from_expansion} to lower-bound the spectral gap of $\plusDownUp_{\ell, \beta, k}$.
    To this end, set $\hat{\ell} = k - \lceil \frac{\alpha}{2} n \rceil$ and note that by \Cref{thm:spectral_independence_fm}, there is some constant $C = C(\Delta, \beta, \gamma)$ such that for all $U \subset V$ with $\size{U} \le \hat{\ell}$, it holds that $\hat{\mu}_{\beta, k}^{U}$ satisfies $C$-spectral independence.
    It follows from \Cref{lemma:SE_from_SI} that $\hat{\mu}_{\beta, k}$ satisfies $(\zeta_0, \dots, \zeta_{k-2})$-spectral expansion for
    \[
        \zeta_m \coloneqq \begin{cases}
            \frac{C - 1}{k - m - 1} \text{ if } m \le \hat{\ell} \\
            1 \text{ otherwise }
        \end{cases}.
    \]
    Next, define $\Gamma_m$ for $0 \le m \le k-1$ as in \Cref{thm:gap_from_expansion}.
    Observe that trivially $\sum_{m = 0}^{k - 1} \Gamma_m \le k = \gamma n$.
    Further, note that $m \mapsto \Gamma_m$ is non-increasing and therefore $\sum_{m = \ell}^{k - 1} \Gamma_m \ge (\hat{\ell} - \ell + 2) \Gamma_{\hat{\ell} + 1} \ge \Gamma_{\hat{\ell} + 1} \frac{\alpha}{2} n$. 
    Hence, the claim follows from applying \Cref{thm:gap_from_expansion} once we prove that $\Gamma_{\hat{\ell} + 1}$ is bounded below by a positive constant that only depends on $\Delta$, $\beta$ and $\gamma$.
    Using the fact that $\ln(1 - x) \ge - \frac{1}{1 - x}$ for all $x < 1$, we get
    \[
        \ln \Gamma_{\hat{\ell} + 1} 
        = \sum_{j = 0}^{\hat{\ell}} \ln\left( 1 - \frac{2 \zeta_j}{1 + \zeta_j}\right) 
        \ge - \sum_{j = 0}^{\hat{\ell}} \frac{2 \zeta_j}{1 - \zeta_j}
        \ge - \sum_{j = 0}^{\hat{\ell}} \frac{2 (C - 1)}{k - j - C} 
        \ge - \frac{2 (C - 1) (\hat{\ell} + 1)}{k - \hat{\ell} - C}.
    \]
    Since $\hat{\ell} \in O(n)$ and $k - \hat{\ell} \ge \frac{\alpha}{2} n$, this proves the claim.
\end{proof}

\section{Proofs for slow mixing}
\label{appendix:slow}

\subsection{Tree Recursions and Non-Reconstruction}

\begin{customprop}{\bf \ref{thm:fixpt}}
    For $\beta > \beta_u$, the following hold:
    \begin{enumerate}
        \item If $|\log \lam| > \log \lam_u$, then $\eqref{eqTreeRecursion}$ has a unique fixed point. It is stable and hence corresponds to the global maximizer of $f_{\Delta,\beta,\lam}$.  This maximizer is $\eta^+_{\Delta,\beta,\lam} = \eta^-_{\Delta,\beta,\lam}$.
        \item If $|\log \lam| = \log \lam_u$, then \eqref{eqTreeRecursion} has two distinct fixed points, one of which is stable and corresponds to the global maximizer of $f_{\Delta, \beta, \lam}$. The other corresponds to an inflection point of $f_{\Delta, \beta, \lam}$. 
        \item If $|\log \lam| < \log \lam_u$, then $\eqref{eqTreeRecursion}$ has three distinct fixed points. The largest and the smallest are both stable, corresponding to the only two local maxima of $f_{\Delta,\beta,\lam}$.  When $\lam >1$, $\eta^+_{\Delta,\beta,\lam}$ is the unique global maximizer; when $\lam<1$, $\eta^-_{\Delta,\beta,\lam}$ is the unique global maximizer; when $\lam=1$ then $\eta^+_{\Delta,\beta,\lam},\eta^-_{\Delta,\beta,\lam}$ are both global maximizers.
    \end{enumerate}
\end{customprop}

\begin{proof} The number of fixed points for the three different regimes of $\lam$ is a classical result that can be found in, e.g., \cite[Lemma 12.27]{georgii2011gibbs}.

    Let
    $$h(x) = \frac{\lam(x e^\beta + 1)^{\Delta-1}}{(x + e^{\beta})^{\Delta-1}} \,.$$
    By rearranging and substituting $d = \Delta - 1$ and $r = x^{1/d}$, we can see that the fixed points of the recursion are exactly the roots of the polynomial
    $$\hat{h}(r) = r^{d+1} - \lam^{1/d}e^\beta r^d + e^\beta r - \lam^{1/d} \,.$$
By Descartes' Rule of Signs, $\hat{h}(r)$ must have either 1 or 3 positive real roots (counted with multiplicities). In the case of (1), then, it is clear that the unique fixed point of $h(x)$ must be stable; else, it would contribute two to the roots of $\hat{h}(r)$.

In the case of (2), we must have fixed points $x_0 \neq x_1$ such that $h'(x_0) = 1$ and $h'(x_1) < 1$. By Theorem~\ref{thm:fixpt-correspondence}, both $x_0$ and $x_1$ correspond to critical points of $f_{\Delta, \beta, \lam}$, the latter being a local maximum. If $x_0$ also corresponded to a local maximum of $f_{\Delta, \beta, \lam}$, we would necessarily have a local minimum corresponding to some fixed point $x \notin \{x_0, x_1\}$, which is a contradiction, and similarly if $x_0$ corresponded to a local minimum. Hence, it must correspond to an inflection point.

In the case of (3), we utilize computations from the proof of \cite[Proposition 2.4]{guolu2018}. Specifically, we consider    
    $$h'(x) = \frac{d(e^{2\beta}-1)h(x)}{(e^{\beta}x+1)(x+e^{\beta})}.$$
    \cite{guolu2018} shows that there exist $0 < x_0 < x_1$ (defined in terms of $\beta$ and $\Delta$) that satisfy $\frac{d(e^{2\beta}-1)x}{(e^{\beta}x+1)(x+e^{\beta})} = 1$, and in particular $\frac{d(e^{2\beta}-1)x}{(e^{\beta}x+1)(x+e^{\beta})} > 1$ if and only if $x_0 < x < x_1$. Moreover, in the non-uniqueness regime for $\lam$, we must have $h(x_0) < x_0$ and $h(x_1) > x_1$. 

    Since $h(0) = \frac{\lam}{e^{\beta d}} > 0$, there must be a fixed point in the interval $(0, x_0)$ and since $\lim_{x \to \infty} h(x) = \lam e^{d\beta} < \infty$, there must also be a fixed point in the interval $(x_1, \infty)$. By the above observations, both must be stable fixed points. 

    The nature of the maximizers of $f_{\Delta, \beta, \lam}$ follows from Theorem~\ref{propIsingTree}.
\end{proof}

\end{document}